\documentclass[article]{elsarticle}

\usepackage{lineno}
\usepackage{amssymb}
\usepackage{cool}
\usepackage{float}
\usepackage{caption}
\usepackage{subcaption}
\usepackage{ragged2e}
\usepackage{booktabs}
\usepackage{soul}
\usepackage{color} 
\usepackage[normalem]{ulem}
\usepackage[hidelinks]{hyperref}
\journal{Journal of Fluids \& Structures}

\begin{document}

\begin{frontmatter}



\title{Investigation of a submerged fully passive energy-extracting flapping foil  operating in sheared inflow}
%\title{Investigation of a submerged  passive energy-extracting hydrofoil  operating in sheared inflow}


\author[inst1]{Konstantinos Theodorakis}
\ead{kostantinostheo1@gmail.com}
\author[inst1]{Dimitris Ntouras}
\ead{ntourasd@fluid.mech.ntua.gr}
\author[inst1]{George Papadakis\corref{mycorrespondSparse Matrices Version, Tech. rep., Department of Computer Science &ingauthor}}
\ead{papis@fluid.mech.ntua.gr}
\affiliation[inst1]{organization={National Technical University of Athens, School of Naval Architecture \& Marine Engineering},%Department and Organization
            addressline={9, Heroon Polytechniou Str}, 
            city={Athens},
            postcode={15780}, 
            country={Greece}}

\cortext[mycorrespondingauthor]{Corresponding author}
\begin{abstract}
 In this work a fully passive energy harvesting foil is studied computationally. An in-house $2^{nd}$ order finite volume CFD solver, MaPFlow, is strongly coupled with a rigid body dynamics solver to investigate the foil operation under uniform and sheared inflow conditions with/without free surface. The mesh follows the airfoil motion using a radial basis function (RBF) mesh deformation approach. Initially, MaPFlow predictions are compared to experimental and numerical results available in the literature, where reasonable agreement is found. Next, one-phase simulations are considered for linearly sheared inflow for various shear rates. Results suggested that foil performance can be enhanced under sheared inflow conditions. Finally, two-phase simulations taking into account the free surface, for both uniform and sheared inflow, are considered. Predictions indicate a significant deterioration in performance of the system when the foil operates under the free surface due to the interaction of the shed vorticity with the free surface. 
\end{abstract}

%%\item Research highlight 1
%%\item Research highlight 2


\end{frontmatter}

\section{Introduction}
Humans have been {\color{black} drawing inspiration} from nature since the dawn of civilization. Nature follows an evolution{\color{black}ary} path where{\color{black}by} only the species with optimal characteristics  survive in this ever changing competitive and hostile environment. {\color{black}In particular,} sea animals exhibit an interesting mechanism of underwater thrust production, {\color{black}which facilitates either their self-propulsion or the harvest of kinetic energy from incoming flows to enhance their forward motion.}  Their secret lies in {\color{black}their ability to exploit} the incoming flow to produce vortex pairs which shed behind them \cite{Anderson1998}, \cite{Wang2000}, \cite{Triantafyllou2005}. This vortex formation creates a {\color{black}downstream jet}  and as a result generates thrust. 
%Vorticity can be either existent in the flow (for example in fish herds vorticity is produced from fish in the first rows) or is produced by a suitable bending and fluttering of the animals body \cite{Zhu2002}, \cite{Liao2003}. 
This ingenious idea of propulsion {\color{black}challenges} the use of traditional means of propulsion such as the screw propeller, since the propulsive efficiency of {\color{black} sea mammals} locomotion (for example whales) can reach up to 85\% \cite{Bose1989}. However the flapping foil motion does not limit its use to thrust production. As a physical mechanism it features the ability to transform energy from one form (kinetic energy of vortices) to another which can be {\color{black}either} thrust or energy harvest. Energy harvesting has attracted {\color{black}extensive}  attention since ocean and river currents {\color{black}present a unique avenue for exploring options of} feasible renewable energy production. 
	%%%%%%%%%%
	%%%%%%%%%%% Active Foil
	%%%%%%%%%%
Early implementations for energy harvesting \cite{McKinney1981} used an active foil with two degrees of freedom (DOF) in the heave and pitch directions in order to execute a sinusoidal motion. For such {\color{black} design}, the pitching axis location is a key parameter for the synchronization of the vortices according to \cite{KinseyDumas2008} while high efficiencies of up to $40\%$ were recorded in this active foil set up \cite{Kinsey2011}. However the mechanical complexity and maintenance costs of the actuators and the kinematic constraints constitute a major challenge for energy harvesting using active control. \par
	
	%%%%%%%%%%
	%%%%%%%%%%% Semi-passive Foil
	%%%%%%%%%%
	
The idea of a semi-passive foil was then introduced by \cite{Shimizu2008} and \cite{Peng_and_Zhu2009} where only the pitch was actively controlled and heave was left unconstrained. A damper was attached to the heave motion to allow for {\color{black}the recovery of} energy through its motion. This approach had the ambition of simplifying the mechanical system while preserving an efficiency comparable to {\color{black}that of} the fully active foil. A physical, full-scale model was tested in \cite{Stingray2002}, however {\color{black}it once again resulted } in high maintenance cost{\color{black}s}. 
	
	%%%%%%%%%%
	%%%%%%%%%%% Passive Foil
	%%%%%%%%%%
	
{\color{black}In an effort} to overcome the above challenges, in \cite{Zhu_flow_induced2009} they attempted numerical simulations on a fully-passive flapping foil, employing dampers and springs in both {\color{black} DOF's}. Four different motion regimes emerged depending on the configuration of the mass, springs and dampers. It was found that with the right tuning of those parameters a periodic motion could be achieved with high energy content in the heave direction and only minimum energy consumption in the pitch direction. 
{\color{black} Critical for the periodic motion of the airfoil is the Leading Edge Vortex (\textit{LEV}) detaching from the foil at large angle of attack (AoA). The \textit{LEV} constitutes a low pressure zone and thus creates an aerodynamic moment acting towards changing the direction of the pitch angle.}

%\st{The self-induced oscillations are initiated due to the static instability of the foil, owning to the the misalignment of the pitching axis pivot point P and the CoG}. The self-sustained oscillations are preserved due to the Leading Edge Vortex (\textit{LEV}) detaching from the foil owing to the deep dynamic stall created at large pitch angles. \st{at the upper and bottom position of the heave motion.} {\color{black}This \textit{LEV} constitutes a low pressure zone and thus creates an aerodynamic moment acting towards changing the direction of the pitch angle}. 
%The motion developed was not only self-sustained but also self-induced even when no \st{initial} {\color{black} external} perturbation was present \cite{Veilleux2017}. {\color{black} When the foil is positioned parallel to the incoming flow, the torsional spring moment forces the foil to a slight pitch deviation in order to reach an equilibrium position. However due to the static instability of the foil the pitch angle continues to grow until the occurence of a deep dynamic stall.}  
	
	%%%%%%%% Passive foil optimization
	
Since structural parameters of a fully-passive foil are the only way to indirectly control its behavior and thus its hydraulic efficiency, various optimization efforts have been carried out employing both experiments and numerical simulations. 
	%The parametric space in each effort is of primal importance since some parameters may be neglected or thoroughly examined. 
In \cite{Dumas2017} they attempted an extensive numerical optimization research, focusing on mass, spring and damper properties of a NACA0015 foil while preserving the pitching axis constant at one third of the chord downstream of the Leading Edge. A different approach was followed by \cite{Duarte2019} where an experimental optimization was carried out. In this case, a fully-passive foil was constructed and the parametric space was focused on the pitching axis position and the pitching stiffness.
	
Expanding the research on passive foils, recent studies have concentrated {\color{black}on} more realistic operating environments in order to assess the feasibility of energy extraction {\color{black}at} a practical level. Simulations at a higher Reynolds number should be encouraged, since it has been observed that efficiency increases in higher Reynolds numbers \cite{Simpson2008}. {\color{black} More recently, the influence of the free--surface effects were examined for low \cite{Deng2022} and high Reynolds number \cite{Zhu2022}. In these publications parametric studies were conducted with respect to the submergence depth and the Froude number. Their results indicate that when the foil was operating near the free--surface the energy harvesting efficiency was reduced, while the increasing Froude number appear to have a favorable role.}  Additionally, in \cite{Liu2013}, \cite{Shoele2013}  deformable foil shapes are considered, introducing energy exchange between the foil deformation and the fluid with results {\color{black}demonstrating} that large deformations may also increase efficiency. On the other hand 3D simulations of the flow field have predicted a drastic decrease in efficiency in cases where the AR of the foil decreases and the flow diverges from the 2D ideal case \cite{Simpson2008}. An overestimation of the efficiency is thus calculated by 2D simulations and researchers are encouraged to bear that in mind. \par

An important factor to be addressed in the installation of a passive foil in shallow waters is the non uniform velocity profile of the inflow. Researchers have idealized this profile as a linear shear profile leading to more simplistic simulations. 
 Such a simple case of linear shear has been examined for all three cases of a flapping foil : active \cite{Cho2014}, semi-active \cite{Liu2019} and fully passive foil \cite{Zhu2012}. Interesting conclusions have arisen indicating that the parametric region in which energy extraction is possible is growing when mild shear is present. In larger shear gradients this region shrinks and eventually collapses \cite{Zhu2012}. Although efficiency and power coefficients decrease when operating in a shear flow, \cite{Zhu2012} demonstrated that similar quality of performance can also be achieved from shear flows, even {\color{black}noting} an increase of 9\% in some cases. {\color{black}Nonetheless}, free surface and how it affects performance and efficiency is not taken into account in  most of the studies available in the literature.

 {\color{black}In the present work, the effect of the sheared inflow on the operation of a passive foil} is investigated. A lightweight fully passive foil is considered,  operating in sheared inflow with and without the free surface. Initially, uniform inflow is considered and {\color{black}a} comparison is made with {\color{black}the} available  experimental and numerical results from \cite{Duarte2021}. Afterwards,  we investigate  how sheared inflow affects the passive foil performance. Finally, we consider a fully passive hydrofoil operating in uniform and sheared inflow  near the free surface region. In that case, apart from the generated waves, the shed vorticity interacts with the free surface effectively altering the performance of the flapping foil. 
 
The structure of the present work is as follows. In \autoref{sec:num} the in-house CFD solver MaPFlow is described. In \autoref{sec:numsetup} the basic modelling parameters are described. {\color{black} Section \ref{sec:res_disc} presents the results of the study. Specifically,} in  \autoref{sec:Duarte_case} the numerical methodology employed is compared against available data in the literature for uniform inflow conditions. In \autoref{sec:shear} the case of the fully passive foil is examined in sheared inflow  for various shear slopes. In \autoref{sec:free} two-phase simulations are considered{\color{black},} taking into account the {\color{black} presence of the} free-surface. Uniform and sheared inflow are investigated and a drop in  performance of the foil operating in such conditions is observed. Finally, Section  \ref{sec:conclusions} {\color{black}presents} the basic conclusions {\color{black}of this examination}.
	
\section{Numerical Methodology \label{sec:num}}
Numerical investigation of Fluid Structure Interaction (FSI) problems is performed by combining two separate computational algorithms. Firstly, a flow solver is utilized to describe the fluid motion, and secondly a dynamic solver that computes the structure's response under the flow excitation. In the present work, results from both one-- and two-- phase flows are presented. The next paragraph {\color{black}considers} the general case of  unsteady incompressible flow of two immiscible fluids. The governing equations and the discretization process are presented. Afterwards, the details of the dynamic solver are described and the coupling of the two solvers is discussed. 
This methodology is implemented as part of the CFD code MaPFlow. The in-house code is developed in NTUA  \cite{Papadakis2014},\cite{Diakakis2019}, {\color{black}and has proved capable of handling} both compressible and purely incompressible flows on arbitrary polyhedral meshes. The code is able to perform in a multi-processing environment utilizing the MPI protocol, while the grid partitioning is performed using the Metis Library \cite{Karypis2013}.
\subsection{Governing equations}
 Two-phase incompressible flows are solved using the artificial compressibility method (ACM) \cite{Chorin1967} , coupled with the Volume of Fluid (VoF) approach \cite{Hirt1981}. The ACM solves the unsteady system of equations by utilizing the dual--time stepping technique \cite{JAMESON1991}, where at each real time iteration a pseudo--steady state problem is solved. This is accomplished by augmenting the original unsteady system of equations by pseudo--time derivatives of the unknown variables. Convergence is accomplished once these derivatives approach to zero and thus the original system of equations is obtained. ACM assumes a relation between the pressure and the density field during pseudo time. The blending is performed by introducing a numerical parameter $\beta$, that is $\frac{\partial\rho}{\partial p}\big|_\tau=\frac{1}{\beta}$. This free parameter for typical free surface flows takes values between 5 and 10 \cite{Ntouras2020},\cite{Dudley2002}. The governing system of equations is described by Equation \eqref{int_govEqs}.
 \begin{equation}
        \Gamma \frac{\partial}{\partial \tau} \int_{D_{i}} \vec{Q}\mathrm{d}D  + 
        \Gamma_{e} \frac{\partial }{\partial t} \int_{D_{i}} \vec{Q}\mathrm{d}D + 
        \int_{\partial D_{i}} \left( \vec{F}_{c} - \vec{F}_{v}\right) \mathrm{d}S 
        = \int_{D_{i}} \vec{S}_{q} \mathrm{d}D
        \label{int_govEqs}
\end{equation}

The above system of equations expresses the change of the primitive variables $\vec{Q}$ inside a control volume $D_i$ with boundary $\partial D_i$, in time $t$. The vector $\vec{Q}=[p,\vec{\upsilon},\vec{\alpha_l}]^T$, includes the pressure $p$, the 3-dimensional velocity vector $\vec{\upsilon}$, and the volume fraction $\alpha_l$. The volume fraction $\alpha_l$ indicates the presence of either the liquid phase with density $\rho_l$, or the presence of the gaseous phase with density $\rho_g$. Using the volume fraction the density of the mixture can be found as $\rho_m = \alpha_l\rho_l+(1-\alpha_l)\rho_g$.\par
Although the system of equations is casted 
in primitive form, in order to advance the solution in time, the conservative form of the equations is used. The variable transformation between the primitive variables $\vec{Q}$ and the conservative variables $U=[0,\rho_m\vec{\upsilon},\alpha_l]^T$, is performed using the transformation matrix $\Gamma_e$, which is given in Equation \eqref{Matrices_G}.\par
 Applying the standard ACM to multi-phase flows, especially when large density ratios are accounted, the system of equations becomes poorly conditioned \cite{Venkateswaran2001}. This happens because the eigenvalues of the system scale with local density of the flow. In order to alleviate this behavior the preconditioning matrix $\Gamma$ of Kunz \cite{Kunz2000} is used to re-scale the fictitious time derivatives and allow for the efficient time marching of the solution.

\begin{equation}
	\begin{array}[h]{ccc}
		\Gamma =
		\begin{bmatrix}
			\frac{1}{\beta\rho_{m}} & 0 & 0\\
			0 & \rho_{m} I_{3\times 3} & \vec{\upsilon} \Delta \rho\\ 
			\frac{\alpha_{l}}{\beta\rho_{m}} & 0 & 1
		\end{bmatrix}
		&,&\Gamma_{e} =
		\begin{bmatrix}
			0 & 0 & 0 \\
			0 &\rho_{m} I_{3 \times 3}& \vec{\upsilon} \Delta \rho\\
			0 & 0 & 1
		\end{bmatrix}
	\end{array}
	\label{Matrices_G}
\end{equation}

In Equation \eqref{Matrices_G}, $\Delta \rho$ is the difference between the densities of the liquid and the gas, $\Delta \rho=\rho_l-\rho_g$ and $I_{3x3}$ is the 3 by 3 identity matrix.\par
The surface term of the integral equation includes the convective fluxes $\vec{F}_c$ and the viscous fluxes $\vec{F}_v$. Both vectors are presented in Equation \eqref{vec_fluxes}. In this Equation, $V_n=\vec{\upsilon}\cdot\vec{n}$ is the fluid velocity projected onto the surface normal $\vec{n}=(n_x,n_y,n_z)$ and $\Delta V$ is the difference between the velocity $V_n$ and the projected grid velocity $V_g=\vec{\upsilon}_{vol}\cdot\vec{n}$.

\begin{equation}
	\begin{array}[h]{ccc}
		\vec{F}_{c} = 
		\begin{bmatrix} 
		{V}_{n} \\ \rho_m u \Delta V + pn_x \\ \rho_m v \Delta V + pn_y\\ \rho_m w \Delta V + pn_z \\ \alpha_l \Delta V
		\end{bmatrix} &, &
		\vec{F}_{v} = 
		\begin{bmatrix}
		0 \\
		\tau_{xx}n_x +\tau_{xy}n_y +\tau_{xz}n_z \\
		\tau_{yx}n_x +\tau_{yy}n_y +\tau_{yz}n_z \\
		\tau_{zx}n_x +\tau_{zy}n_y +\tau_{zz}n_z \\
		0 \\
		\end{bmatrix}
	\end{array}
	\label{vec_fluxes}
\end{equation}

The viscous stresses $\tau_{ij}$, using the Boussinesq approximation are computed as
\begin{equation}
	\tau_{ij} = \left( \mu_{m}+\mu_{t} \right) \left( \frac{\partial u_i}{\partial x_j} + \frac{\partial u_j}{\partial x_i} \right) - \frac{2}{3}\rho_m\delta_{ij}k
	\label{viscStress}
\end{equation}
where $\mu_m$ is the viscosity of the mixture, $\mu_t$ is the turbulence viscosity, $k$ is the turbulent kinetic energy and $\delta_{ij}$ is the Kronecker's symbol.\par
For the turbulence closure, the k-$\omega$ SST model of Menter is employed \cite{Menter1994}.
In case of free surface flows, it has been noted that the turbulence models tend to overproduce turbulence viscosity in the vicinity of the free surface \cite{Larsen2018},\cite{Kamath2015}. In order to suppress the turbulent viscosity near the free surface Devolder  et al. \cite{Devolder2017} introduced a source term in the equation of the turbulent kinetic energy. This source term is activated near the free surface and scales with the local viscosity and the gravity vector.
%\begin{equation}
%	G_{b} = -\frac{\nu_{t}}{\sigma_{t}}\frac{\partial \rho}{\partial x_{j}}g_{j}
%	\label{buoyTerm}
%\end{equation}

In order to perform simulations of numerical wave tanks, the lateral boundaries of the domain are equipped with damping zones that absorb any outgoing perturbation and make sure that no reflections occur due to the boundary conditions. 
In MaPFlow, this is performed by defining forcing zones near the boundaries of the computational domain. Forcing zone technique drives the numerical field to the desired solution by adding source terms to the governing equations. In MaPFlow, source terms are added only to the momentum equations. The form of the source terms is presented in Equation \eqref{SUnwt}. The damping is performed by eliminating the vertical component of the velocity vector. that is in a 2D simulation $\vec{\upsilon}_{tar}=(u,0)$, where u is the local velocity in the x-direction.
\begin{equation}
	\vec{S}_{nwt}= C_{nwt}\rho_{m}\left( \vec{\upsilon}-\vec{\upsilon}_{tar} \right)
	\label{SUnwt}
\end{equation}

The coefficient $C_{nwt}$ is used to smoothly variate the influence of the forcing terms from the start $x_s$ of the forcing zone to its end $x_e$, at the boundary of the computational domain. The smooth transition is regulated by the factor $\alpha_{nwt}$ and a function $f_{nwt}$, which is defined inside the forcing zones--see Equation \eqref{Cnwt}.
\begin{equation}
	C_{nwt}	= \alpha_{nwt} f_{nwt}(x_r)\:, \:\: x_{r}=\frac{x_{s}-x}{x_{s}-x_{e}}
	\label{Cnwt}
\end{equation}

In \cite{Ntouras2020} and \cite{Peric2015} the influence of the various parameters of the coefficient is examined. In the present work, an exponential form of the $f_{nwt}$ is chosen.
\begin{equation}
   f_{nwt}(x_r) = \frac{\exp\left( x_{r}^{n} \right)-1}{\exp\left( 1 \right)-1}
   \label{Fnwt}
\end{equation}
\subsection{Discretization}
The discretization of the equations is performed using the finite volume method. Given a computational mesh at the geometric center of each cell a control volume $D_i$ is defined, with its boundaries being the faces (or the edges in 2D) of the cell, $\partial D_i=\sum_{f=1}^{N_f}\Delta S_f$, with $N_f$ being the total number of the cell's faces and $\Delta S_f$ the area of the face $f$. The surface terms of the equations are approximated using the midpoint rule, while the volume integrals are considered constant in each $D_i$. This process leads to the following definition of the spatial residual $\vec{R}_{D_i}$.
\begin{equation} 
	\vec{R}_{D_{i}} \simeq 
	\sum_{f}^{N_f}\left( \vec{F}_{c} - \vec{F}_{v} \right)\bigg|_{f} \Delta S_{f} - D_{i}\vec{S}_{q} 
	\label{RHSint}
\end{equation}

Employing the ACM, the system of equations obtains a hyperbolic form in pseudo-time. As a result, the convection terms of the incompressible equations can be evaluated by solving  a local Riemann problem at each face. MaPFlow uses the approximate Riemann solver of Roe \cite{Roe1981}, where the eigenvalues are scaled using the the preconditioining matrix of Kunz. The viscous fluxes are approximated with a second order central differentiation scheme, supplemented with a directional derivative to account for the skewness of the mesh.\par

%% Reconstruction
In two-phase flows the reconstruction schemes should be adjusted considering the specific features of the flow. In this work, the surface tension has not taken into account thus the velocity and pressure are continuous functions in space. For the velocity field, a standard piecewise linear interpolation scheme, without limiter can be used \cite{Ntouras2020}. However, the gradient of the pressure field, under the influence of the gravitation forces, appears a jump discontinuity at the interface of the two fluids. A standard second order approximation would lead to the development of the so--called "parasitic currents"\cite{Queutey2007}, near the density discontinuity. In order to remedy this behavior, the approach of Queutey et al. \cite{Queutey2007} is followed. This approach notes that, although there is a jump in the gradient of the pressure field, the gradient divided by the density field is continuous, meaning $[\nabla p]\neq 0$, but  $[\frac{\nabla p}{\rho}]=0$. Taken this into consideration, they proposed a density base interpolation scheme that follows these features of the flow.\par
Furthermore, special care must be taken to the reconstruction of volume fraction field. Due to the free surface discontinuity, the gradient of the interpolation scheme would become undetermined. One way to prevent this, is to use limiter functions, which are activated in regions of large gradients turning the scheme to a first order accurate, resulting however to excessive smearing of the free surface. Another way to approximate the field without the use of gradients, is by adopting a family of interpolation schemes that are based on the Normalised Variable Diagram of Leonard \cite{Leonard1988}. These schemes, depending of the flow characteristics, can regulate their behavior accordingly. In order to maintain stability, the can switch from a second order accurate scheme to an upwind (1st order) approximation, but under certain conditions they can switch to a downwind approximation, which would lead to an artificial compression of the discontinuity. In the present work, in two phase simulation the BICS scheme \cite{Wackers2011} is used.\par
Fluid--Structure Interaction (FSI) simulations require an effective time discretization process, that will accurate march the solution in time and will also ensure that no errors are introduced from the mesh deformation.
Specifically, in MaPFlow, the solution is marched in time implicitly by employing the dual-time stepping technique. At each real time iteration a (pseudo--) steady problem is solved. Equation $\eqref{int_govEqs}$, by introducing the finite volume technique and the definition of Equation \eqref{RHSint}, takes the following form
\begin{equation}
	\Gamma \frac{\partial \left(\vec{Q}^{*}D_{i}\right)}{\partial \tau} + \vec{R}^{*} = 0
	\label{steady}
\end{equation}
where $\vec{R}^{*}$ is the unsteady residual defined as
\begin{equation}
	\vec{R}^{*} = \vec{R}_{D_{i}} \left( \vec{Q}^{*} \right) + \Gamma_{e}\frac{\partial \left(\vec{Q}^{*}D_{i}\right)}
	{\partial t}
	\label{starR}
\end{equation}

Equation \eqref{steady} defines a steady problem that is solved iteratively at each real timestep. The iterative procedure is initialized by setting as $\vec{Q}^*=\vec{Q}^n$ and convergence is accomplished once $\dfrac{\partial \cdot}{\partial \tau}\rightarrow0$ or $\vec{R}^*\rightarrow0$, and thus the variables of the new time iteration are obtained $\vec{Q}^*=\vec{Q}^{n+1}$.\par
The unsteady term is discretized as a series expansion of successive levels backwards  in time (BDF schemes) \cite{Biedron2005}.
\begin{equation}
    \begin{aligned}
        \frac{\partial \left(\vec{Q}D_{i}\right)}{\partial t} = \frac{1}{\Delta t}\left[\varphi_{n+1}\left( D_{i}\vec{Q}\right)^{n+1} + \varphi_{n} \left( D_{i}\vec{Q}\right)^{n}\right.\\
        \left. +\varphi_{n-1}\left(D_{i}\vec{Q}\right)^{n-1}+\varphi_{n-2}\left(D_{i}\vec{Q}\right)^{n-2}+\dots \right]
	\end{aligned}
	\label{unstDiscr}
\end{equation}

When a control volume is deformed, the Geometric Conservation Law (GCL) should be satisfied. GCL is the expression of the mass conservation law applied to a constant density and velocity field. 
\begin{equation}
	\frac{d}{dt}\int_{D_{i}\left( t \right) }^{}\mathrm{d}D = \oint_{\partial D_{i}\left( t \right)}^{}\vec{u}_{vol}\cdot\vec{n}\mathrm{d}S
	\label{GCL}
\end{equation}

Using a similar discretization strategy as before, the GCL is expressed as
\begin{equation}
	\frac{1}{\Delta t} \left[\left(\varphi_{n+1}D_{i}^{n+1} + \varphi_{n}D_{i}^{n} + \varphi_{n-1}D_{i}^{n-1} +\varphi_{n-2}D_{i}^{n-2} \right) + \ldots \right]=
	\vec{R}_{GCL}^{n+1}
	\label{gclDiscr}
\end{equation}
where the residual of the GCL is defined as
\begin{equation}
	\vec{R}_{GCL}^{n+1} = 
	\sum_{f}^{N_f}\left(V_{g}\Delta S\right)_{f}^{n+1}
	\label{Rgcl}
\end{equation}

To ensure that the GCL is satisfied, Equation \eqref{gclDiscr} is applied directly to
the discretization of the unsteady term and thus Equation \eqref{unstDiscr} becomes
\begin{equation}
	\begin{aligned}
		\frac{\partial \left(\vec{Q}D_{i}\right)}{\partial t} =\vec{Q}^{n}\vec{R}_{GCL}^{n+1} +\frac{1}{\Delta t} \left[\varphi_{n+1}\left(  \vec{Q}^{n+1}-\vec{Q}^{n} \right)D_{i}^{n+1} \right.\\ 
		\left. +\varphi_{n-1}\left( \vec{Q}^{n-1}-\vec{Q}^{n} \right)D_{i}^{n-1} + \varphi_{n-2}\left( \vec{Q}^{n-2}-\vec{Q}^{n} \right)D_{i}^{n-2}+\ldots \right]
	\end{aligned}
	\label{unstDiscr3}
\end{equation}

In MaPFlow two successive levels of solution are retained, yielding a second order accurate scheme in time. The~fictitious time derivative of equation is discretized using a first-order 
backward difference~scheme
\begin{equation}
	\frac{\partial\left( \vec{Q}^{*}D_{i} \right)}{\partial \tau} = D_{i}^{n+1}\frac{\vec{Q}^{*,k+1}-\vec{Q}^{*,k}}{\Delta \tau}
	= D_{i}^{n+1}\frac{\Delta \vec{Q}^{*,k}}{\Delta \tau} 
	\label{pseudoDscr}
\end{equation}

To facilitate convergence the local time stepping technique is used.~The~local pseudo-timestep is determined by
\begin{equation}
	\Delta \tau = CFL\frac{D_{i}}{\hat{\Lambda}_{c,i}}
	\label{localDT}
\end{equation}
where $\hat{\Lambda}_{c,i}$ is the convective spectral radii and it is defined by
\begin{equation}
	\hat{\Lambda}_{c,f} = \sum_{f=1}^{N_{f}}
	\left(\big|V_{n}-\frac{V_{g}}{2}\big|+c \right)_{f} \Delta S_{f}
	\label{convSpectr}
\end{equation}
, where $c$ is the artificial sound speed and is give by
\begin{equation}
        c = \sqrt{\beta + \left( V_{n} - \frac{V_{g}}{2} \right)^{2}}
        \label{artifSS}
\end{equation}

The above discretization process results to a linear system of algebraic equations for each pseudo--time iteration. The system is solved using a Gauss-Seidel iterative method supplemented with the reverse Cuthill-Mckee reordering scheme.
\subsection{Rigid Body Dynamic Solver}

FSI problems take into account the response of the structure under the excitation of the flow. The present study focuses on 2D flows with two degrees of freedom, namely the vertical translational motion (heave) and the rotational (pitch). A structural system subjected to dynamic forces is modeled as
\begin{equation}
    M\ddot{x}(t) + C\dot{x}(t) + K x(t)= \vec{F}_{tot}(t)
    \label{MCK}
\end{equation}
where $x$ is the 2D vector of displacements, $M$ the mass matrix, $C$ the damping matrix and $K$ the stiffness matrix. The vector $F_{tot}=(F_2,M_3)$ includes the total excitation forces and moments of the system and are computed based on the flow characteristics as 
\begin{equation}
    \begin{gathered}
        F_2 = \oint_{\partial B} pn_y + \left(\overline{\tau}\cdot\vec{n}\right)n_y \mathrm{dS} \\
        M_3 = \oint_{\partial B} \left(p\vec{n} + \overline{\tau} \cdot\vec{n}\right) \times
        \vec{r}\mathrm{dS} 
    \end{gathered}
    \label{excForces}
\end{equation}

In the above equation $\partial B$ is the boundary of the body, $\overline{\tau}$ viscous stress tensor, $\vec{n}$ is the outwards pointing normal of the surface boundary and $\vec{r}$ is the vector to the center of rotation.\par
\begin{figure}[H]
\centering
\includegraphics[width=0.7\textwidth]{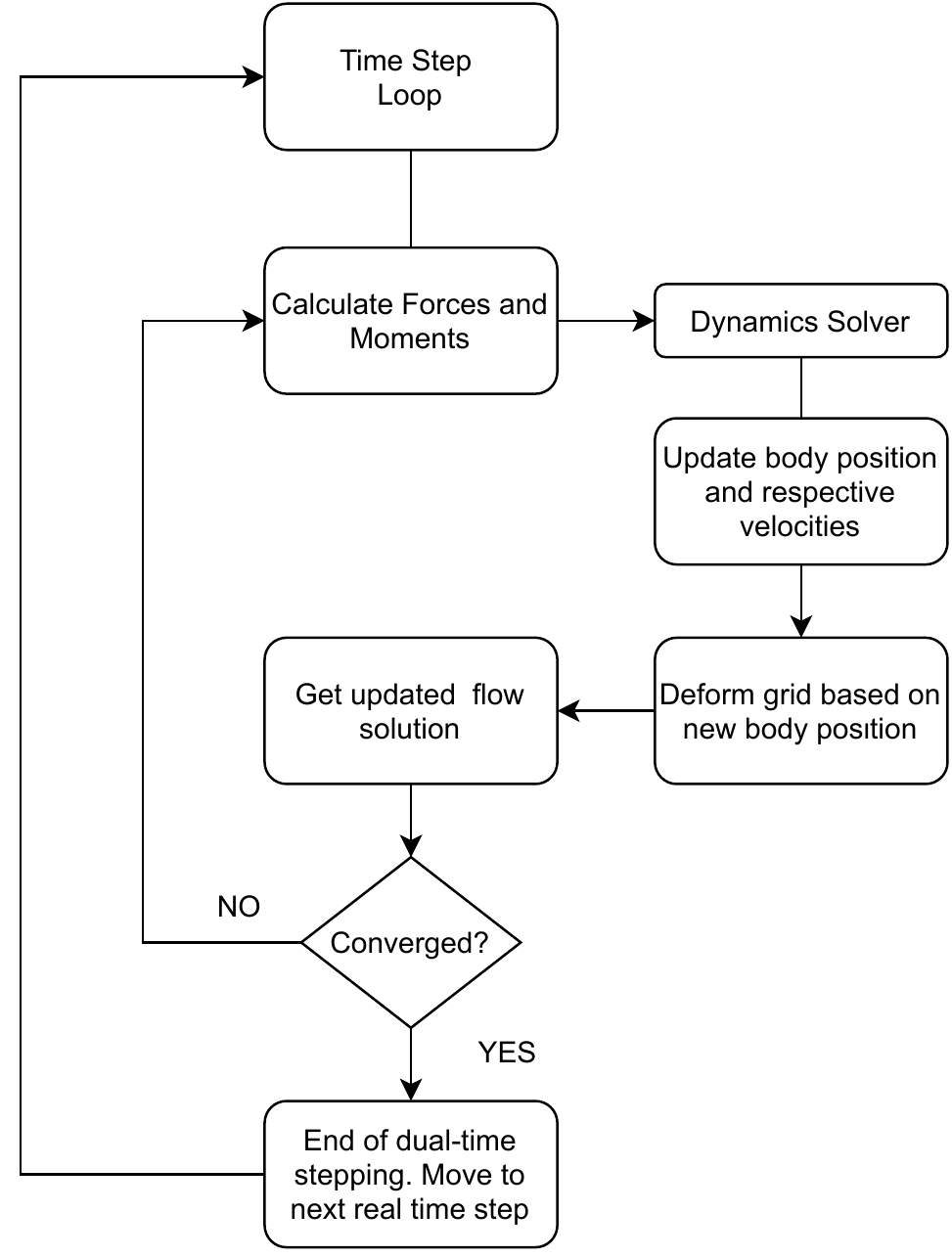}
	        \caption{Flow chart of the fluid-rigid body dynamics coupling. For each real time step, internal iterations between the dynamics and the flow solver are made to ensure a stong coupling between the two.}
	        \label{fig:rbddeform}
\end{figure}

Equation \eqref{MCK} is a non--linear equation. The linearization of the equation is performed during the pseudo-timesteps. After each pseudo--iteration, the flow solver provides the forces and moments to the dynamic one. Integrating numerically \eqref{MCK} the new position of the body is computed and the next pseudo-iteration is performed, until both solvers converge. This iterative procedure between the two solvers, implies a strong coupling  {\color{black}of} the fluid dynamics and the structural {\color{black}problems}. A schematic representation of the algorithm followed in the FSI problems is presented in figure \ref{fig:rbddeform}. The numerical integration of equation \eqref{MCK} is performed using the Newmark--$\beta$ method. Finally, the mesh follows the airfoil motion using a RBF mesh deformation technique (presented in the Appendix).

\section{Basic Numerical Setup \label{sec:numsetup}}

 Duarte et al.  in \cite{Duarte2019} constructed and tested a fully passive foil to use as a hydrokinetic turbine for low energy flows similar to currents. In this experimental test case, springs and dampers were attached to a NACA0015 foil which was then mounted to rails. The configuration was dragged in a water tank. The foil developed a natural non-sinusoidal heave and pitch motion, {\color{black}which allowed} for energy extraction through the viscous dampers. Energy was primarily extracted from the heave motion while the pitch motion aided on the right positioning of the foil. Additionally, a series of numerical simulations on the same passive foil set-up are considered in  \cite{Duarte2021}. 

%We denote $\overrightarrow{x}$ as the flow direction, $U_\infty$. The foil  as a NACA0015 foil considered to be a rigid body participating in a combined heaving and pitching motion under the influence of the hydrodynamic forces. The heave motion is carried out in the $\overrightarrow{y}$ direction while the pitch motion in the $(P,\overrightarrow{z})$ axis direction, where $P$ is a point on the chord. 

\par
\begin{figure}
     \centering
     \includegraphics[width=0.7\textwidth]{ {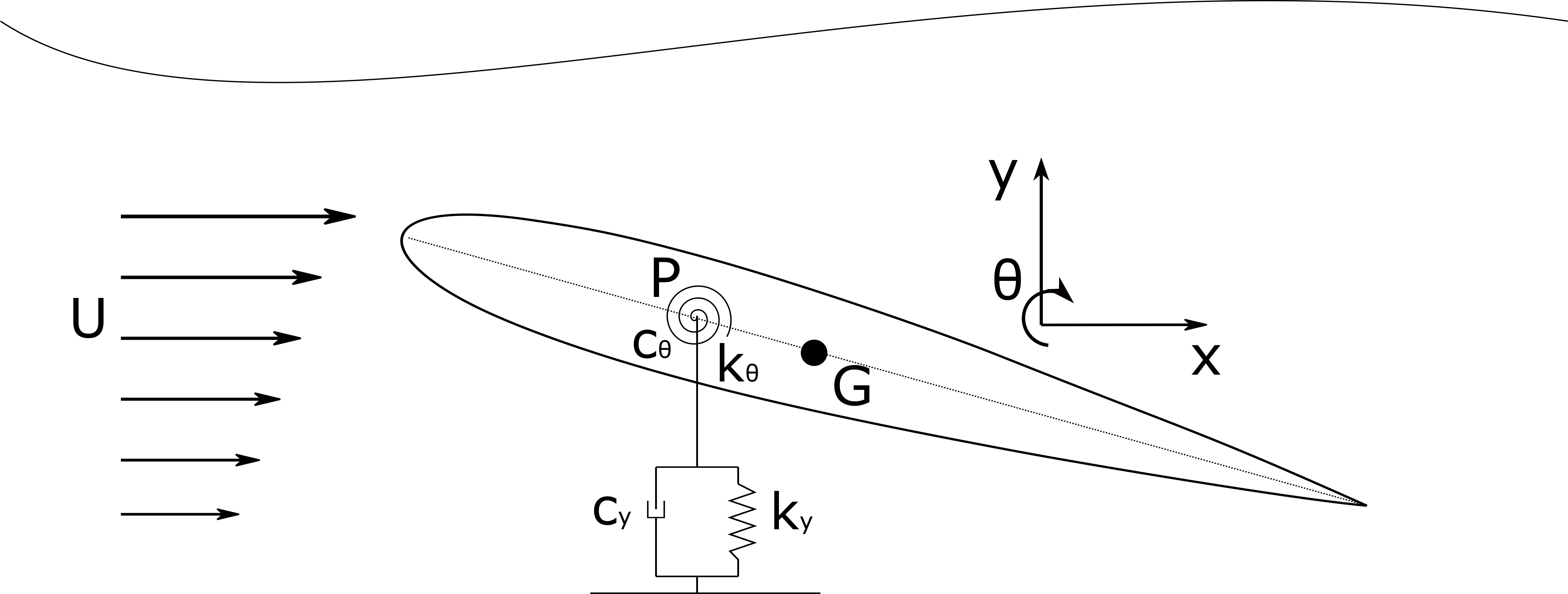}} 
     \caption{The basic setup used in this work. In general the foil can be submerged below the free surface and operate in sheared inflow  }
     \label{fig:passive_foil}
  \end{figure}

 The basic setup of such a configuration  in  sheared inflow operating near the free surface, is illustrated in \autoref{fig:passive_foil}. The rigid NACA0015 foil utilizes dampers and springs in both DOFs. The stiffness and viscous damper coefficients are denoted as  $k_y$ and $c_y$ in the heave direction and $k_\theta$ and $c_\theta$ in the pitch direction  respectively. Dampers and springs are considered  linear in all cases examined. The point $P$ which defines the pitching axis lies on the chord line in a distance $l_\theta$ from the Leading Edge. The distance from $P$ to the Center of Gravity ($G$) is denoted as $\lambda_g$. A positive value of $\lambda_g$ positions P upstream from G. The incoming flow  is aligned with the  x-axis but can vary in the vertical direction due to shear.

The motion of the passive foil emerges from the hydrodynamic forces acting on the foil surface together with the viscous and spring forces. Two coupled differential equations emerge from Newton{\color{black}'s} second law, with respect to the point $P$ and the pitching axis (as denoted in  \cite{Duarte2019},\cite{Duarte2021}):
	  
	 \begin{align}
	   &m_y \ddot{y}+c_y \dot{y} + k_y y +\Lambda(\dot{\theta}^2sin\theta-\ddot{\theta}cos\theta)=F_y \\
	   &I_{\theta} \ddot{\theta} + c_{\theta} \dot{\theta} + k_{\theta} \theta -\Lambda(\ddot{y}  cos\theta) = M_{\theta}
	    \label{rbd_eq}
	  \end{align}
  , where $m_y$ is the heaving mass and $I_{\theta} $ the moment of inertia of the foil with respect to the pitching axis, $(P,\overrightarrow{z})$. The heaving mass, $m_y$, may differ from the pitching mass, $m_{\theta}$, since some mechanical components may participate only in one of the two motions. In the right part of the equations, $F_y$ is the hydrodynamic force acting on the heave direction and $M_\theta$ is the hydrodynamic moment {\color{black}with} respect to the pitching axis $(P,\overrightarrow{z})$.   \par
  
  The above equations are expressed {\color{black}with} respect to the point $P$ which in the general case differs from $G$. This gives rise to the so called Static Imbalance, $\Lambda$, which couples the two equations. $\Lambda$ is defined as the product :
  \begin{align}
   \Lambda = \lambda_g  m_{\theta} 
\end{align}
  
  Static imbalance occurs from the misalignment of the pitching axis and the center of gravity and arises from our freedom to choose an optimum pitching axis which may or may not coincide with it. As a result it couples the two equations and therefore allows for energy transfer between the two DOFs.
  {\color{black}  The non-dimensional form of the physical parameters is used in order to compare our results with the reference results from the bibliography. Those parameters are gathered in the table below along with the definition of their non-dimensional form.   

  \begin{table}[H]
    \centering
    \caption{Definition of non-dimensional parameters used in the present study}
    {\color{black} \begin{tabular}{cccccc}
    	Parameter           & Definition   &  &  &   Parameter            & Definition  \\ \hline \hline
    	$Re$           & {\Large $\frac{U_{\infty}c}{\nu}$}               &  &  &  $m_\theta^*$           & \Large{$\frac{m_\theta}{\rho b c^2}   $} \vspace{2pt}   \\[2pt]  \hline 
        $m_y^*$  & \Large{$\frac{m_y}{\rho b c^2}$}  &  &  &  $\Lambda$ & $\lambda_g$  $m_{\theta}$ \vspace{2pt}              \\ \hline
        $k^*_{\theta}$ & \Large{$\frac{k_\theta}{\rho U_\infty^2 b c^2}$} &  &  &  $c^*_{\theta}$ & \Large{$\frac{c_\theta}{\rho U_\infty b c^3}$} \vspace{2pt} \\ \hline
        $k^*_{y}$      & \Large{$\frac{k_y}{\rho U_\infty^2 b}$}          &  &  &  $c^*_{y}$      & \Large{$\frac{c_y}{\rho U_\infty b c}$} \vspace{2pt}   \\ \hline
        $I^*_{\theta}$ & \Large{$\frac{I_{\theta}}{\rho b c^4}$}          &  &  &  $l^*_{\theta}$ & \Large{$\frac{l_{\theta}}{c}$} \vspace{2pt} \\ \hline
    \end{tabular}}
\end{table}

  }

\par The damping coefficients serve the cause of harvesting energy from the foils motions and since this energy can be quantified through $c_\theta$ and $c_y$ we proceed to define an efficiency coefficient to assess the foil performance. The hydraulic efficiency ($\eta$) is therefore defined as :
    
    \begin{align}
    	\eta=\frac{1}{\Delta t} \int_{t_0}^{t_0 +\Delta t} \frac{c_y  \dot{y}^2 + c_{\theta}\dot{\theta}^2 }{\frac{1}{2} \rho U_{\infty}^3 S  } dt
    	\label{eq:efficiency_passive}
    \end{align}
    
    , where $S$ is the maximum area of the cross section swept by the foil, being the product of the foils span, $b$, and the total heave distance scanned by the foil. {\color{black} The total distance scanned by the foil is defined as the sum of the heave amplitude of the pitching axis pivot point, P, in the positive and negative directions, as it is illustrated in Figure \ref{fig:total_heave_scanned}.
    
    \begin{figure}[H]
     \centering
     \includegraphics[width=0.5\textwidth]{ {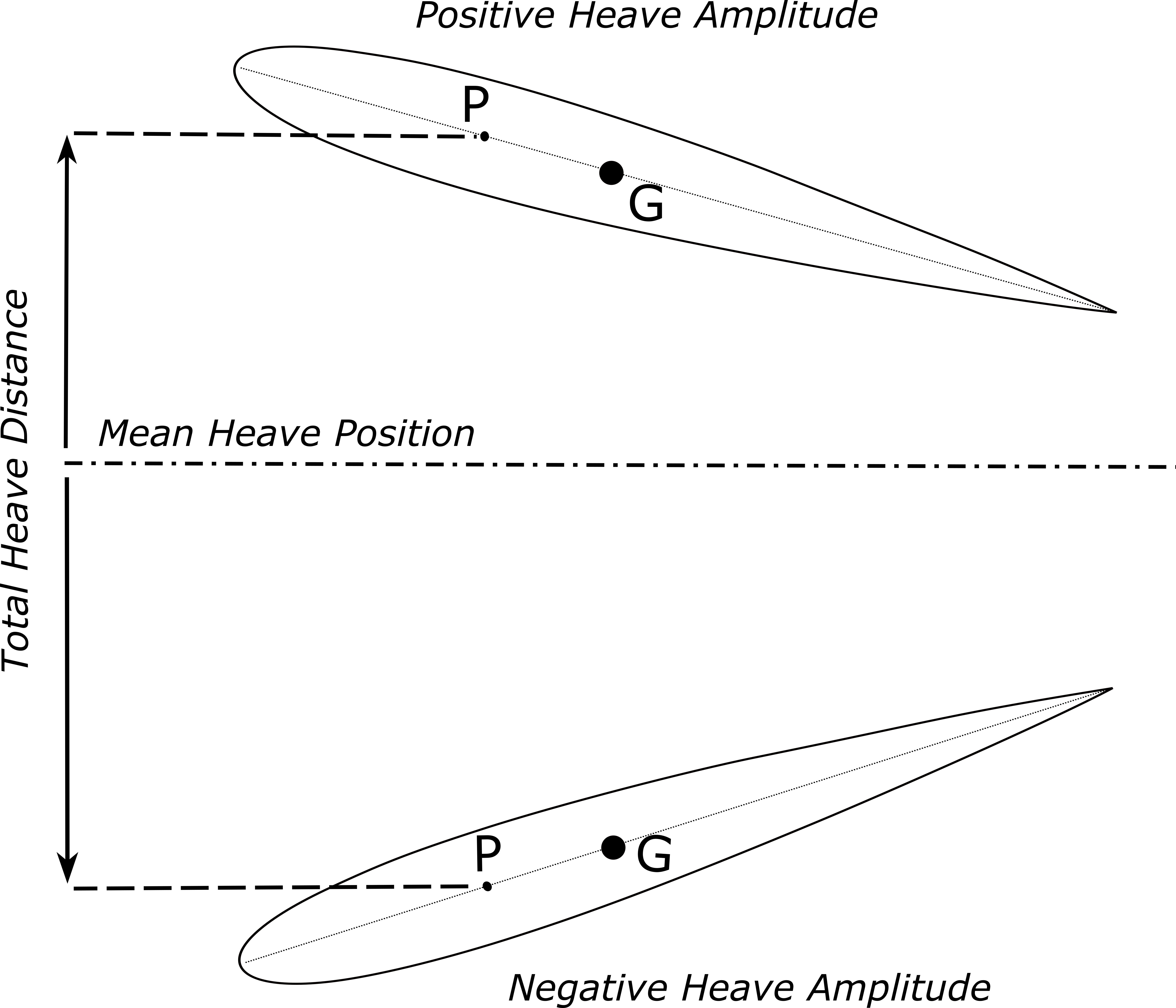}} 
     \caption{Definition of the total heave distance scanned by the foil}
     \label{fig:total_heave_scanned}
  \end{figure}}
    
    Another efficiency metric can be defined, normalizing the hydraulic power in terms of the projected surface of the foil. Consequently the mean power coefficient ($\overline{C}_p$)  is defined as:
     \begin{align}
    	\overline{C}_p=\frac{1}{\Delta t} \int_{t_0}^{t_0 +\Delta t} \frac{c_y  \dot{y}^2 + c_{\theta}\dot{\theta}^2 }{\frac{1}{2} \rho U_{\infty}^3 b c  } dt
    	\label{eq:cp_passive}
     \end{align}
    
%A representation of the set up can be viewed below :
%	    \begin{figure}%
%	        \centering
%	        \includegraphics[scale=0.7]{{figures/Duarte_validation/duarte_fully_passive_foil_setup2.JPG} }
%	        \caption{Experimental set up of Duarte passive foil }
%	        \label{fig:duarte_setup}
%	\end{figure}

   % Experimental data were acquired and were later compared to numerical results \cite{Duarte2021} based on the same set up.
    
 Finally, the main parameters in non-dimensional form  regarding the mass-spring-damper system can be found below (based on  \cite{de2019conception}) .
\begin{table}[H]
    \centering
    \caption{Definition of parameters used in the present study}
    \begin{tabular}{cccc}
    	Parameter           & Value      & Parameter            & Value    \\ \hline \hline
    	$Re$                & $6 \cdot 10^{4}$  &  $U_\infty$   & 1       \\ \hline 
    	$m*$                & 0.92              &  $\Lambda$           & 0.00065          \\ \hline
        $k^*_{\theta}$      & 0.071             &  $c^*_{\theta}$      & 0.052            \\ \hline
        $k^*_{y}$           & 0.72              &  $c^*_{y}$           & 0.93             \\ \hline
        $I^*_{\theta}$      & 0.0563            &  $l^*_{\theta}$      & 0.33             \\ \hline
    \end{tabular}
\end{table}

For the results presented the chord of the NACA0015 foil was set to $c=0.1$ and is considered to operate in water.
    %\vspace{1cm}
    %, where $l^*_{\theta}$ is the non-dimensional distance between the Pitching axis and the Leading Edge.
    %\subsection{Mesh Dependency}
\section{Results \& Discussion} \label{sec:res_disc}
In this section numerical results using the aforementioned setup are presented. Initially, the passive foil is considered to operate in uniform inflow and predictions are compared to the literature \cite{Duarte2019},\cite{Duarte2021}. Afterwards operation in sheared inflow is considered with and without taking into account the free surface. For all the cases presented bellow, at least 50 oscillating cycles where considered while the  fluid-rigid body dynamics coupling converged when the error of acceleration was below $10^{-09}$.
%Finally the SST $k-\omega$ turbulence model was employed for all simulations.

\subsection{Fully passive foil operating in uniform inflow\label{sec:Duarte_case}}

 Prior to comparing MaPFlow predictions to the ones available in the literature  a grid-independence study is conducted. In order to better capture the generated vorticity two mesh refinement zones are used. The first one is a rectangular box beginning $12c$ upstream and ending $33c$ downstream from the airfoil. The height of this box is $12c$ in each direction. The second, circular refinement zone, is located near the airfoil with a radius of $3c$. Three successively refined grids were generated \{color{red, with}  cell sizes  of $10\%,\  5\%\ and\ 2\%$ of chord length, $c$, \{color{respectively}.
    
\begin{figure}[H]
    	\begin{subfigure}{0.48\textwidth}
    		\includegraphics[width=\textwidth]{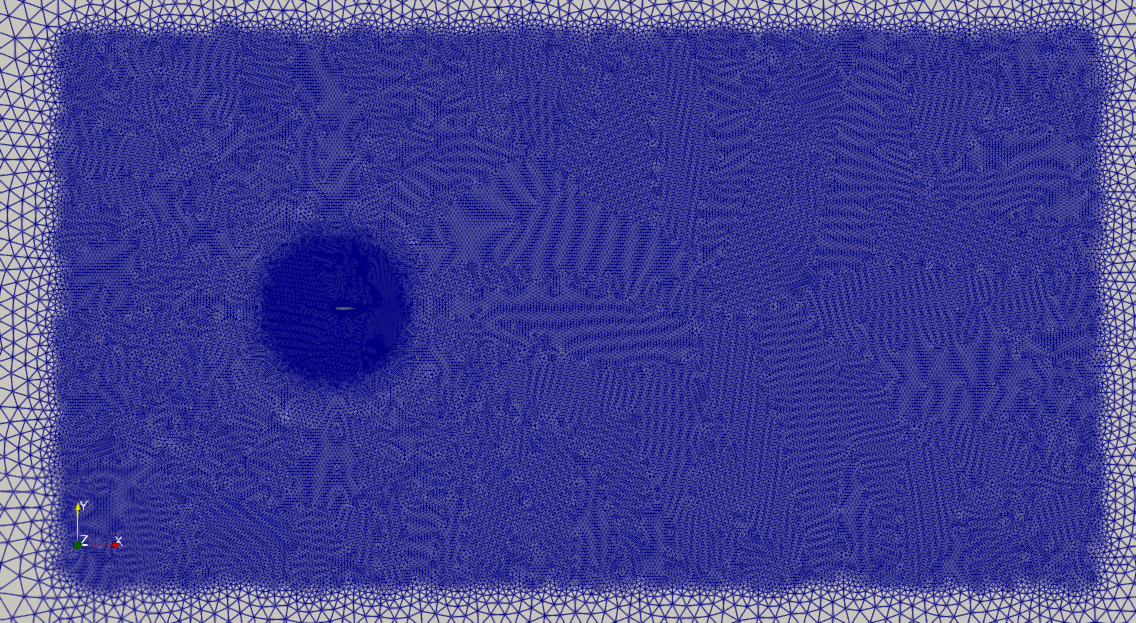}
    		\caption{Rectangular Refinement zone}
    	\end{subfigure}
    	%\vspace{0.5cm}
    	\hfill
    	\begin{subfigure}{0.48\textwidth}
    		\includegraphics[width=\textwidth]{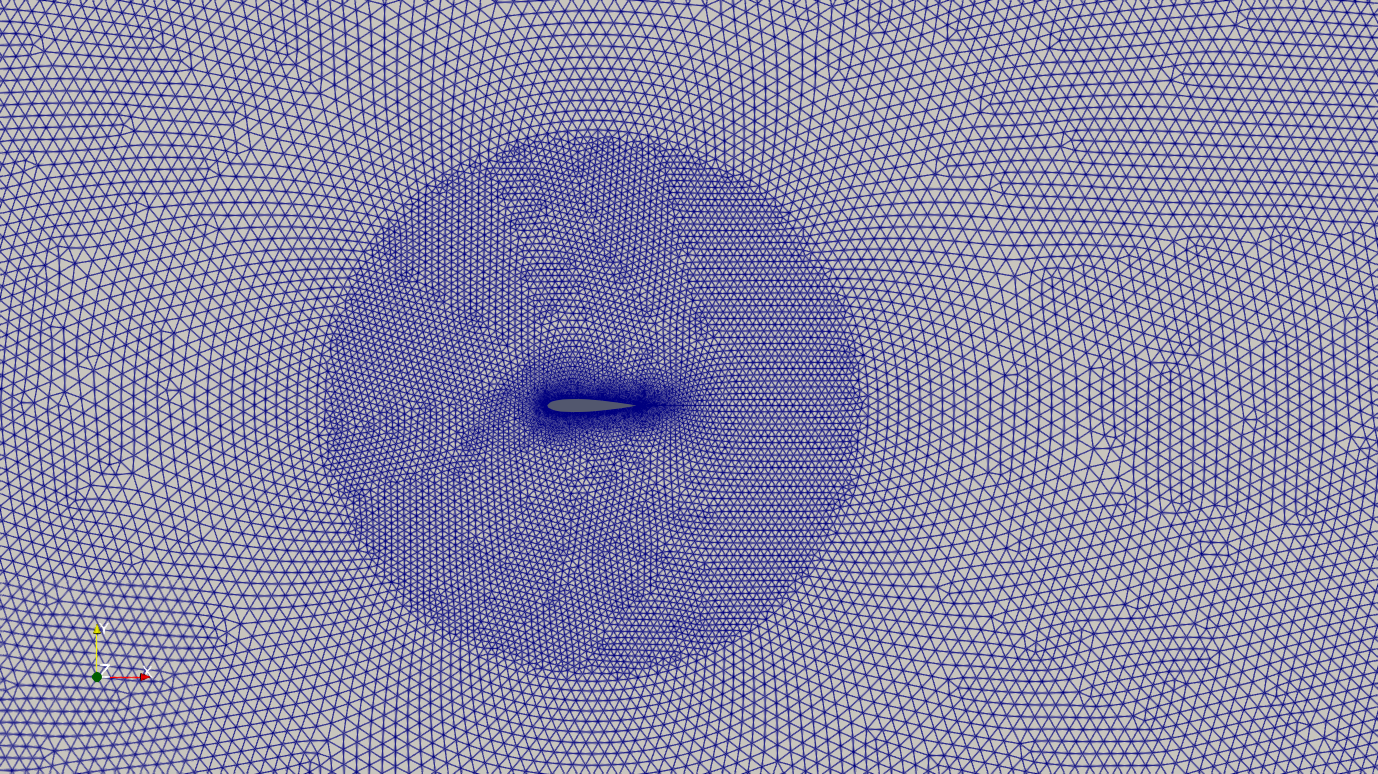}
    		\caption{Circular refinement zone}
    	\end{subfigure}
 %   	\begin{subfigure}{0.48\textwidth}
 %   		\includegraphics[width=\textwidth]{figures/Duarte_validation/mesh_screenshot_size_5.pn%g}
 %   		\caption{Inner circle cell size 5}
 %   	\end{subfigure}
 %   	\hfill
 %   	\begin{subfigure}{0.48\textwidth}
 %   		\includegraphics[width=\textwidth]{figures/Duarte_validation/mesh_screenshot_size_2.pn%g}
 %   	\caption{Inner circle cell size 2}
 %   	\end{subfigure}
    	\caption{Overview and refinement zones of domain }
    \end{figure}
    The relative errors with respect to the finest mesh are calculated for the amplitudes of heave and pitch motions and are presented in Table \ref{tab:grid}.
\begin{table}[H]
    \centering
    \caption{Grid-independence study results}
    \begin{tabular}{cccccc}
    {\label{tab:grid}}
      & cells & $\Delta A_{\theta}$    & $\Delta A_{y}$ & \color{black} $\Delta \overline{C}_p$ & \color{black} $\Delta\eta$       \\
      \hline \hline   \vspace{0.2 cm}
    	Finer mesh (size 2\%c)  & 300000 & -                    &  -    & \color{black} - & \color{black} -                 \\ \hline \vspace{0.2 cm}
    	Fine mesh (size 5\%c)   & 160000  & -0.239 \%            &  0.673 \% & \color{black} 0.12 \% & \color{black} -5.35 \%  \\ \hline  \vspace{0.2 cm}
    	Coarse mesh (size 10\%c)& 120000  & -0.497 \%            &  0.557 \%  & \color{black} -1.35 \% & \color{black}  -2.91 \%      \\ \hline \vspace{0.2 cm}
    \end{tabular}
\end{table}

Following the grid-independence study a time step independence study is conducted using the fine mesh mentioned above. The time discretization is based upon the damped natural frequency ($f_d=1.15Hz$) of the heaving motion, defined as :
\begin{align}
	f_{d} = f_n \sqrt{1-\zeta^2}, \ f_n=\frac{1}{2\pi}\sqrt{\frac{k_y}{m_y}}, \ \zeta=\frac{c_y}{2\sqrt{k_y m_y}}
\end{align}

Three different timesteps are considered, $\Delta t$=0.0001,\ 0.0002 and 0.0004s. They correspond to $\approx$ 2150, 4300 \ and\ 8600 time-steps per period. The table below suggests satisfactory convergence for the intermediate time step of $\Delta t = 0.0002$s.

\begin{table}[H]
    \centering
    \caption{Time step independence study results}
    \begin{tabular}{ccccc}
    {\label{tab:dt}}
     Timestep & $\Delta A_{\theta}$    & $\Delta A_{y}$ & \color{black} $\Delta \overline{C}_p$ & \color{black} $\Delta\eta$    \\
      \hline \hline   \vspace{0.2 cm}
    	$\Delta t=0.0001$  & -                    &  -  & \color{black} - & \color{black} -           \\ \hline \vspace{0.2 cm}
    	$\Delta t=0.0002$    & 0.13\%   &   0.0057\%    & \color{black} 1.338 \% & \color{black} -1.657 \% \\ \hline  \vspace{0.2 cm}
    	$\Delta t=0.0004$    & -0.38\%    &-0.{\color{black}{}2}98\%      & \color{black} 1.383 \% & \color{black} -1.838 \%  
    	\\ \hline \vspace{0.2 cm}
    \end{tabular}
\end{table}
% check post processing percentages again 

\subsubsection{Comparison with experimental and numerical results}

\begin{figure}[H]
	\centering
		\begin{subfigure}{0.49\textwidth}
		\includegraphics[width=\textwidth]{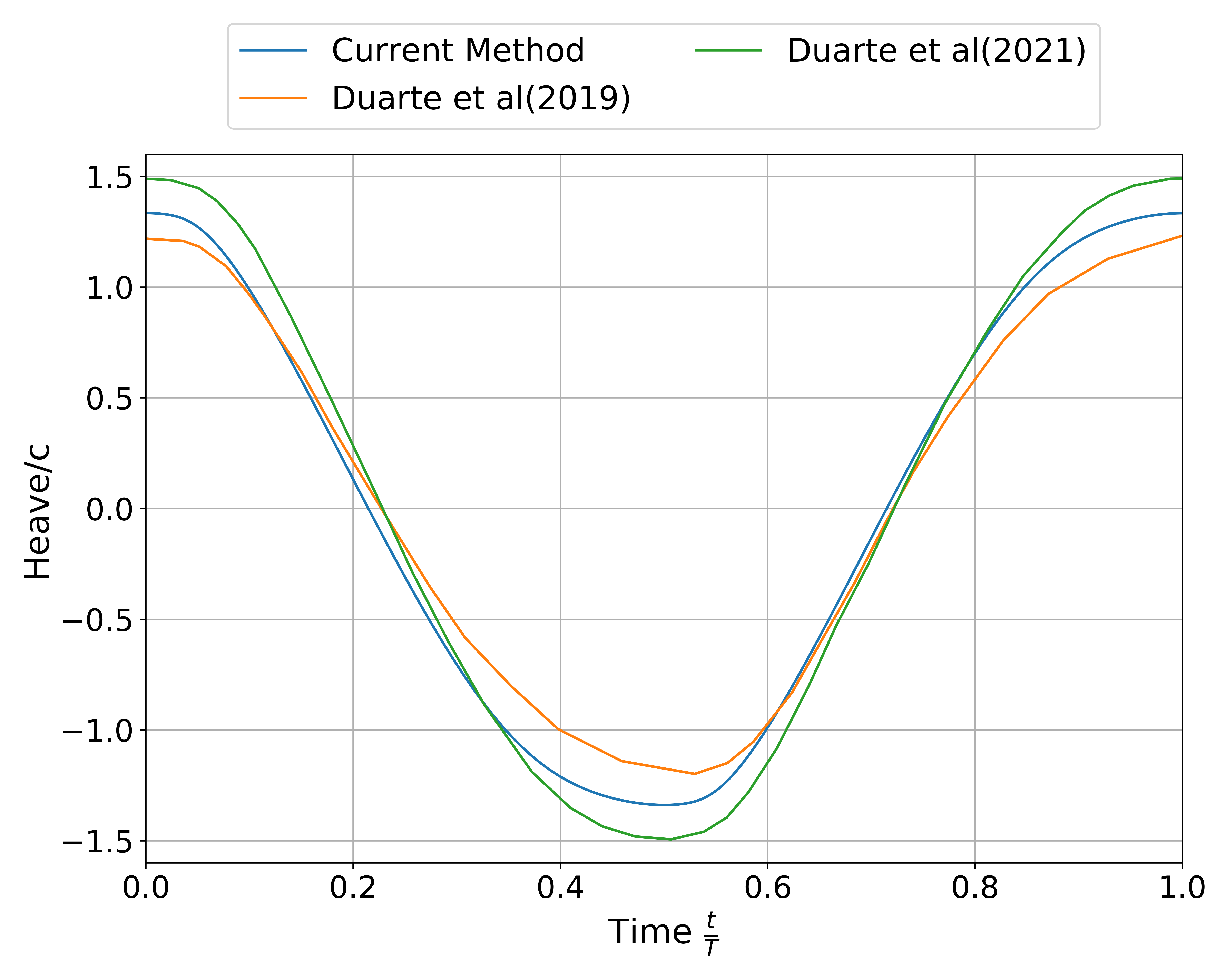}
	\end{subfigure}
	\hfill
	\begin{subfigure}{0.49\textwidth}
		\includegraphics[width=\textwidth]{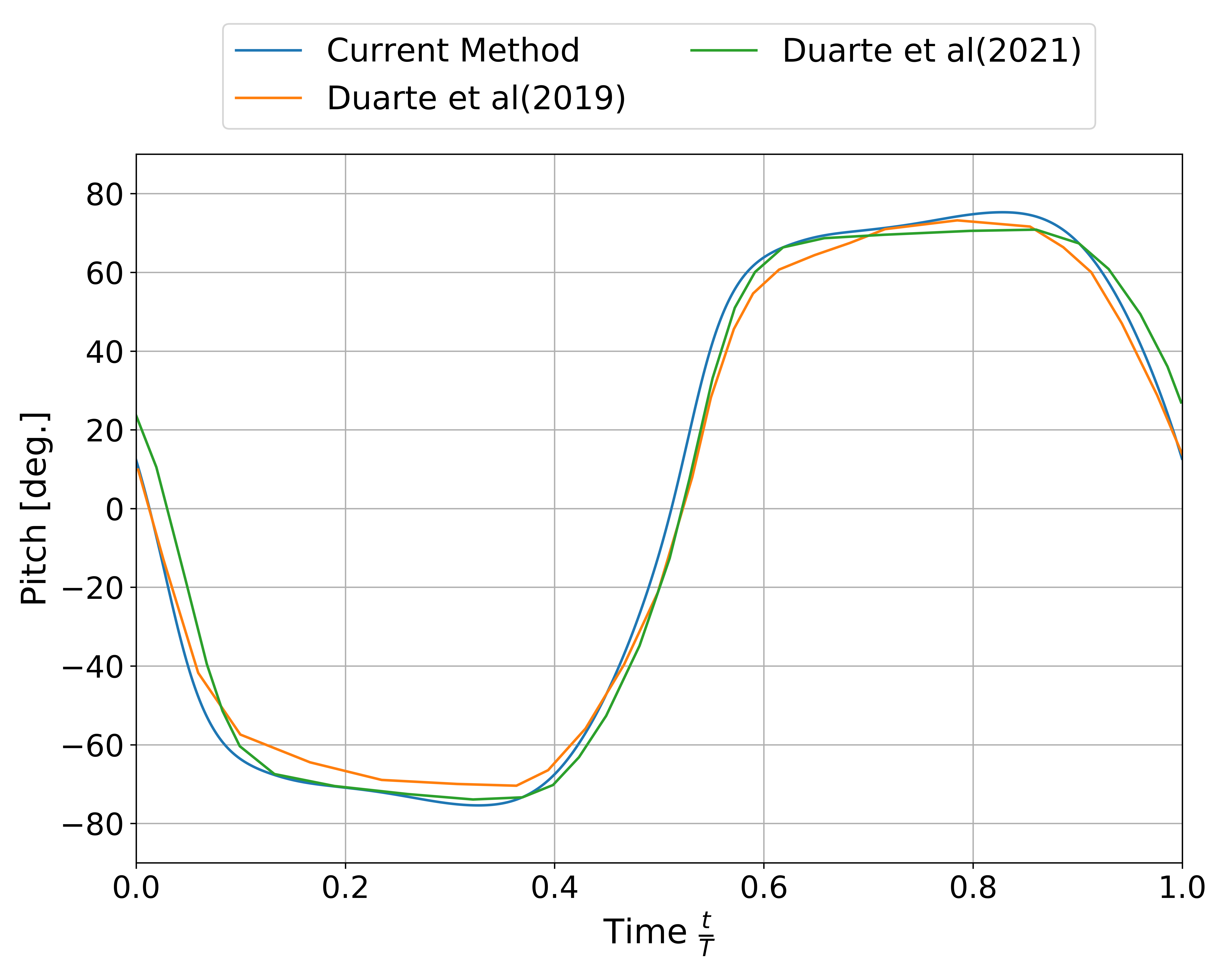}
	\end{subfigure}
	\caption{Heave and Pitch  comparison with experimental (Duarte et al 2019 \cite{Duarte2019}) and numerical (Duarte et al 2021 \cite{Duarte2021}) results {\label{fig:kinecomp}}}
\end{figure}

The above results are compared to the experimental results acquired by \cite{Duarte2019} and the numerical simulation performed by  \cite{Duarte2021}. The comparison is based on the two kinematic quantities $(A_{\theta}\  and\ A_y)$ and the two energy coefficients described above. The pitch and heave signal over a period can be seen in Figure \ref{fig:kinecomp}. It is evident that the comparison in both heave and pitch between the current methodology and the results available in the literature {\color{black}are} in very good agreement.

A quantitative comparison is presented in \autoref{tab:duartecomp}. The error between the current simulations and measurements are in line between the published numerical results. The error in maximum predicted pitch amplitude ($A_\theta$) is  ~ 5\% while the error for the maximum heave amplitude ($A_y$) is ~10\%. For the power coefficient ($\overline{C}_p$) and efficiency ($\eta$) the error is ~5\%.
\begin{table}[H]
\caption{Errors between current simulation, experimental \cite{Duarte2019} and numerical reference results \cite{Duarte2021}. $A_\theta$, $A_y$ denote the maximum amplitude in pitch and heave respectively, $\overline{C}_p$ is the power coefficient and $\eta$ the efficiency.
\label{tab:duartecomp}}  
{\small{
\begin{tabular}{ccccc}
\centering
  & $A_\theta$ & $A_y$ &  $\overline{C}_p$  & $\eta$  \\
  \hline \hline
 Exper.   & 71.81 $^{\circ}$           & 1.216 $c$              &    1.09            & 0.44              \\ \hline
Numer.Ref  & 72.38 $^{\circ}$ (+0.79\%) & 1.491 $c$ (+22.5 \%) &    1.21 (+11.31\%) & 0.405 (7.97 \%)   \\ \hline
Current             & 75.34 $^{\circ}$ (+4.91\%) & 1.336 $c$  (+9.81\%)  &   1.13  (+4.03\%)  & 0.419 (-4.81 \%)  \\ \hline

\end{tabular}}
}
\end{table}

\subsection{Passive foil operating in sheared inflow{\label{sec:shear}}}

The fully passive foil examined by Duarte  \cite{Duarte2021} is now considered to operate in a sheared inflow. A comparison is presented between uniform and linear shear inflow to identify the trends in both the kinematics and the efficiency of the foil. The shear profile (for the x-axis velocity component) is dictated by \autoref{eq:shear} :

\begin{equation}
	U(y)=U_{\infty}Ky
	\label{eq:shear}
\end{equation}

, where $K$ is a parameter controlling the shear rate. Two shear rate groups of mild and strong shear are applied to the flow. Finally, the foil is initially positioned vertically, so that $U_0=1 m/s$ at the foil level at the start of the simulation.

%$K$ is derived from the non dimensional $K_{nondim}$ which is mentioned as $K$ in \cite{Cho2014} using the following formula :

%\begin{equation}
%	K=K_{nondim}*\frac{U_{\infty}}{c}
%\end{equation}

 %   \begin{figure}[H]
 %       \centering
 %       \includegraphics[width=\textwidth]{figures/duarte_shear/X-vel ocity_shear_duarte_K03.png}
 %       \caption{X-component of velocity in regards to Y-coordinate}
 %       \label{fig:duarte_shear_profile}
 %   \end{figure}

%In \autoref{fig:duarte_shear_profile} the velocity in $X$-direction increases from the bottom to the top of the computational domain in a linear way according to the shear ratio. The step-like pattern in high and low values of $Y$ Position is due to the bigger cells used in those  areas making the velocity difference between cells large enough to be noticed. In the center part where the foil oscillates the velocity increments are small enough to approximate a continuous shear profile. 

\begin{figure}[H]
		%	\centering
		%%%%%%%%%%%%%%%%%%%%%%%%% uniform
		\begin{subfigure}[b]{0.49\textwidth}
			\includegraphics[width=\textwidth]{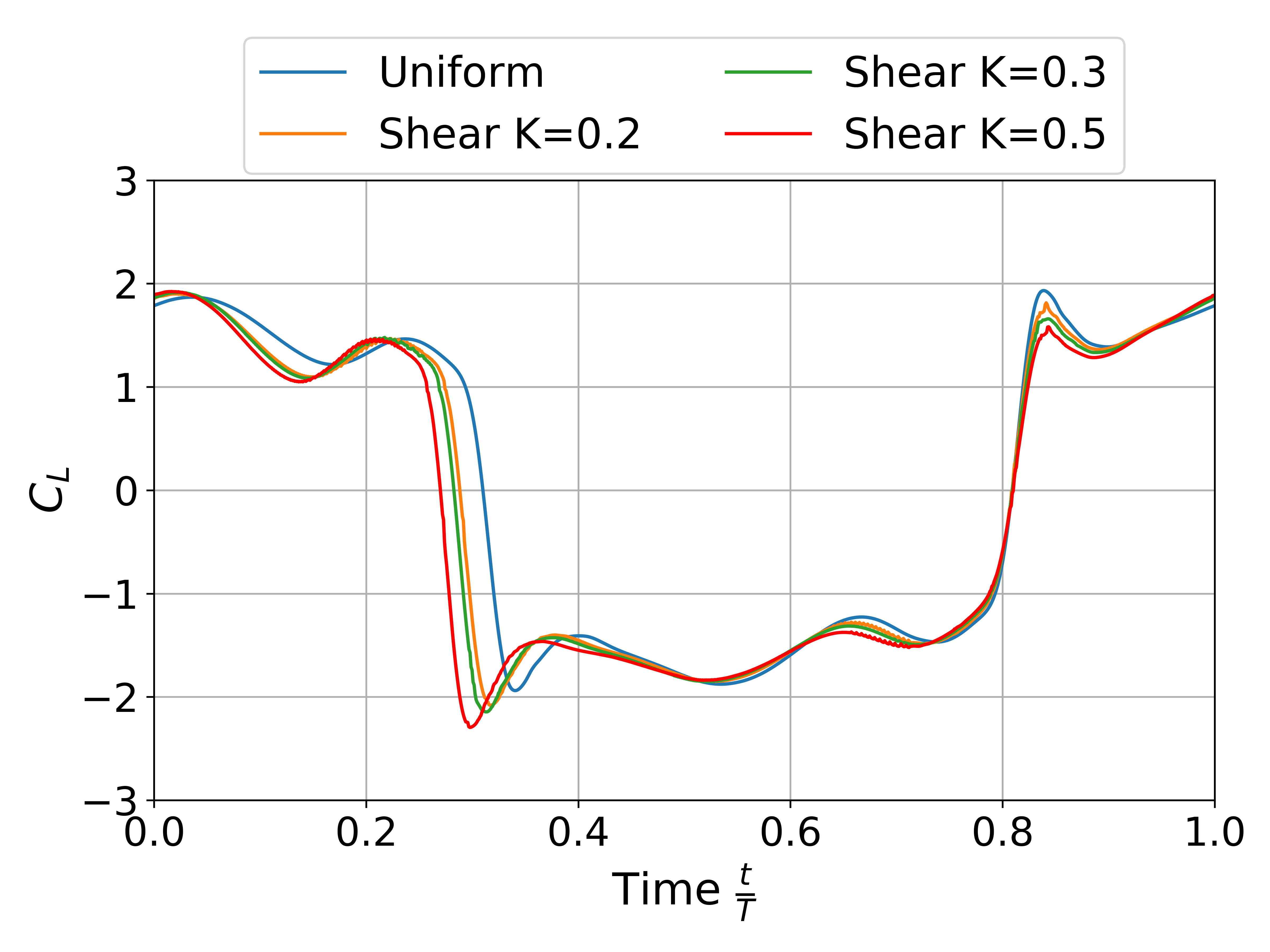}
			%\caption{$Uniform\ \ flow$}
			\vspace{0.01 cm}
		\end{subfigure}
		\hfill
		\begin{subfigure}[b]{0.49\textwidth}
			\includegraphics[width=\textwidth]{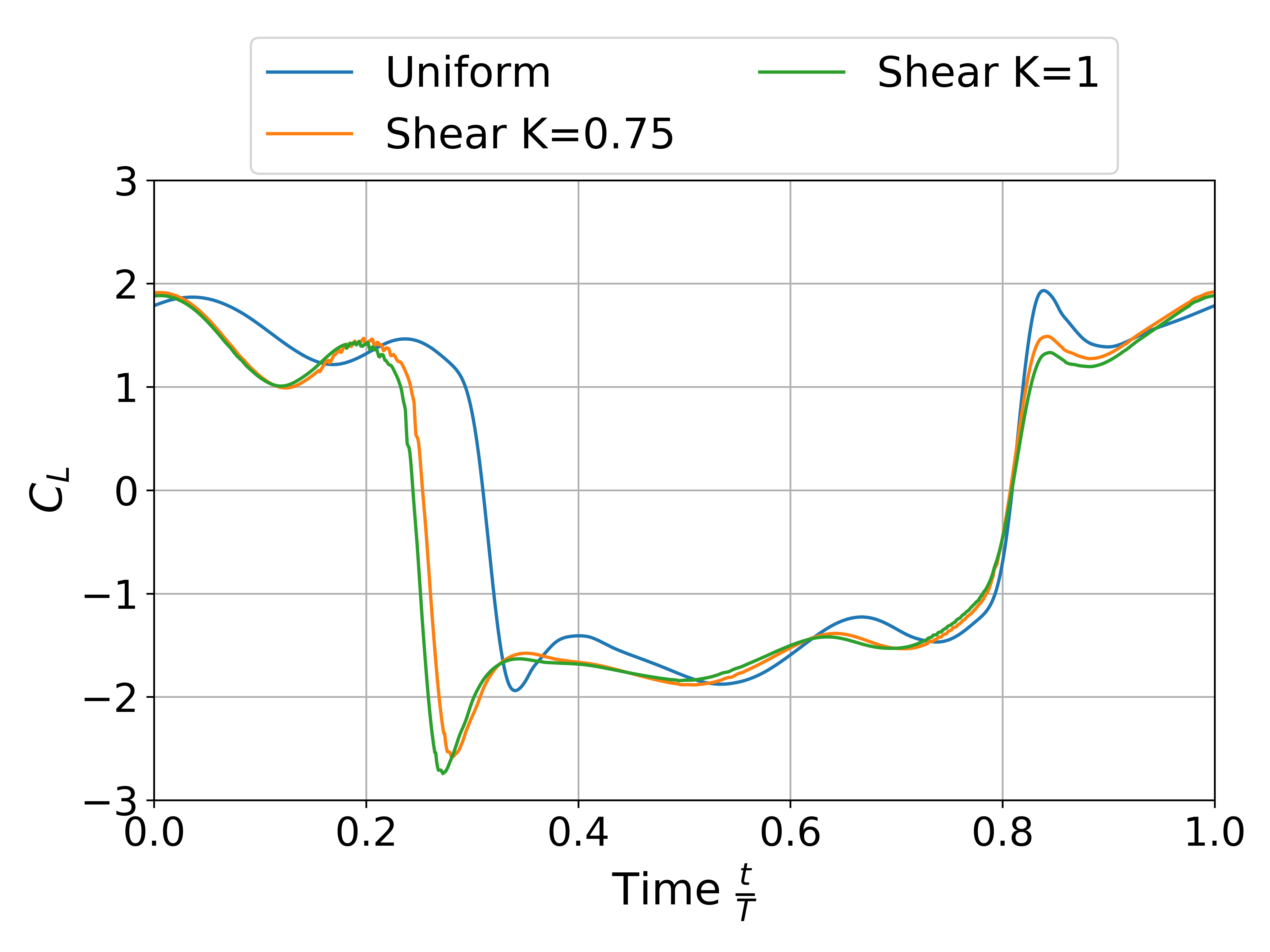}
			%\caption{$ K=0.20$}
			\vspace{0.01 cm}
		\end{subfigure}
		\\
		%%%%%%%%%%%%%%%%%%%%%%% K=0.30
		\begin{subfigure}[b]{0.49\textwidth}
			%	\subfigure[$\frac{Heave}{c}$ comparison ]{%
			\includegraphics[width=\textwidth]{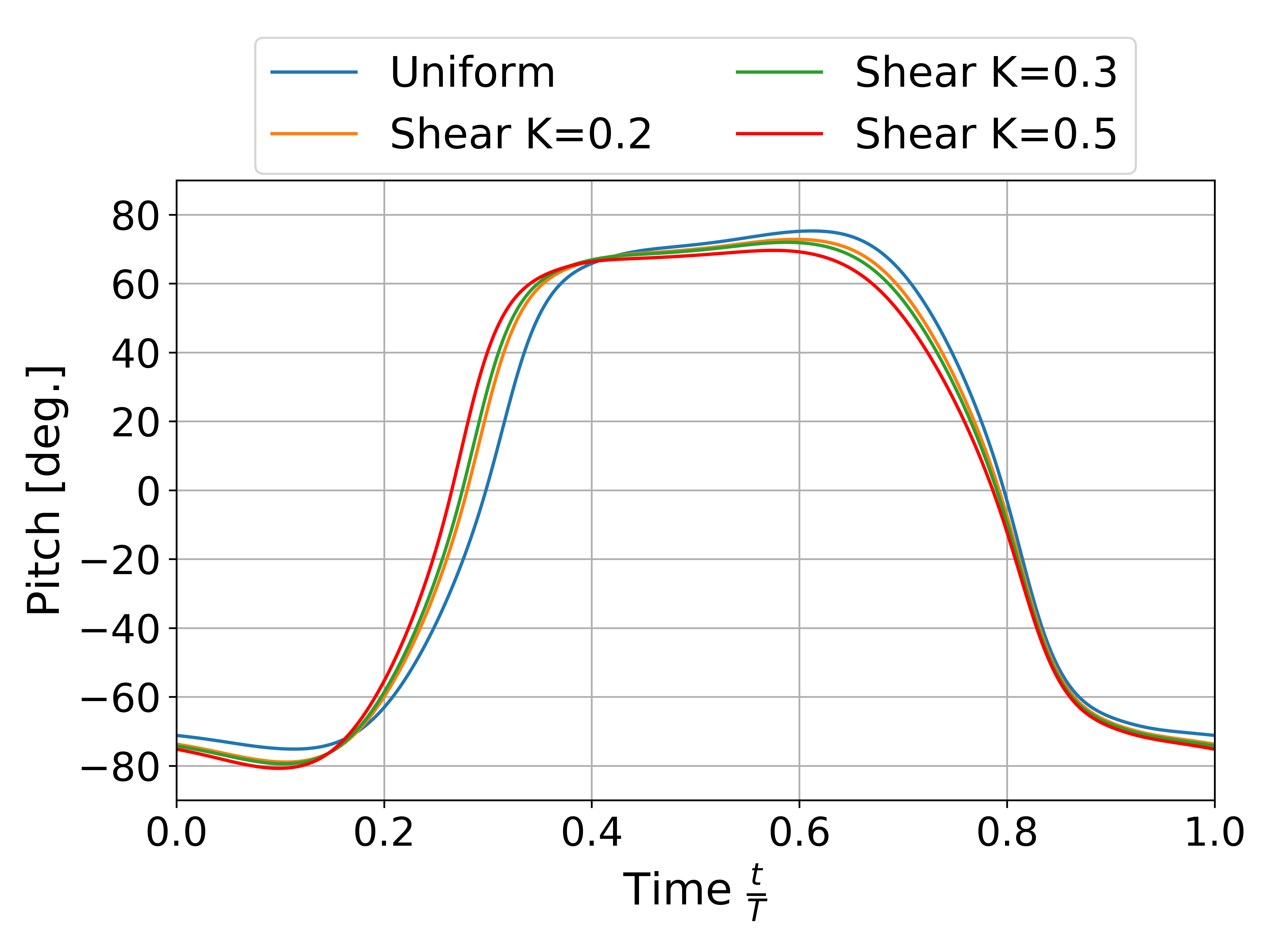}
			%\caption{$K=0.30$}
		\end{subfigure}
		\hfill
			%%%%%%%%%%%%%%%%%%%%%%% K=0.50
			\begin{subfigure}[b]{0.49\textwidth}
				%	\subfigure[$\frac{Heave}{c}$ comparison ]{%
			\includegraphics[width=\textwidth]{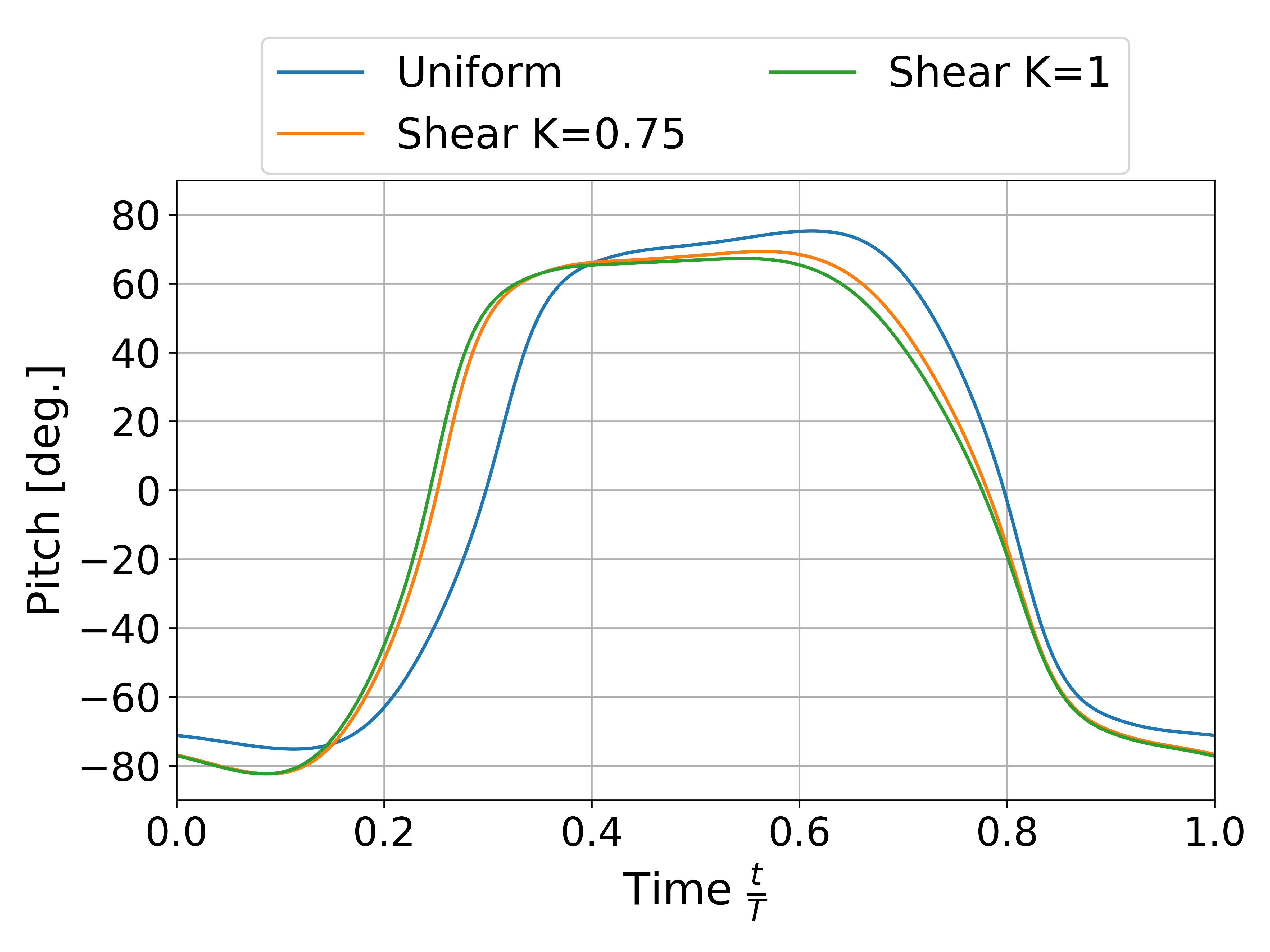}
				%\caption{$K=0.50$}
			\end{subfigure}
		\caption{Lift coefficient $C_L$ (upper) and Pitch (lower) comparison over a period. Mild shear cases are presented in the left while strong shear cases in the right.}
		\label{fig:loads}
	\end{figure}

The pitch of the foil as well as the lift coefficient ($C_L $) are compared during a one period interval in \autoref{fig:loads}. Mild shear cases are presented on the left while strong shear cases  {\color{black}are displayed on} the right. It is evident, that in mild shear conditions the divergence from the uniform case is small. When strong shear is considered differences are noticeable especially in the first half of the period. Due to the stronger shear, the foil pitches earlier compared to the uniform case which has a direct impact on the $C_L$.
Regarding the period of an oscillating  cycle, in the sheared inflow case it differs from the period of the uniform flow. The differences however are slight for mild shear rates while for strong shear rates they become significant. These are presented in \autoref{tab:period}. It is evident that the oscillating period increases around 8.5\% for the strong shear cases.
\begin{table}[H]
\centering
        \caption{Period increment for different shear rates\label{tab:period}}
		\begin{tabular}{c c c }
		\toprule
		%	    	\hline
		$Shear\ Rate\ (K)$ & $Period$    &  $ \%\ Increment $       \\ \hline \hline
		
		$ 0 $              & 0.8298      & -                        \\ \hline
		$0.20$             & 0.8296      &  - 0.024 \%              \\ \hline
		$0.30$             & 0.8352      &  + 0.651 \%              \\ \hline
		$0.50$             & 0.8368      &  + 0.892 \%              \\ \hline
		$0.75$             & 0.8994      &  + 8.452 \%              \\ \hline
		$1.00$             & 0.8994      &  + 8.452 \%              \\ \hline
		\bottomrule
	    \end{tabular}
    \end{table}

\begin{figure}[H]
\centering
    \begin{subfigure}[b]{\textwidth}
    	\includegraphics[width=\textwidth]{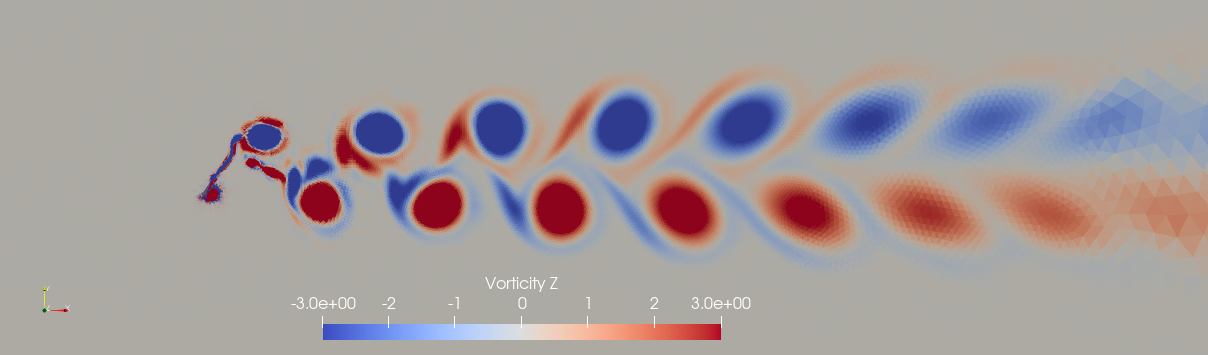}
    		\caption{$Uniform\ Flow$}
    	\end{subfigure}
    \vspace{1 pt}
%   	\begin{subfigure}[b]{\textwidth}
%   		\includegraphics[width=\textwidth]{figures/duarte_shear/vorticity_contour_farfield_shear02.png}
%   		\caption{$Shear\ Flow\ K=0.20 $}
%   	\end{subfigure}
%    	\vspace{1 pt}
    	\\
    	\begin{subfigure}[b]{\textwidth}
    		\includegraphics[width=\textwidth]{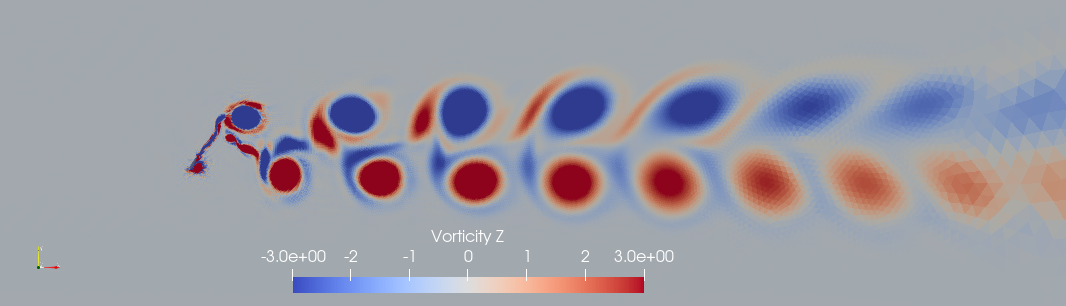}
    		\caption{$Shear\ Flow\ K=0.30$}
    	\end{subfigure}
        \vspace{1 pt}
    	\\
%   	\begin{subfigure}[b]{\textwidth}
%   		\includegraphics[width=\textwidth]{figures/duarte_shear/vorticity_contour_farfield_shear05.png}
%   		\caption{$Shear\ Flow\ K=0.50$}
%   	\end{subfigure}
%   \vspace{1 pt}
%	\end{figure}
%	\begin{figure}[H]\ContinuedFloat
		\begin{subfigure}[b]{\textwidth}
			%	\subfigure[$Pitch$ comparison ]{%
			\includegraphics[width=\textwidth]{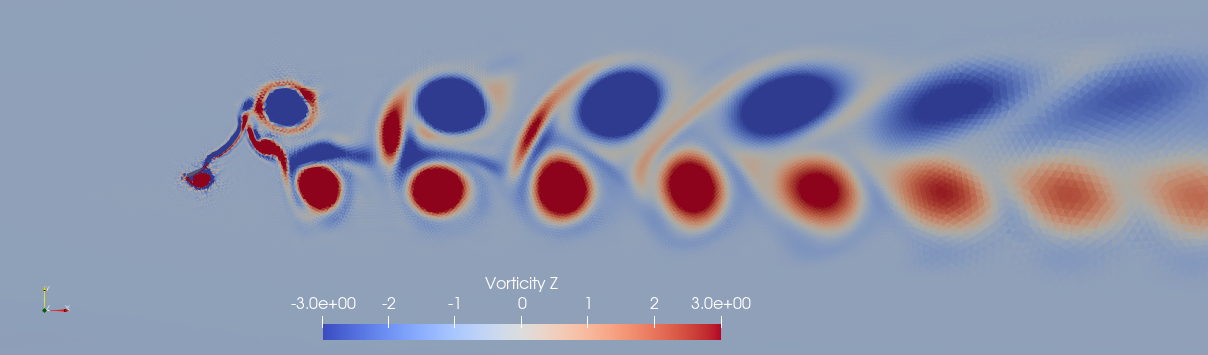}
			\caption{$Shear\ Flow\ K=0.75 $}
		\end{subfigure}
		\vspace{1 pt}
%	\\
%%%%	\begin{subfigure}[b]{\textwidth}
%%%%		%	\subfigure[$\frac{Heave}{c}$ comparison ]{%
%%%%		\includegraphics[width=\textwidth]{figures/duarte_shear/vorticity_contour_K1.png}
%%%%		\caption{$Shear\ Flow\ K=1.00$}
%%%%	\end{subfigure}
%%%%	\vspace{1 pt}
		\caption{Vorticity Contours showing the vortex street produced by different shear rates}
		\label{fig:vort_one_phase}
	\end{figure}

In \autoref{fig:vort_one_phase} the vorticity contours for the two shear cases (a mild and a strong one)  are compared to the uniform inflow. 
As \autoref{fig:vort_one_phase} suggest the vortex patterns in the wake are very similar in the mild shear cases ($K=0.30$), however as the shear rate increases ($K=0.75$) the effect of the shear is evident. Especially in the high shear rate, it can be seen that  the upper vortices (negative vorticity, blue color) convect downstream faster. Additionally, a decrease in the x-direction offset between the upper and lower vortexes is evident. Taking the uniform flow as reference we focus on the second pair of vortexes and note that the upper vortex is slightly delayed  {\color{black}compared to} the lower. In the strong shear case $K=0.75$ we note that in the same pair of vortexes the upper vortex has surpassed the lower one.
%In  the mild shear case $K=0.3$ we note that the same pair of vortexes are positioned in roughly the same longitudinal coordinate as in the uniform inflow.  
	
To further understand the effect of the shear inflow  {\color{black}on} the flapping foil response a Fourier analysis is performed. The analysis is {\color{black} concentrated on $\dot{y}$ since the heave velocity is responsible for the majority of the extracted energy through the viscous dampers.}    

%An interesting conclusion arises from the emergence of a second important frequency visible in the spectrum in Figure \ref{fig:one_phase_spectrum}.
	
	\begin{figure}[H]
		%	\centering
		%%%%%%%%%%%%%%%%%%%%%%%%% uniform
		\begin{subfigure}[b]{0.33\textwidth}
			%	\subfigure[$Pitch$ comparison ]{%
			\includegraphics[width=\textwidth]{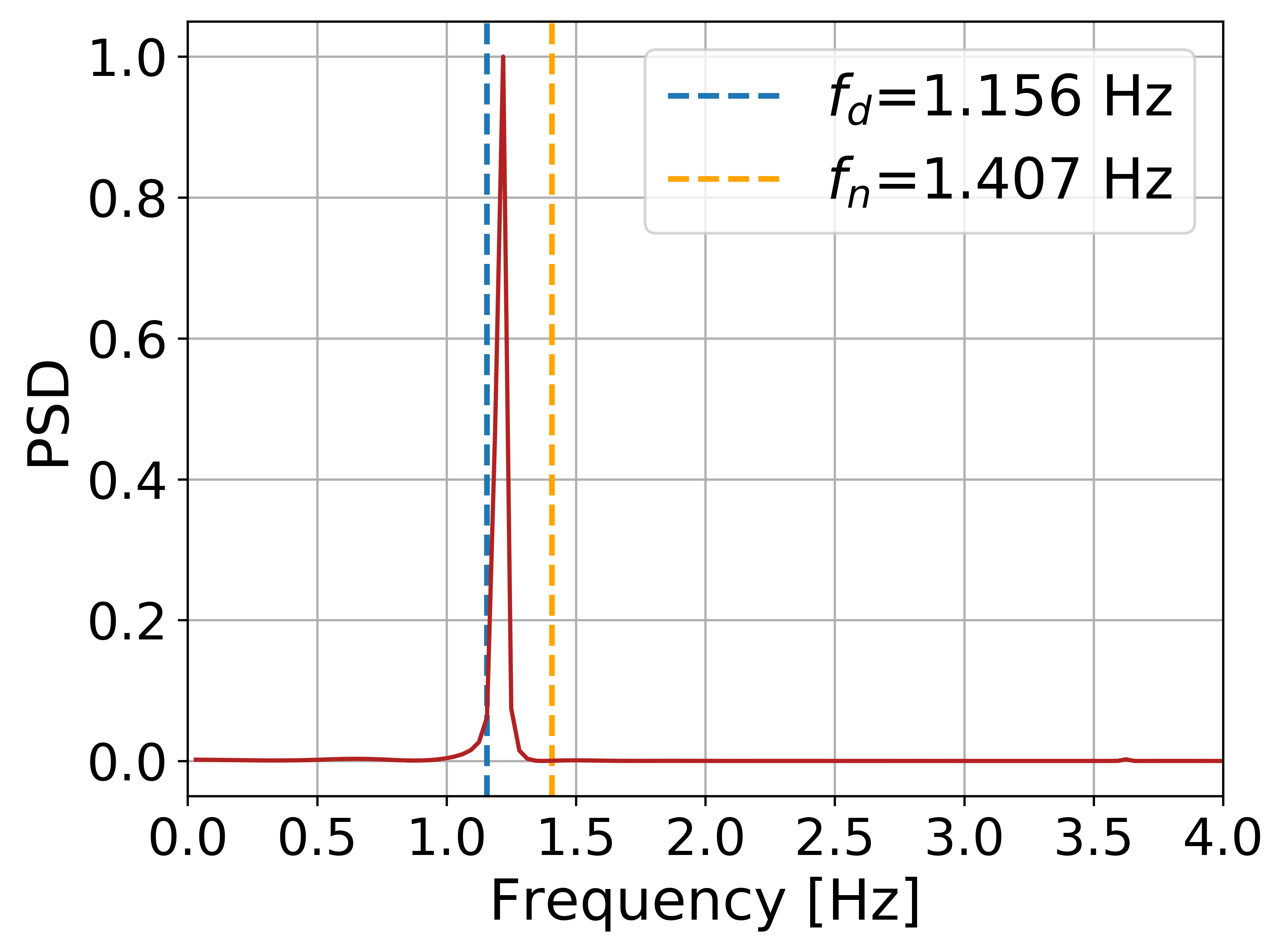}
			\caption{Uniform inflow}
		\end{subfigure}\hfill
		\begin{subfigure}[b]{0.33\textwidth}
			%	\subfigure[$\frac{Heave}{c}$ comparison ]{%
			\includegraphics[width=\textwidth]{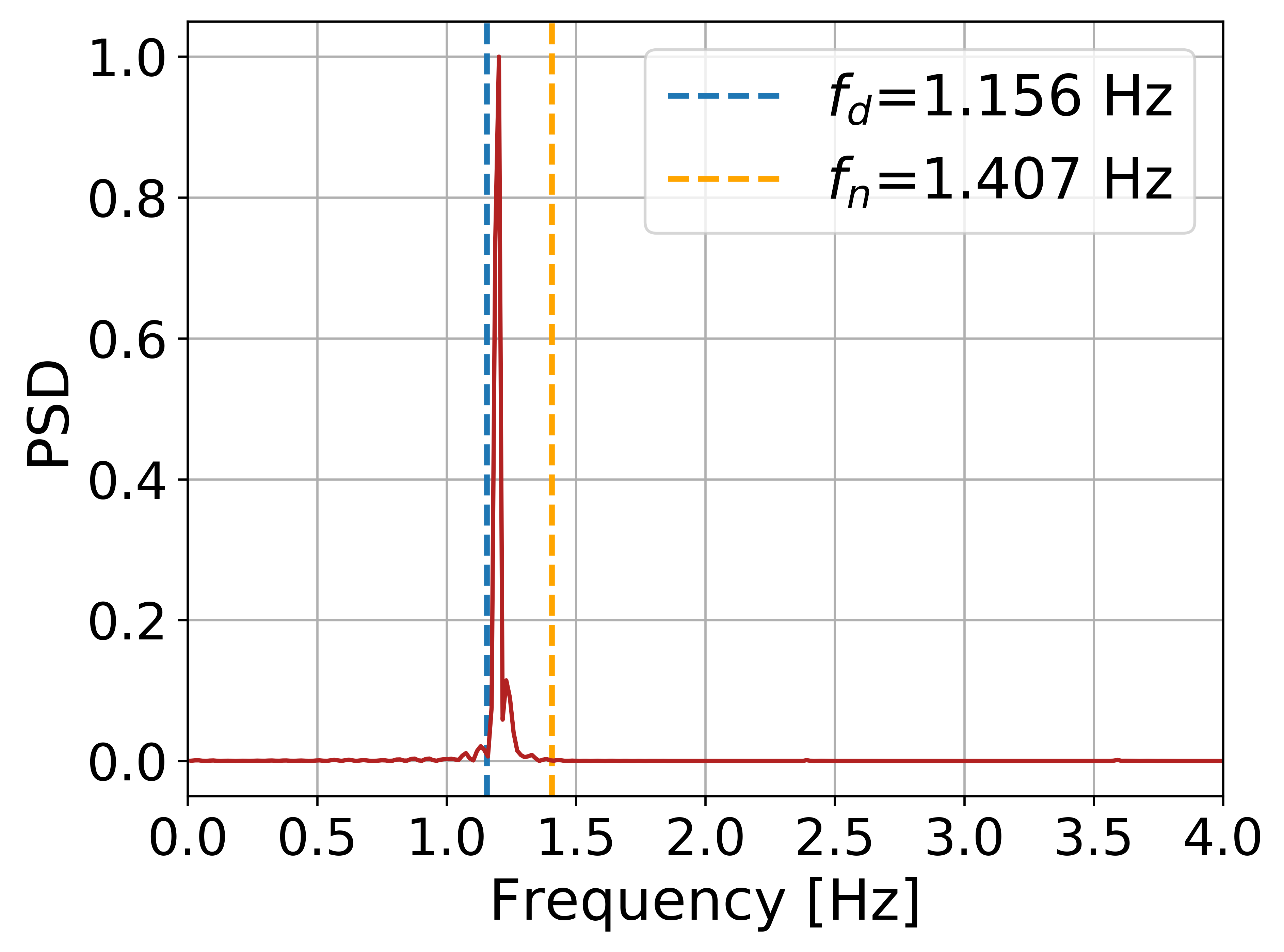}
			\caption{$ K=0.30$}
		\end{subfigure}\hfill
		\begin{subfigure}[b]{0.33\textwidth}
			%	\subfigure[$\frac{Heave}{c}$ comparison ]{%
			\includegraphics[width=\textwidth]{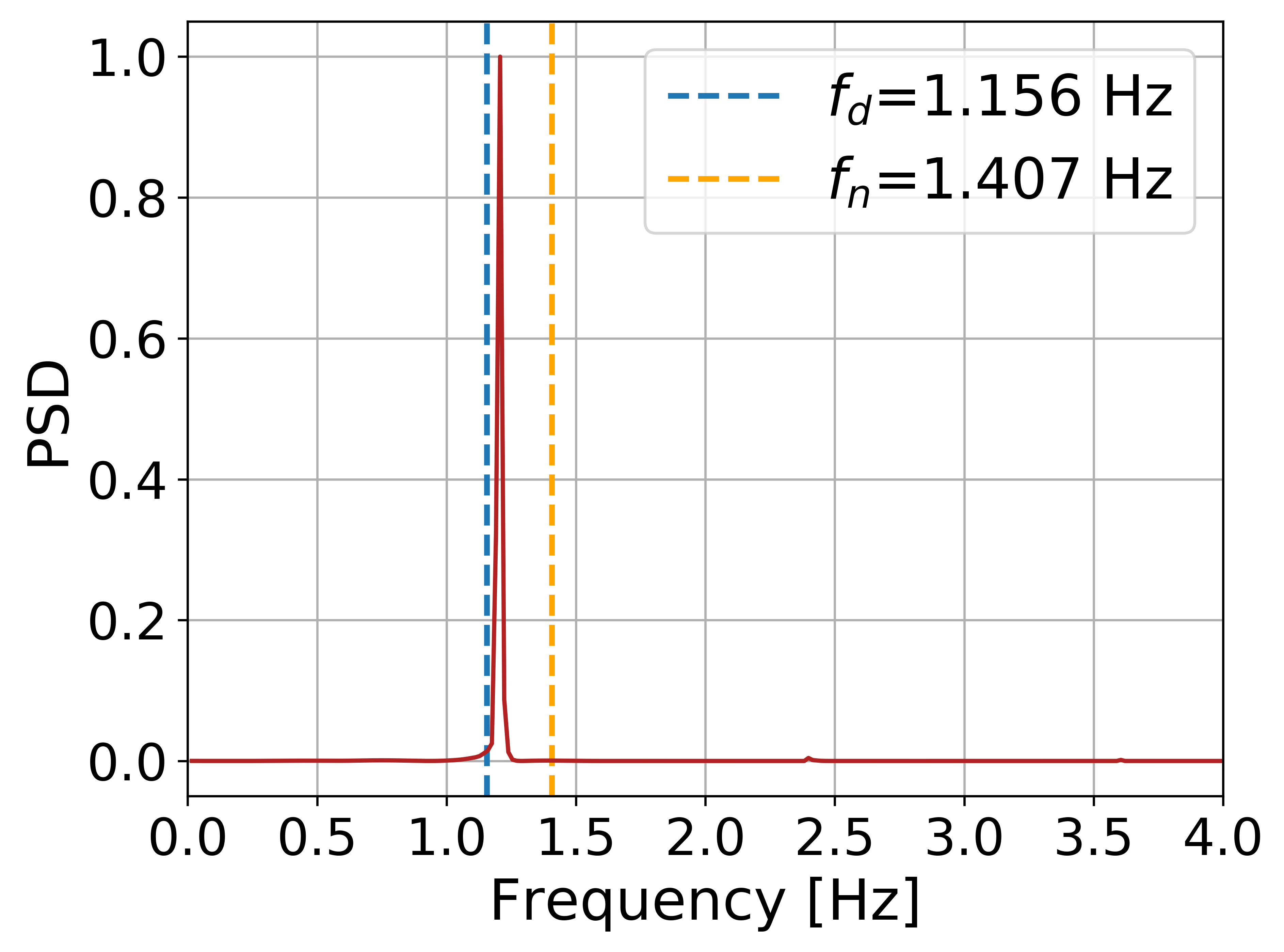}
			\caption{$ K=0.75$}
		\end{subfigure}
		%%%%%%%%%%%%%%%%%%%%%%%%%%%  K=1.00
%%%%	\begin{subfigure}[b]{0.49\textwidth}
%%%%		%	\subfigure[$\frac{Heave}{c}$ comparison ]{%
%%%%		\includegraphics[width=\textwidth]{figures/spectrums/fftK1.png}
%%%%		\caption{$ K=1.00$}
%%%%	\end{subfigure}
		\caption{Power Spectral Density (PSD) plots for heave velocity for various shear rates in one phase simulations. By $f_n$ the natural frequency of the system is denoted and by $f_d$ the damped natural frequency for the heave movement.}
		\label{fig:one_phase_spectrum}
	\end{figure}
	 
	The natural and damped natural frequency of the heave motion is also depicted, since most of the energy is harvested through this motion. We notice that the foil oscillates at approximately the damped natural frequency with only minor shifts when shear is applied.
%In the uniform case the highest spike is at the main frequency of $f=1.2\ Hz$ which is he frequency of the main oscillation the foil executes. A smaller amplitude secondary frequency is present at $f=3.59\ Hz$. Once shear is enforced to the velocity profile, a new frequency emerges at $f=2.39\ Hz$  roughly double as the main frequency.

%The amplitude of this new frequency show a monotonous increase with the shear rate in contrast with the frequency at $f=3.59\ Hz$ which declines after $K=0.75$. .

\begin{figure}[H]
        \begin{subfigure}[b]{0.38\textwidth}
         \includegraphics[width=\textwidth,keepaspectratio]{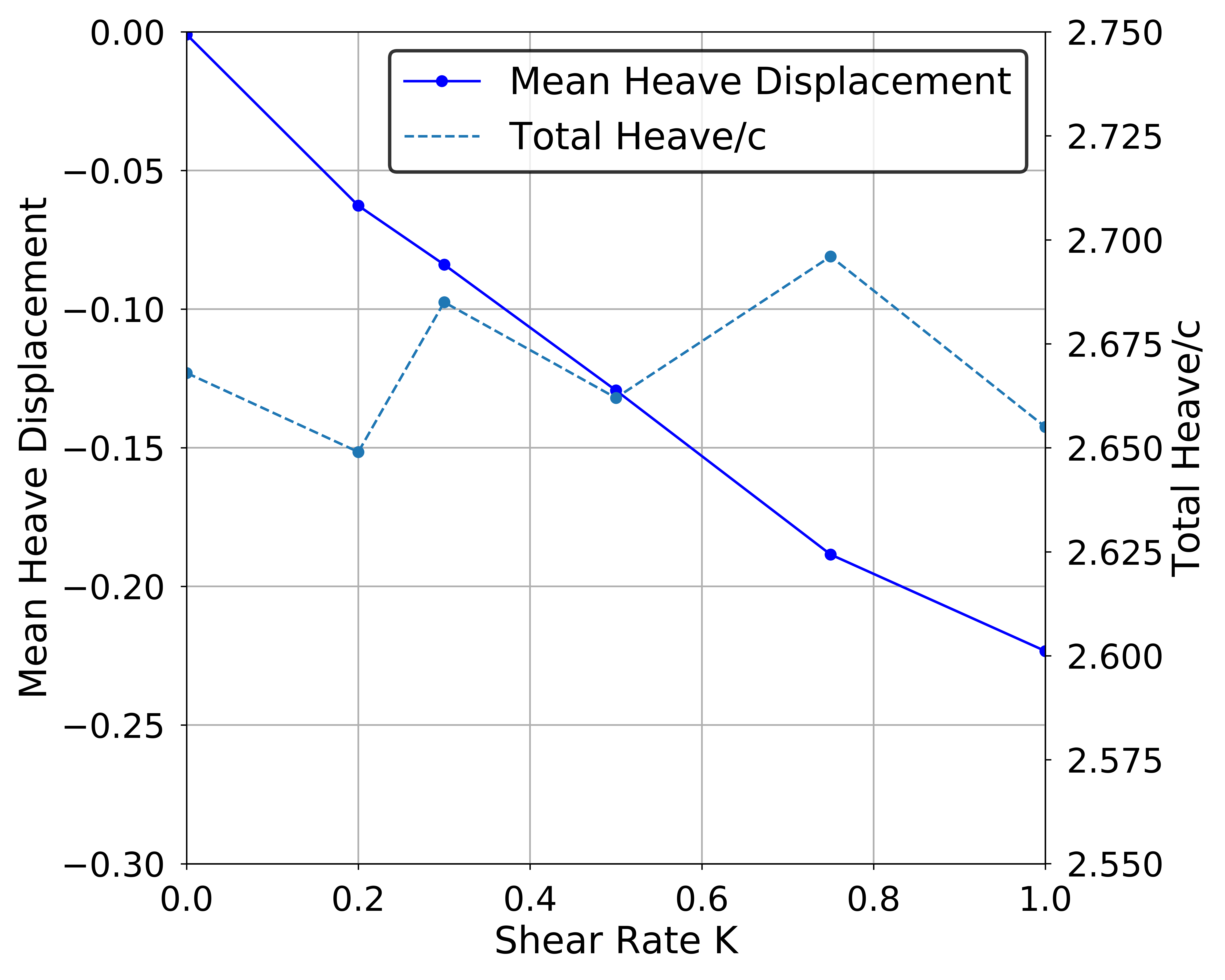}
	    \caption{Shift of mean heave position and total heave distance scanned for different shear rates}
	    \label{fig:heave_displacement}
         \end{subfigure}\hfill
        \begin{subfigure}[b]{0.30\textwidth}
		\includegraphics[width=\textwidth]{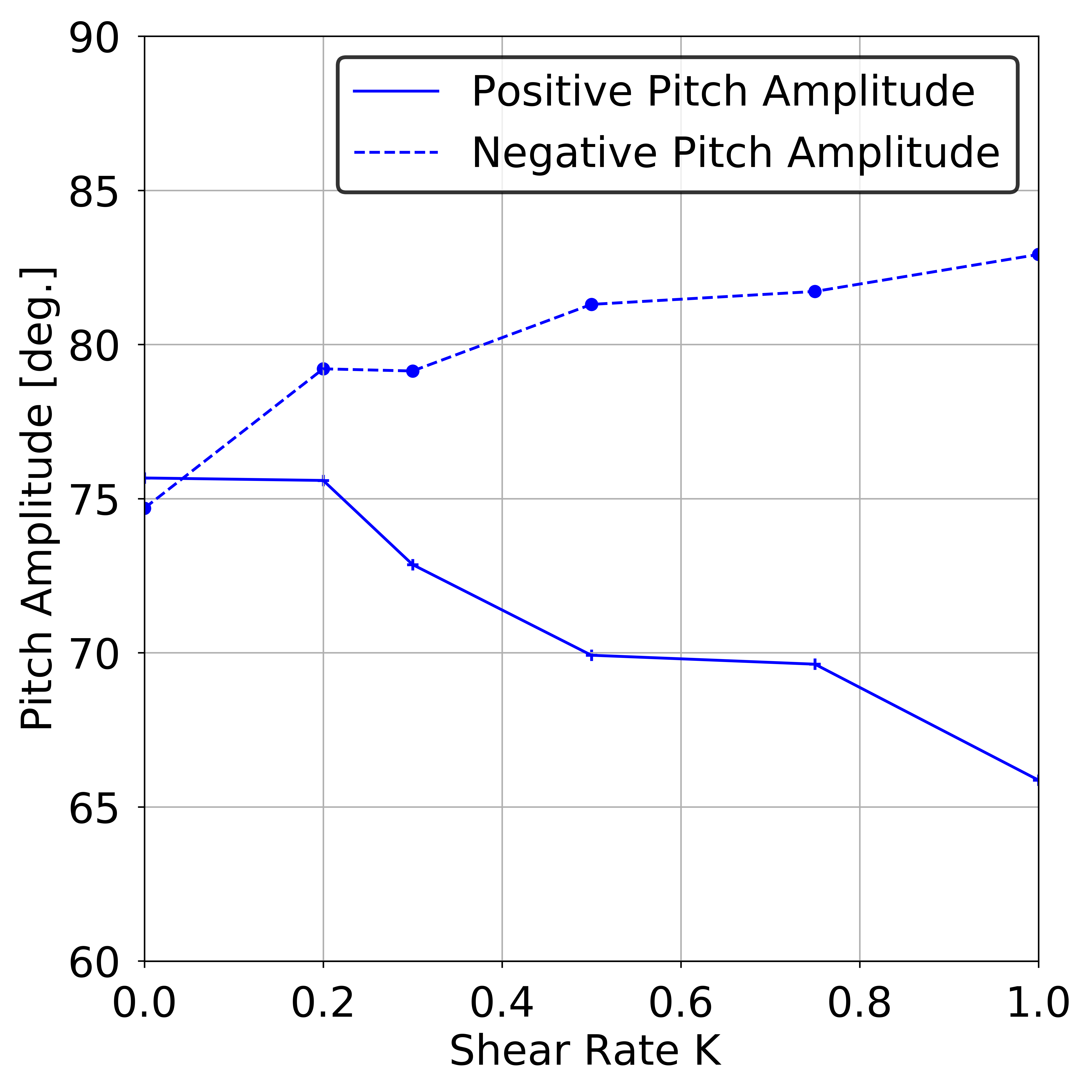}
		\caption{Positive and Negative pitch amplitudes for different shear rates.}
		\label{fig:max_pitch}
	    \end{subfigure}\hfill
	    \begin{subfigure}[b]{0.30\textwidth}
	       \includegraphics[width=\textwidth,keepaspectratio]{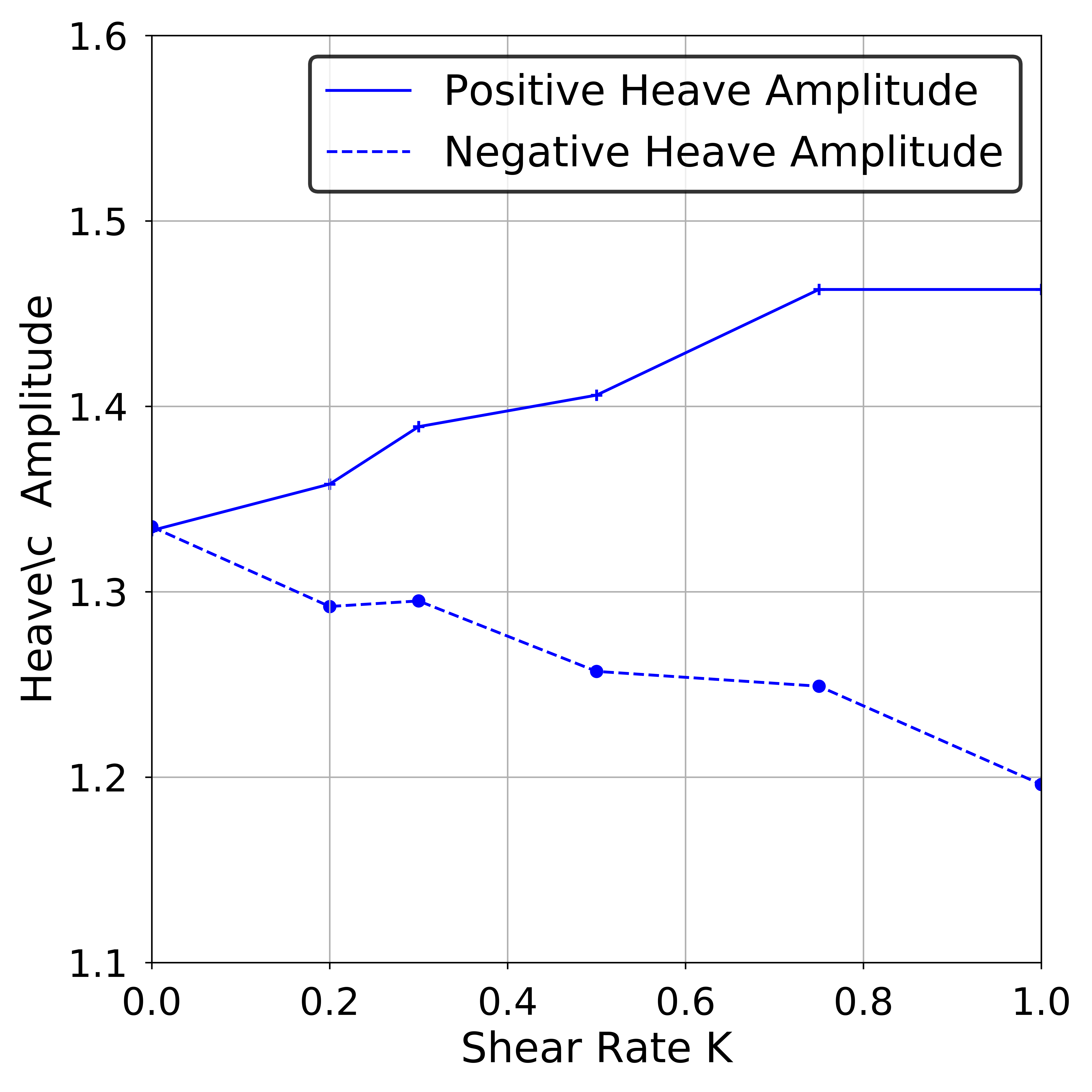}
			%\caption{$ K=0.20$}
			\caption{Positive and Negative heave amplitudes for different shear rates}
			\label{fig:max_min_heave}
	    \end{subfigure}
	    \caption{Comparison of the kinematic quantities of the passive foil for various shear rates in one phase simulations}
\end{figure}

Regarding the heave kinematics, the mean heave displacement  can be seen in \autoref{fig:heave_displacement}. We observe a downward shift of the mean position of the foil, which is in compliance with the results derived by Zhu \cite{Zhu2012}. Similar to Zhu's observation, a passive foil tends to shift vertically (in a mean sense) towards regions with lower velocities.

Additionally, in the sheared case the amplitude of the heave motion is  non-symmetric since larger heave amplitudes are noted in the region with higher velocities (positive heave amplitude). Higher velocities lead to more intense hydrodynamic forces and exaggerated heave responses from the passive foil, \autoref{fig:max_min_heave}. An increase of 9.75\% and a decrease of 10.41\% is noted for the positive and negative heave amplitude respectively between uniform flow and maximum shear rate. This translates to larger maximum than minimum heave in the case of the shear flow and becomes more noticeable in strong shear rates.

Furthermore, the total heave distance greatly affects the performance and efficiency of the foil since when the hydraulic cross section is shrinking the total energy available for extraction is reduced. This distance differs depending on the shear rate and may be smaller or larger than the uniform flow case. The trends for the positive and negative heave amplitudes (\autoref{fig:max_min_heave}) counter balance each other leading to small changes in the total heave distance scanned (see \autoref{fig:heave_displacement}). 

Regarding the pitch motion, the positive pitch amplitude decreases by as much as 12.95\%  while it increases by 11\% for the negative amplitude.  {\color{black}For the sake of clarity}, the counter-clockwise rotation encountered in the negative heave positions is considered as positive (see \autoref{fig:max_pitch}). 
This observation is in compliance with the above conclusions since we expect moderate foil responses in the bottom position where lower flow velocities are encountered and the opposite behavior is expected in the upper position where higher velocities are encountered and foil responses are exaggerated.

\begin{figure}[H]
		%	\centering
		%%%%%%%%%%%%%%%%%%%%%%%%% uniform
		\begin{subfigure}[b]{0.45\textwidth}
			\includegraphics[width=\textwidth,keepaspectratio]{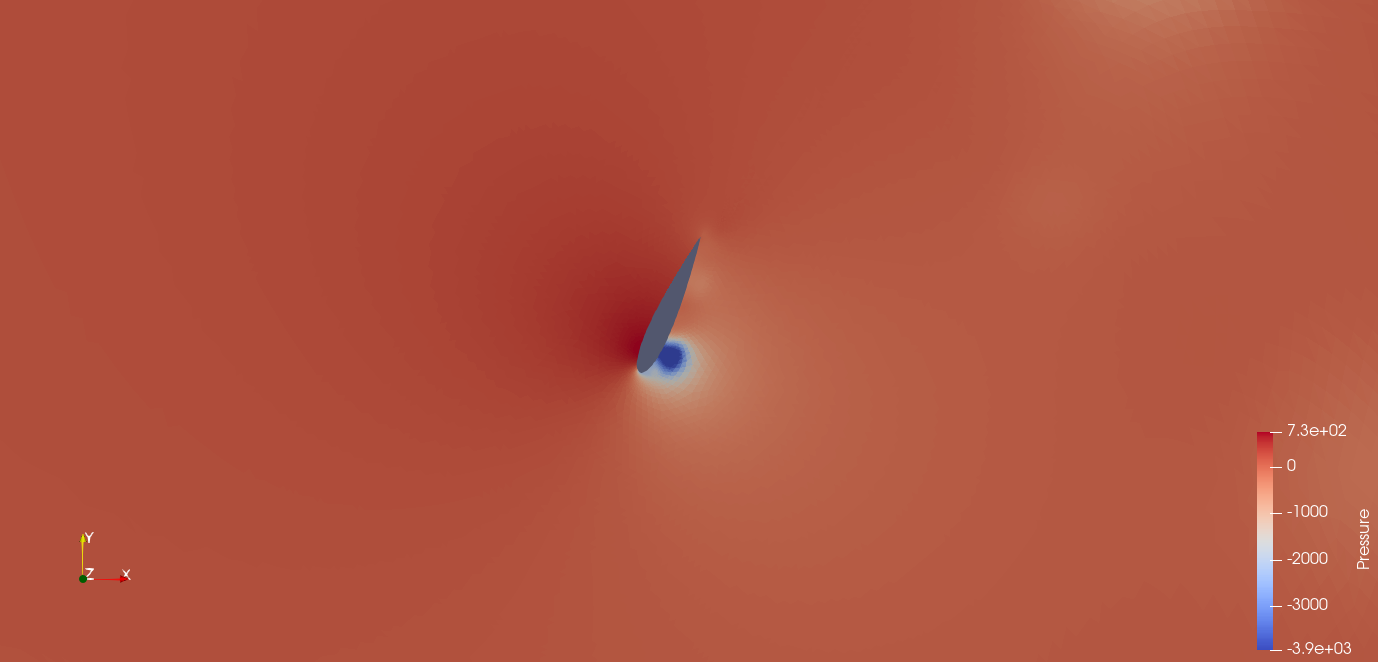}
			%\caption{$Uniform\ \ flow$}
			\vspace{0.01 cm}
		\end{subfigure}
		\hfill
		\begin{subfigure}[b]{0.45\textwidth}
			\includegraphics[width=\textwidth,keepaspectratio]{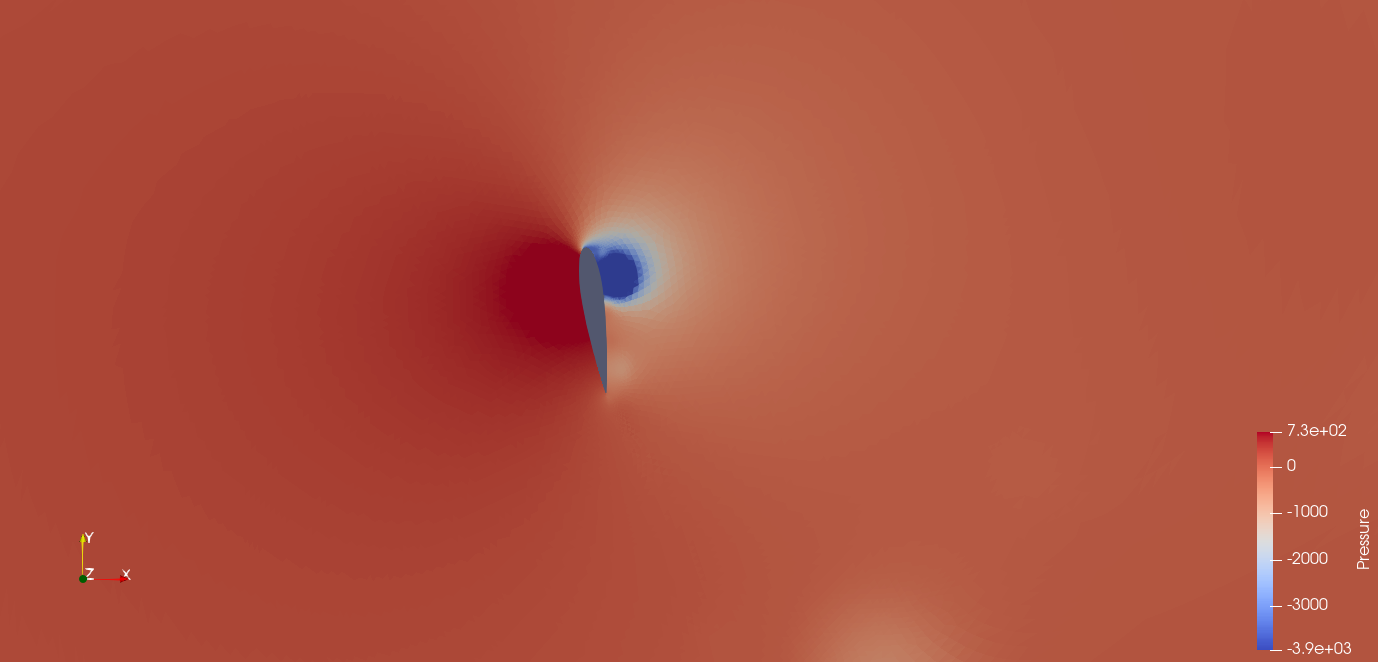}
			%\caption{$ K=0.20$}
			\vspace{0.01 cm}
		\end{subfigure}
		\\
		%%%%%%%%%%%%%%%%%%%%%%% K=0.30
		\begin{subfigure}[b]{0.45\textwidth}
			%	\subfigure[$\frac{Heave}{c}$ comparison ]{%
			\includegraphics[width=\textwidth,keepaspectratio]{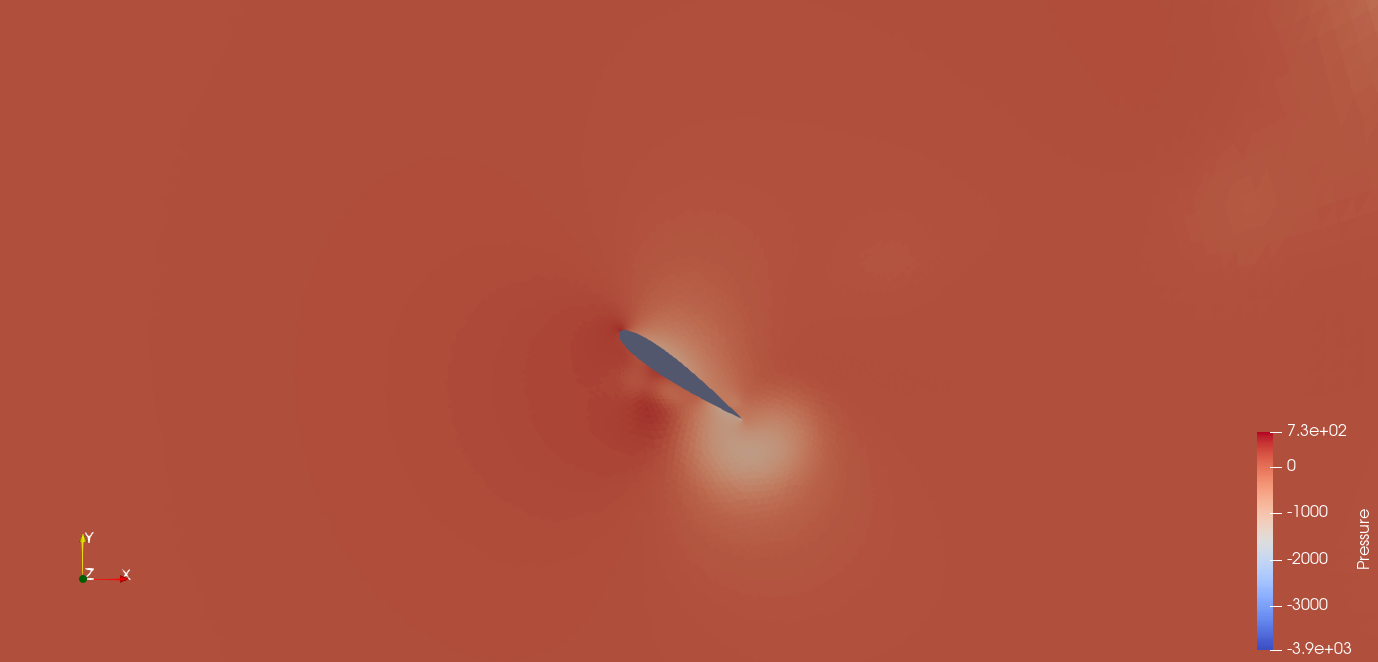}
			%\caption{$K=0.30$}
		\end{subfigure}
		\hfill
			%%%%%%%%%%%%%%%%%%%%%%% K=0.50
			\begin{subfigure}[b]{0.45\textwidth}
				%	\subfigure[$\frac{Heave}{c}$ comparison ]{%
			\includegraphics[width=\textwidth,keepaspectratio]{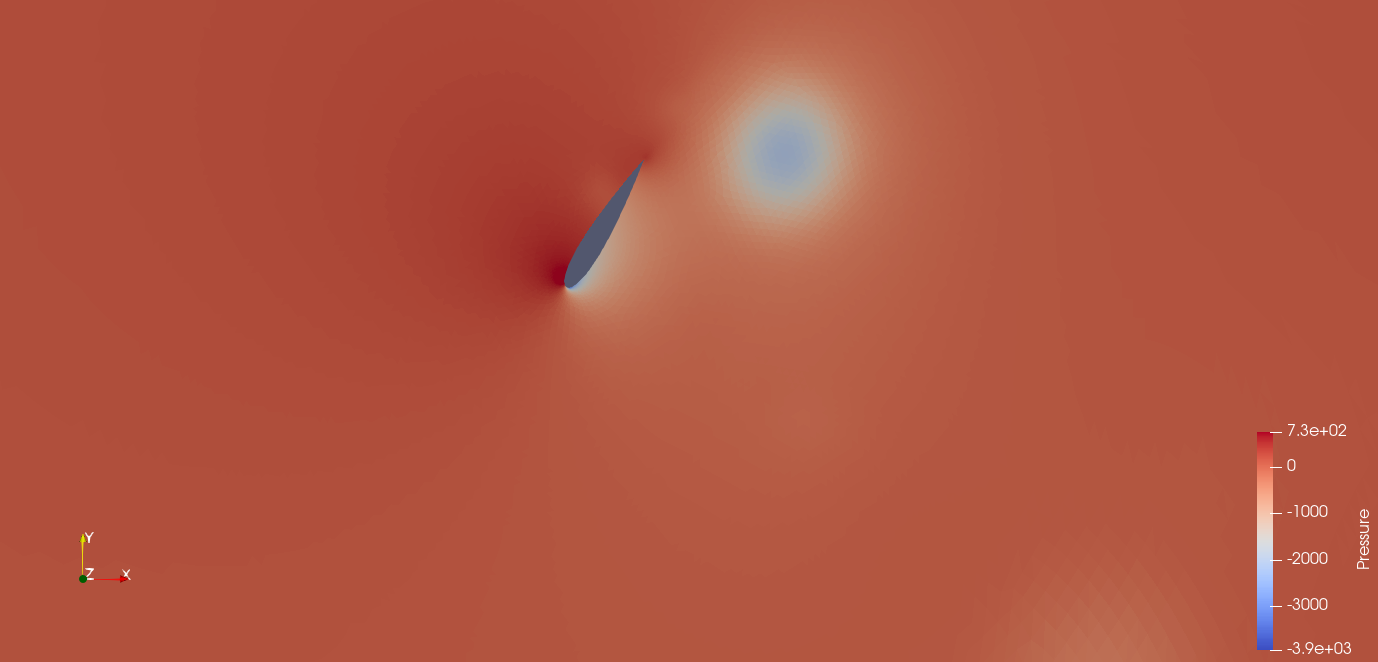}
				%\caption{$K=0.50$}
			\end{subfigure}
		\caption{Left: Pressure contour of the flow field at negative heave amplitude before and after the LEV shedding. Right: Pressure contour of the flow field at positive heave amplitude before and after the LEV shedding. The time interval between the consecutive screenshots is the same between left and right figures.  }
		\label{fig:vertical_shift_LEV}
	\end{figure}

 The effect of shear on the kinematics of the foil can be attributed to the lower velocities encountered in the negative heave amplitude which lead to a slower convection of the LEV. This is illustrated in \autoref{fig:vertical_shift_LEV} where pressure contours for  the maximum negative heave position (left) and the  maximum positive heave position (right) are displayed. When the heave is negative (lower inflow velocity) the low pressure zone of the LEV convects  {\color{black}more slowly} with respect to the maximum positive heave position, thus remaining close to the foil for a longer period of time.

\begin{figure}[H]
	    \centering
		 \includegraphics[width=0.7\textwidth]{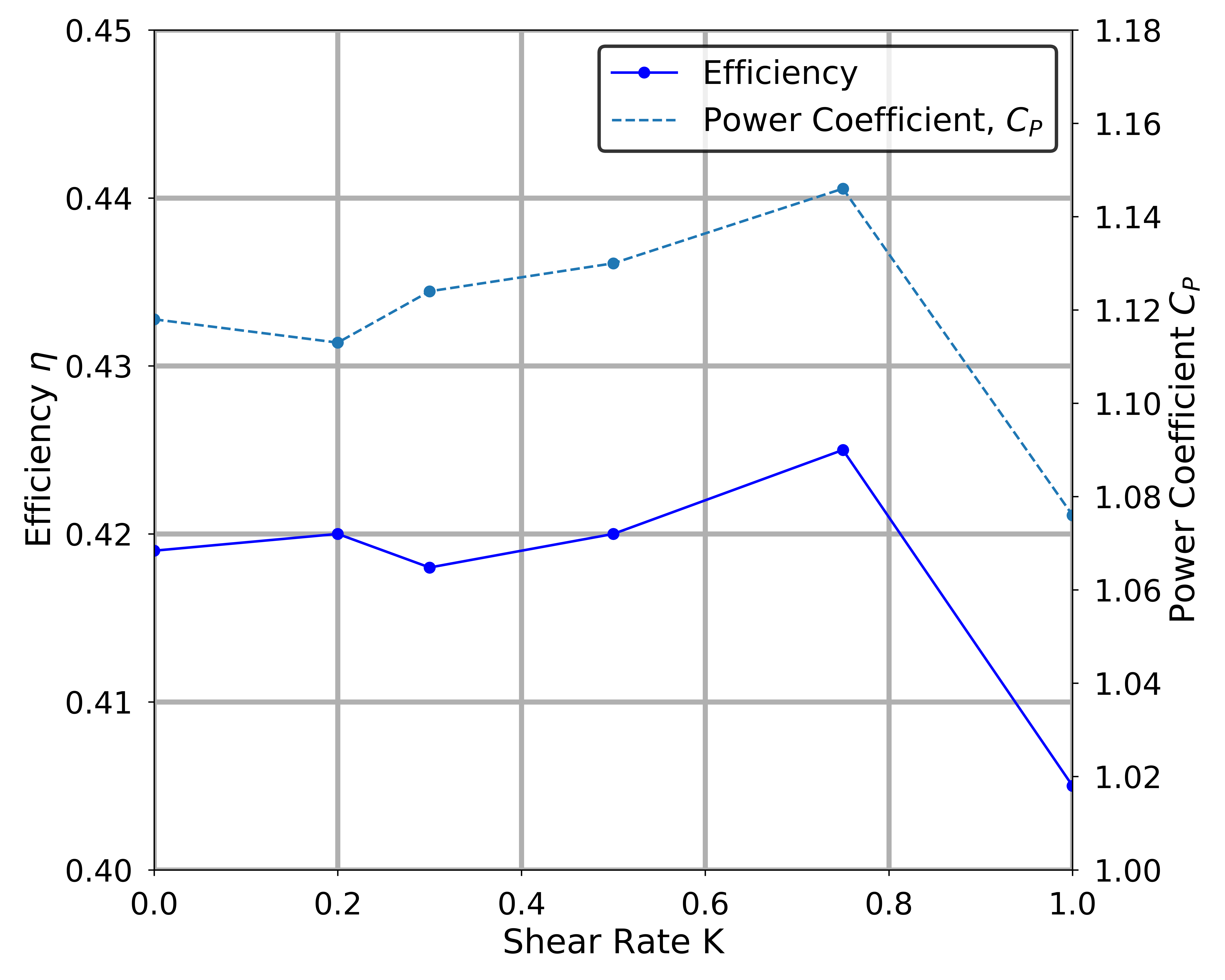}
		\caption{Efficiency and power coefficient for different shear rates}
		\label{fig:eff_cp_one_phase}
	\end{figure}

The efficiency ($\eta$) and the power coefficient ($\overline{C}_p$) are directly affected by the above kinematic parameters. In  \autoref{fig:eff_cp_one_phase}, $\eta$ and $\overline{C}_p$ for the various shear rates can be seen. A peak in both $\eta$ and $\overline{C}_p$ is observed at $K=0.75$. However the performance enhancement is relatively small, yielding an increase of 1.42\% for the efficiency and 2.50\% for the power coefficient.  Performance deteriorates significantly  at the larger shear rate of $K=1.00$ with a decrease of 3.34\% for the efficiency and 3.76\% for the power coefficient.
	% Comment the non-uniform pitch amplitude in the uniform flow case

% for use in caption of a graph % Shift of the mean vertical position of the foil in different shear flows. The profiles with dashed lines indicate the original location of the foil. Those with solid lines show the shifted mean position. Source : \cite{Zhu2012}

%   \begin{figure}[H]  
%   \includegraphics[width=\textwidth]{figures/duarte_shear/heave_pitch_eff/Efficiency_and_cp_to_K_duarte_shear.png}
%    \caption{Efficiency and power coefficient for different shear rates}
%   \label{fig:eff_cp_one_phase}
%   \end{figure}
\subsection{Operation of the passive foil under the free surface in uniform and sheared inflow {\label{sec:free}}}

\begin{figure}[H]
			\centering
		\begin{subfigure}[b]{\textwidth}
			%	\subfigure[$Pitch$ comparison ]{%
			\includegraphics[width=\textwidth]{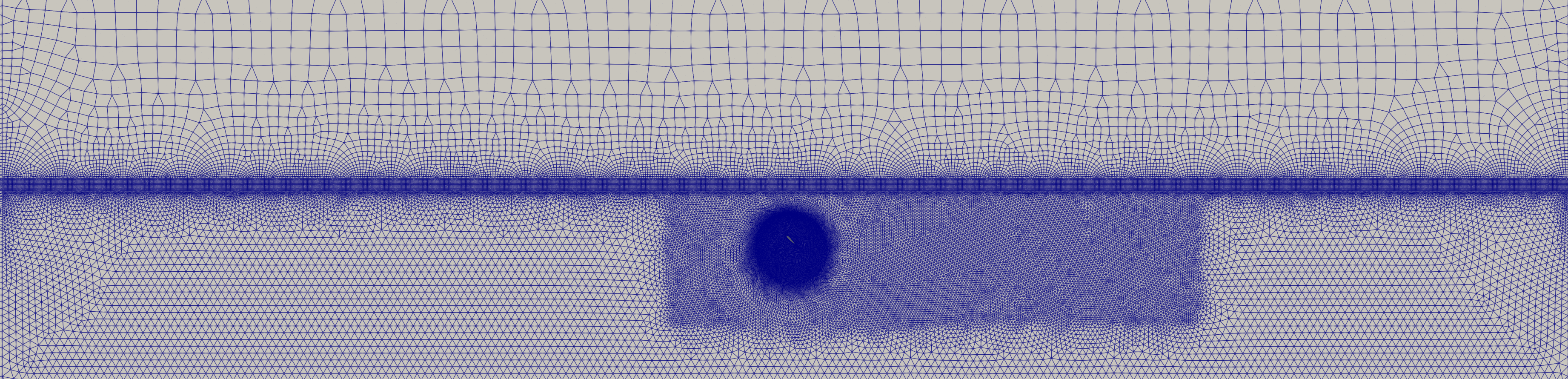}
%			\caption{$Mesh\ Overview $}
		\end{subfigure}
%	\\
%`	\vspace{1cm}
%    \begin{subfigure}[b]{0.49\textwidth}
%    \includegraphics[width=\textwidth]{figures/free_surface_shear/mesh/close_view_free_surface.png}
%    \end{subfigure}
	\caption{Mesh  overview  and  refinement zones for two-phase simulations \label{fig:freemesh}}
   	\end{figure}

    In this section the effect of the free surface is also taken into account. To that end two-phase flow (water-air) is considered. A new grid is  generated since a new refinement region {\color{black} should} be defined in the free surface vicinity. Nevertheless, the grid in the near airfoil regions remains approximately the same as in \autoref{sec:Duarte_case}. Mesh snapshots can be seen in \autoref{fig:freemesh}. The VoF method was utilized to capture the free surface interface. The foil is positioned 5.5$c$ below the free surface and 10$c$ above the sea bed. At the farfield a damping region  is employed to avoid the pollution of the solution with reflections from the generated water waves \cite{Ntouras2020}. The mass-spring-damper configuration of the passive foil remained identical to the previous simulations in order to isolate the effect of the free surface. 
	
The same families of shear rates were chosen as in the previous section, namely a mild and a strong one. The mild and strong shear groups consist of $K=\left[0,\, 0.2,\, 0.3 \right]$ and $K=\left[0.75,\, 1.00 \right]$ respectively. The time-step employed remained identical to the previous simulations $\Delta t = 0.0002$s

\begin{figure}[H]
		%	\centering
		%%%%%%%%%%%%%%%%%%%%%%%%% uniform
		\begin{subfigure}[b]{0.49\textwidth}
			\includegraphics[width=\textwidth]{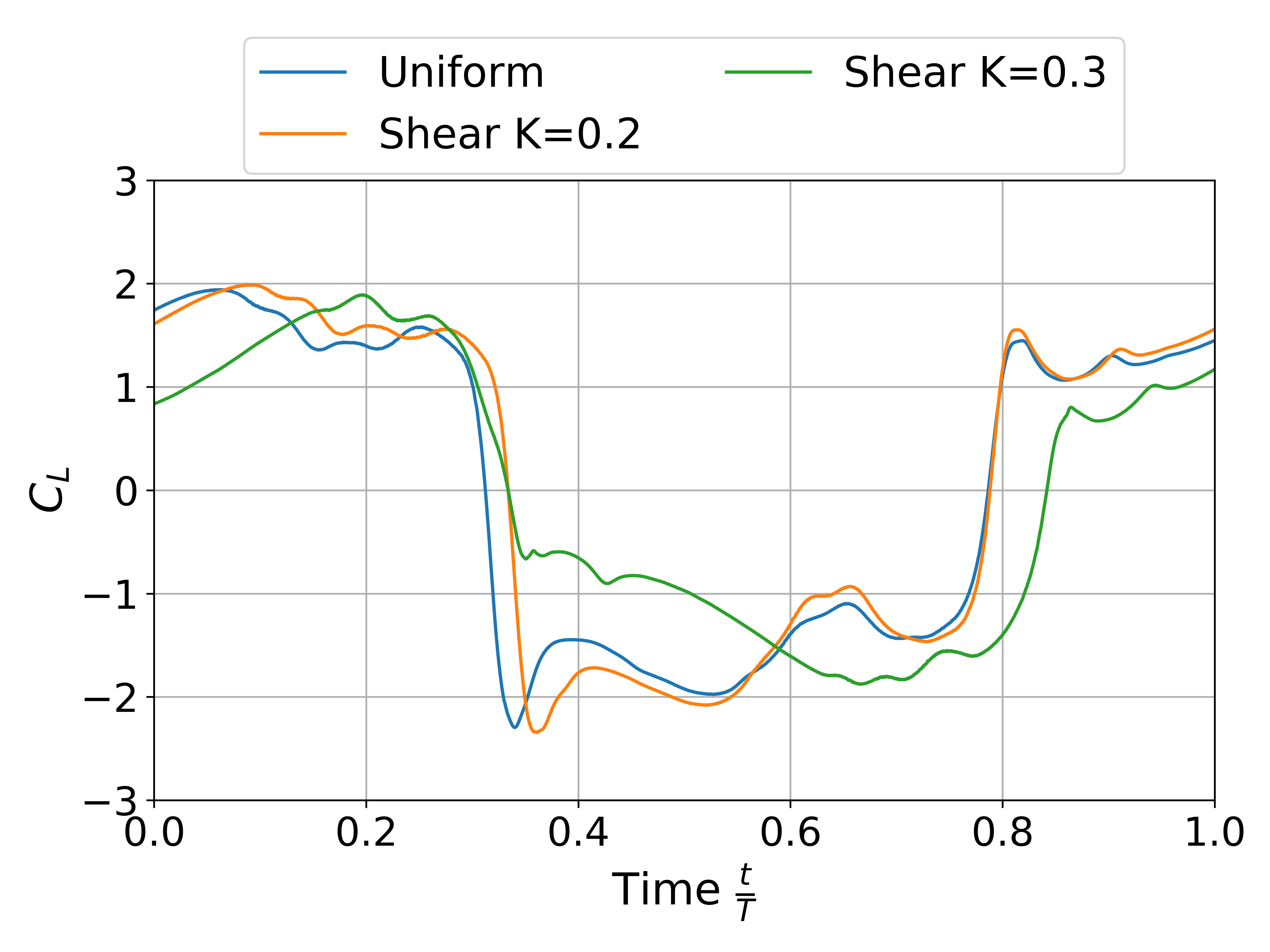}
			%\caption{$Uniform\ \ flow$}
			\vspace{0.01 cm}
		\end{subfigure}
		\hfill
		\begin{subfigure}[b]{0.49\textwidth}
			\includegraphics[width=\textwidth]{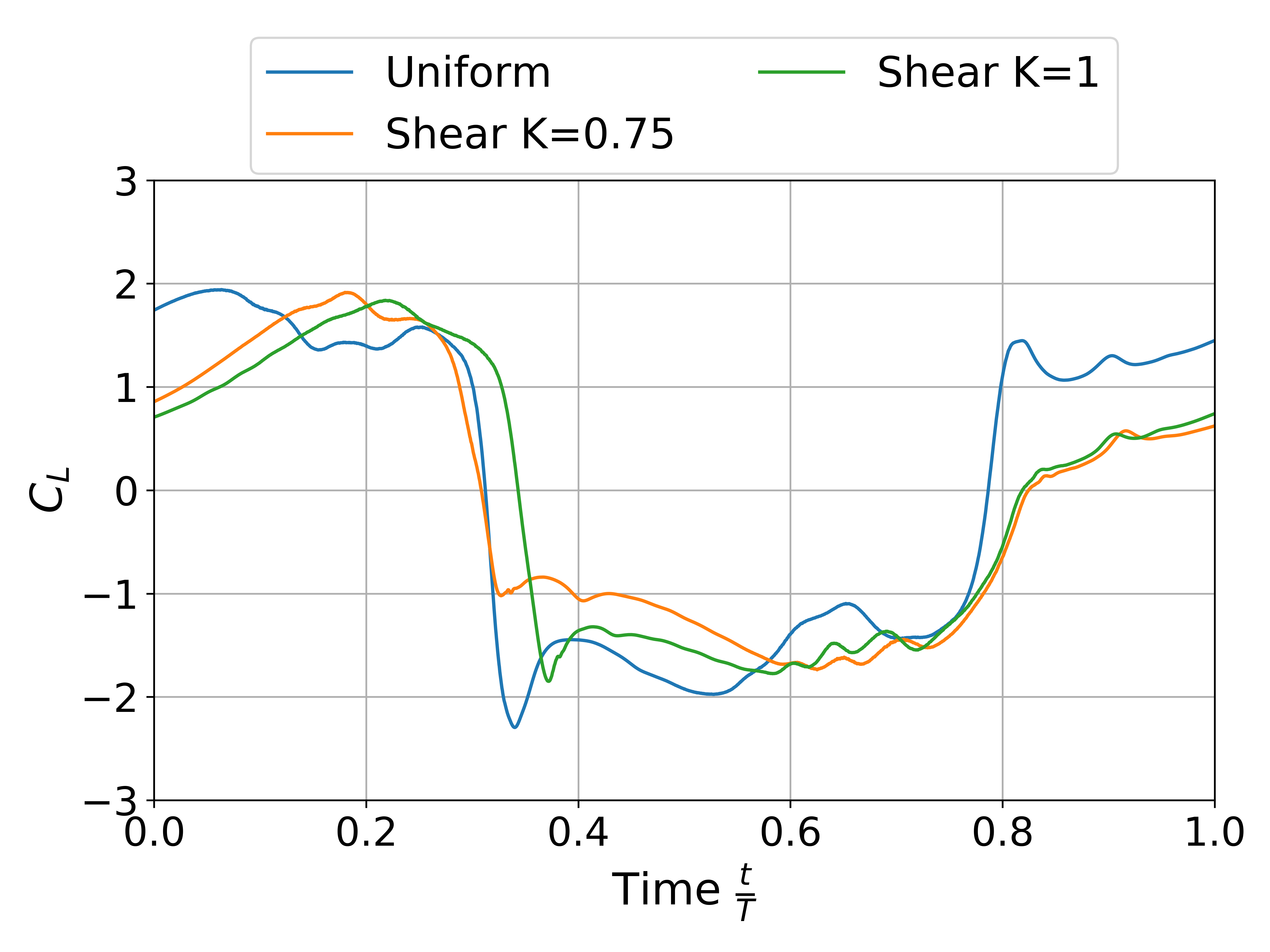}
			%\caption{$ K=0.20$}
			\vspace{0.01 cm}
		\end{subfigure}
		\\
		%%%%%%%%%%%%%%%%%%%%%%% K=0.30
		\begin{subfigure}[b]{0.49\textwidth}
			%	\subfigure[$\frac{Heave}{c}$ comparison ]{%
			\includegraphics[width=\textwidth]{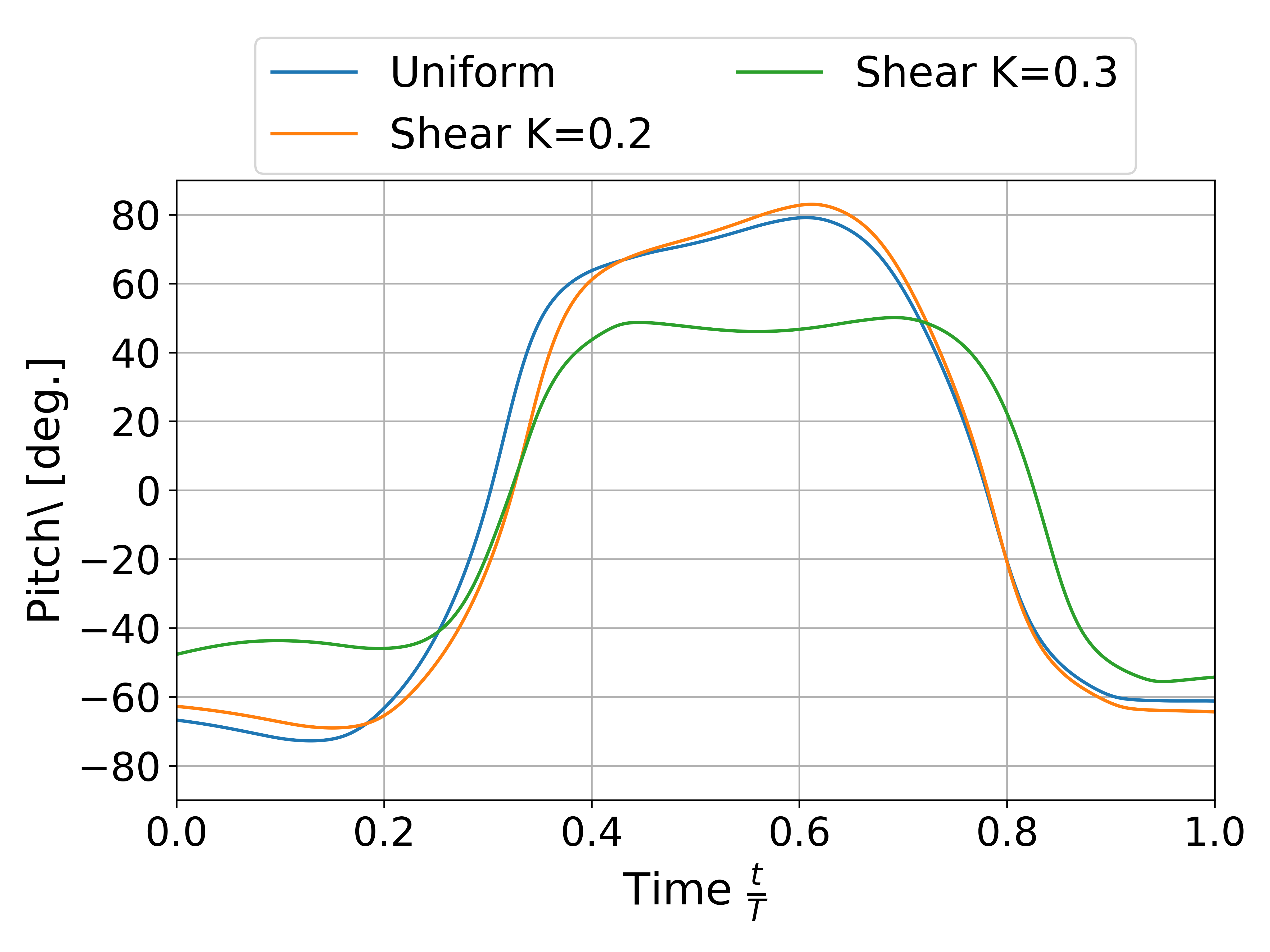}
			%\caption{$K=0.30$}
		\end{subfigure} 
		\hfill
			%%%%%%%%%%%%%%%%%%%%%%% K=0.50
			\begin{subfigure}[b]{0.49\textwidth}
				%	\subfigure[$\frac{Heave}{c}$ comparison ]{%
			\includegraphics[width=\textwidth]{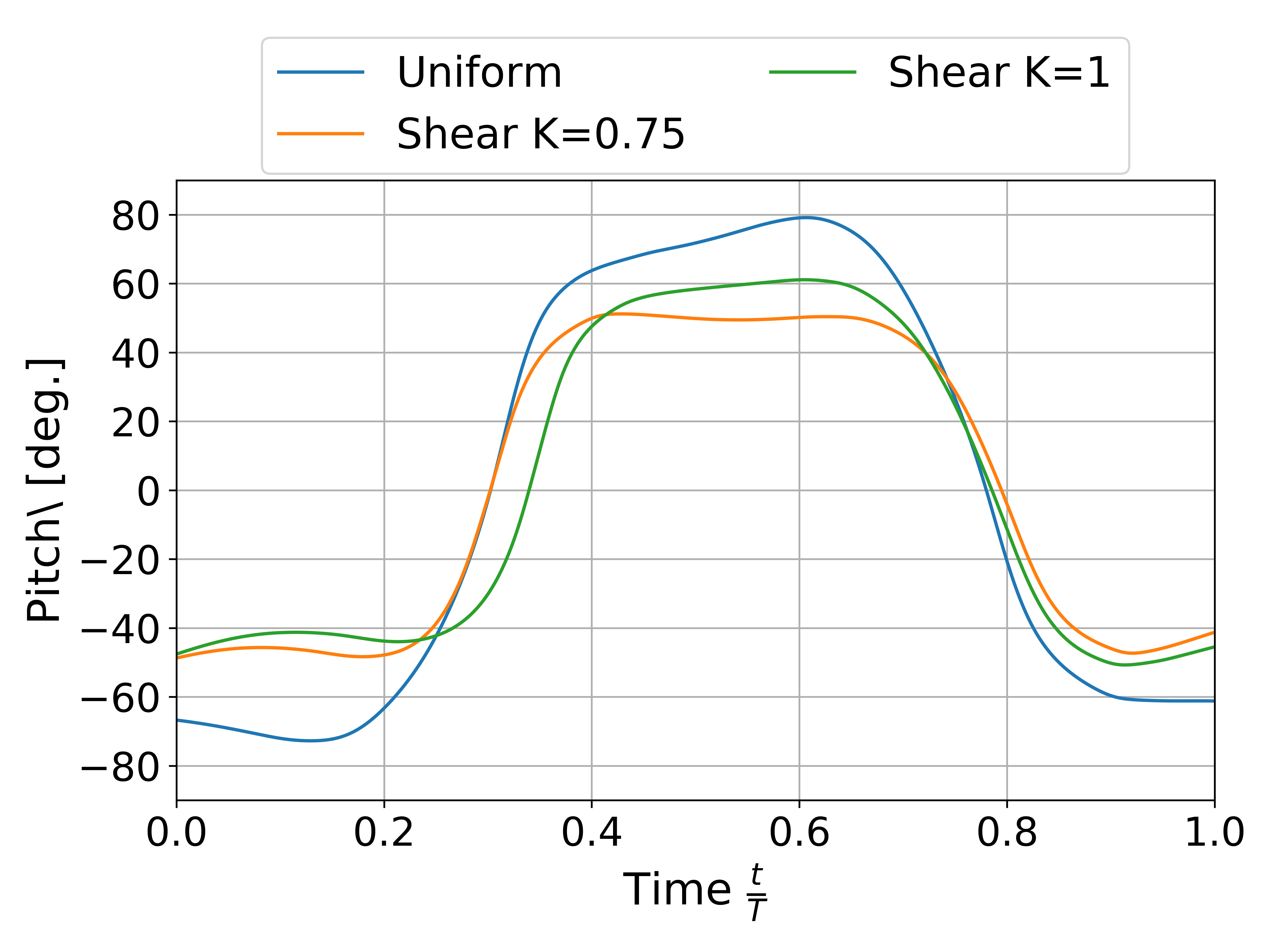}
				%\caption{$K=0.50$}
			\end{subfigure}
		\caption{  Lift coefficient $C_L$ (upper) and Pitch (lower) comparison over a period. Mild shear cases are presented in the left while strong shear cases in the right.}
		\label{fig:fsloads}
	\end{figure}

The pitch of the foil as well as the lift coefficient ($C_L$) for one oscillating cycle is presented in  \autoref{fig:fsloads}. Initially, it is evident that the changes in both the pitch response as well as the $C_L$ are more pronounced in this case as the shear varies. It is noted here that, apart from the shear rate of $K=0.2$, in all other cases the period of each oscillating cycle varied slightly during the simulation. 

\begin{figure}[H]
	\vspace{-1 cm}
	\begin{subfigure}[b]{\textwidth}
		%	\subfigure[$\frac{Heave}{c}$ comparison ]{%
		\includegraphics[width=\textwidth]{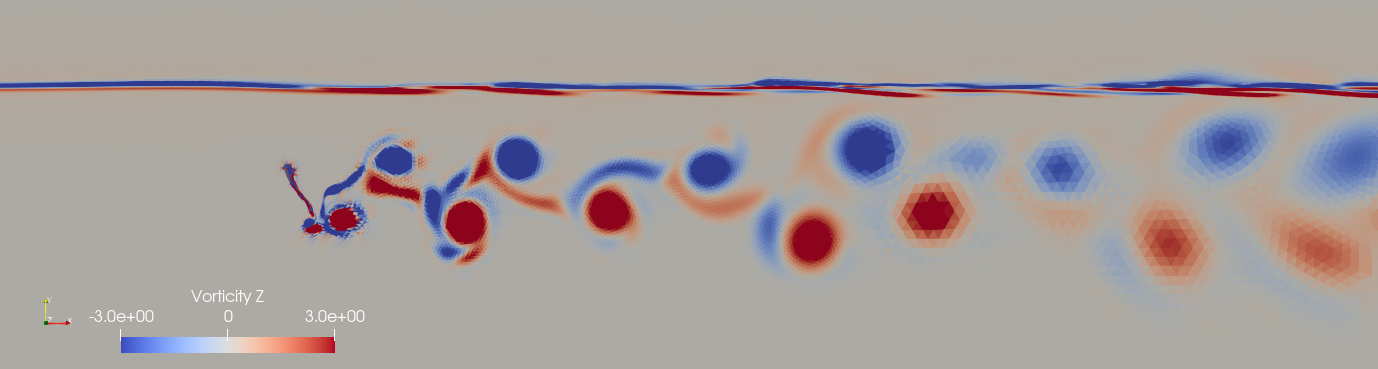}
		\caption{$Uniform\ \ Flow$}
		\vspace{0.5cm}
	\end{subfigure}
	\\
	\begin{subfigure}[b]{\textwidth}
		%	\subfigure[$Pitch$ comparison ]{%
		\includegraphics[width=\textwidth]{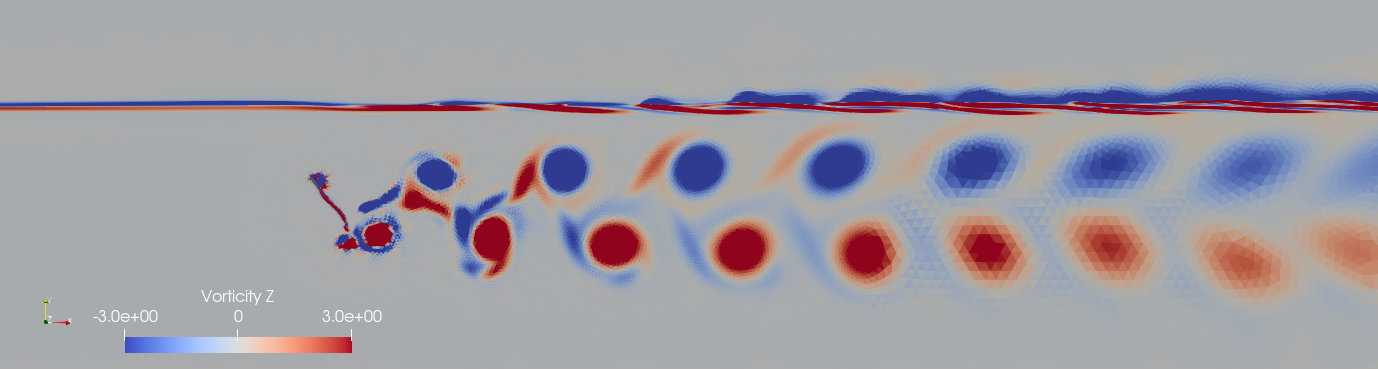}
		\caption{$Shear\ \ Rate\ \ K=0.20$}
		\vspace{0.5cm}
	\end{subfigure}
    \\
   \begin{subfigure}[b]{\textwidth}
	%	\subfigure[$Pitch$ comparison ]{%
	    \includegraphics[width=\textwidth]{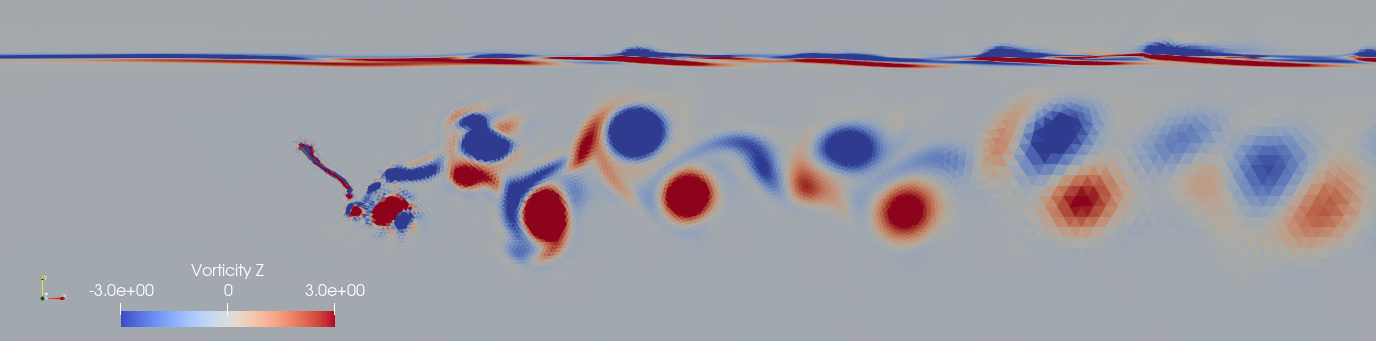}
	\caption{$Shear\ \ Rate\ \ K=0.30$}
	\vspace{0.5cm}
    \end{subfigure}
    \\
  \begin{subfigure}[b]{\textwidth}
	%	\subfigure[$Pitch$ comparison ]{%
	\includegraphics[width=\textwidth]{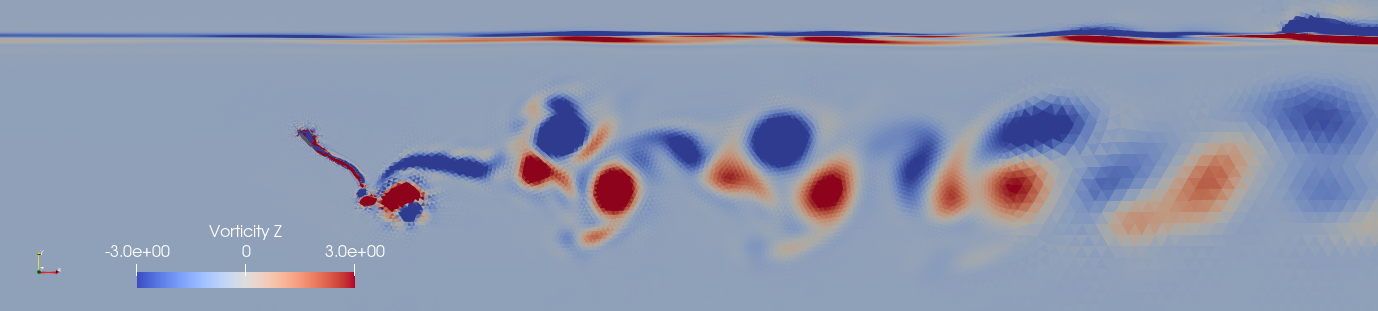}
	\caption{$Shear\ \ Rate\ \ K=0.75$}
	\vspace{0.5cm}
\end{subfigure}
\\
\begin{subfigure}[b]{\textwidth}
	%	\subfigure[$Pitch$ comparison ]{%
	\includegraphics[width=\textwidth]{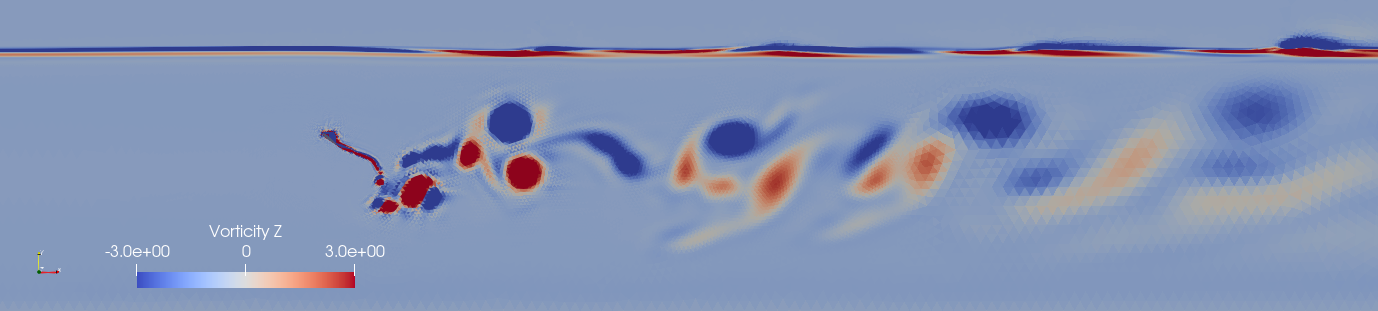}
	\caption{$Shear\ \ Rate\ \ K=1.00$}
\end{subfigure}
	\caption{Vorticity Contours showing the vortex street produced by different mild and strong shear rates in free surface simulations  }
	\label{fig:passive_free_surface_shear_vorticity}
\end{figure}

When the passive foil operates near the water level, the shed vorticity interacts with the free surface. This is evident in \autoref{fig:passive_free_surface_shear_vorticity} where vorticity contours for the various shear rates can be seen. It is clear, that the vortex street in this case has major differences in comparison with the one phase simulation, in  \autoref{fig:vort_one_phase}. The emitted vorticity interacts with the free surface leading to more complex wake structures. 
	
 An interesting observation emerges from the vortex street of the mildest shear rate, $K=0.2$ where order has been reconstituted and a unique shedding frequency dominates the pattern. This smoothing effect of the mild shear is lost in $K=0.3$ and higher shear rates.

\par
\begin{figure}[H]
		%	\centering
		\begin{subfigure}[b]{0.33\textwidth}
			%	\subfigure[$Pitch$ comparison ]{%
			\includegraphics[width=\textwidth]{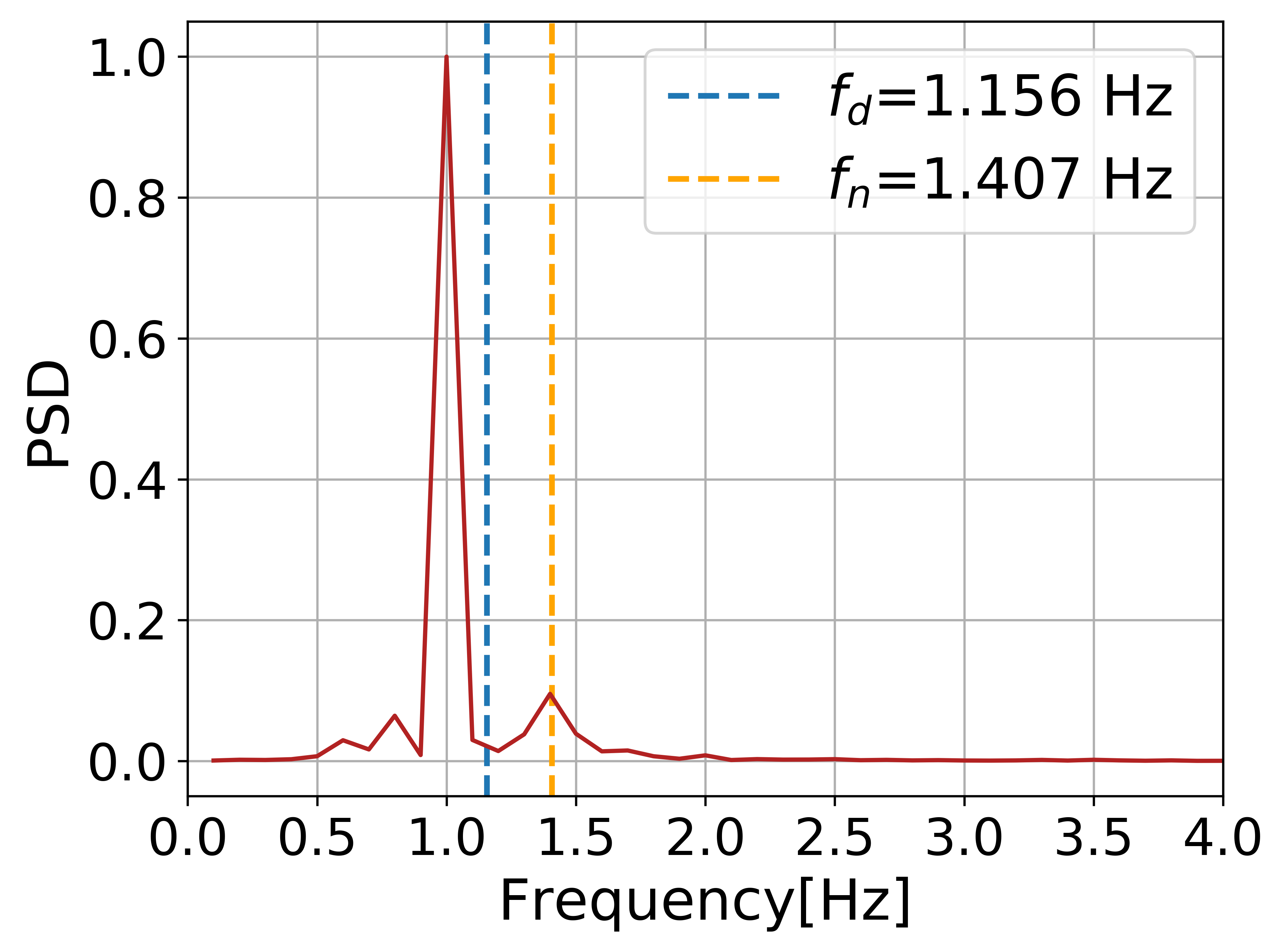}
			\caption{Free surface uniform inflow}
		\end{subfigure}\hfill
		%%%%%%%%%%%%%%%%%%%%%%% K=0.2
		\begin{subfigure}[b]{0.33\textwidth}
			%	\subfigure[$Pitch$ comparison ]{%
			\includegraphics[width=\textwidth]{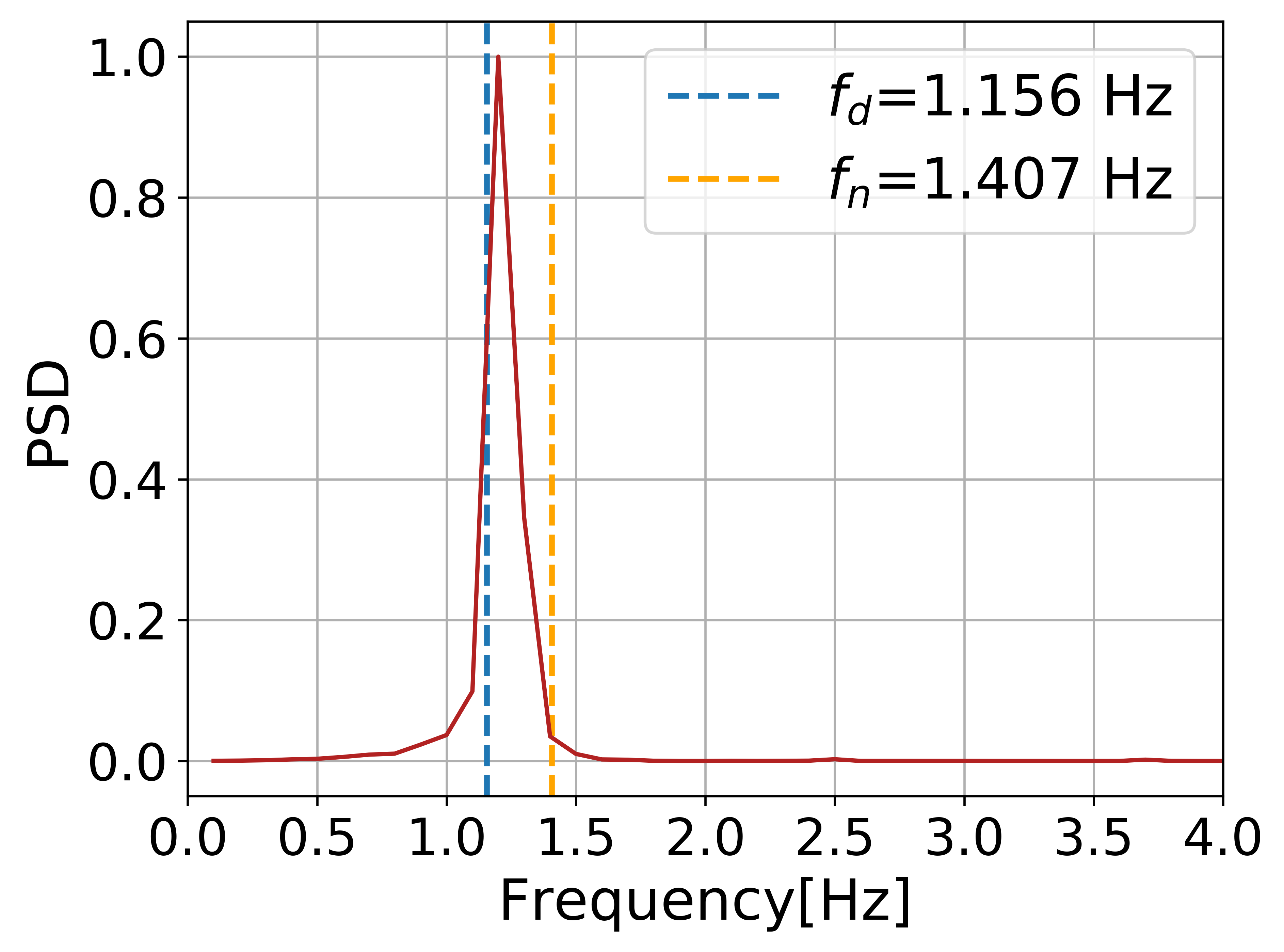}
			\caption{Free surface $K=0.20$}
		\end{subfigure}\hfill
		\begin{subfigure}[b]{0.33\textwidth}
			%	\subfigure[$Pitch$ comparison ]{%
			\includegraphics[width=\textwidth]{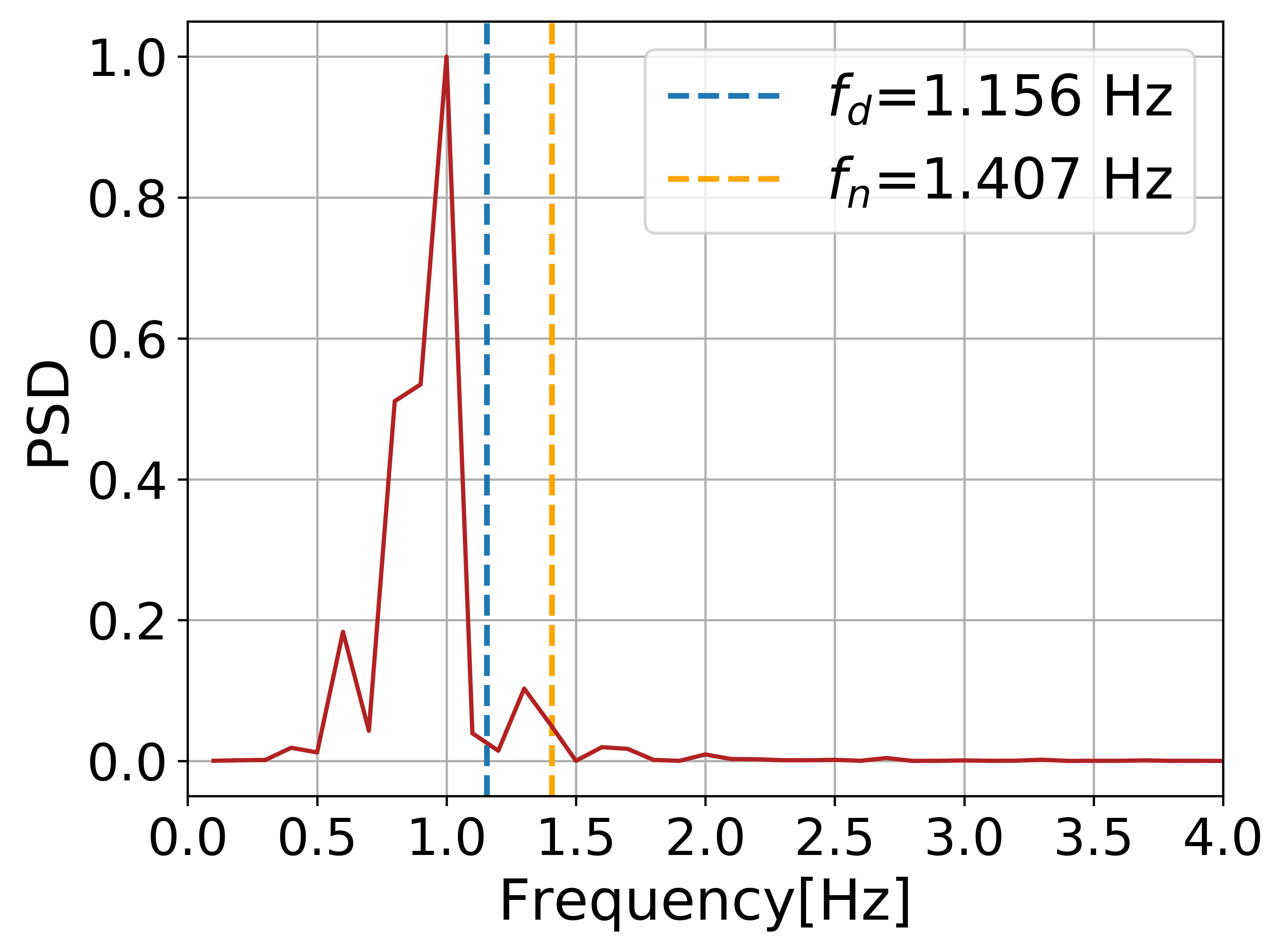}
			\caption{Free  surface  $K=0.30$}
		\end{subfigure}\\ %hfill
			%%%%%%%%%%%%%%%%%%%%%%% K=0.75
		\begin{subfigure}[b]{0.5\textwidth}
			%	\subfigure[$Pitch$ comparison ]{%
			\centering
			\includegraphics[width=0.66\textwidth]{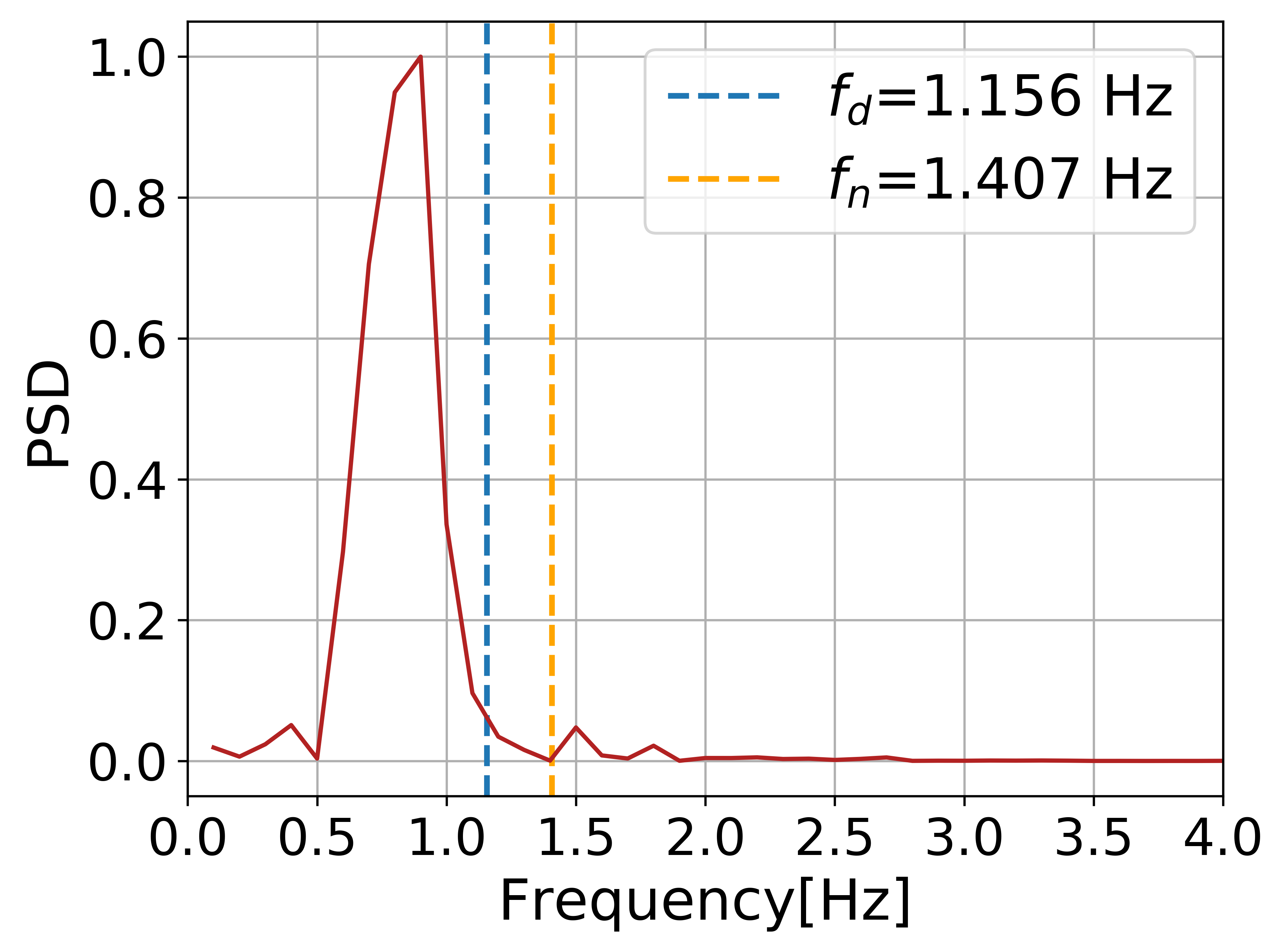}
			\caption{Free  surface $K=0.75$}
		\end{subfigure}\hfill
		\begin{subfigure}[b]{0.5\textwidth}
			\centering
			%	\subfigure[$Pitch$ comparison ]{%
			\includegraphics[width=0.66\textwidth]{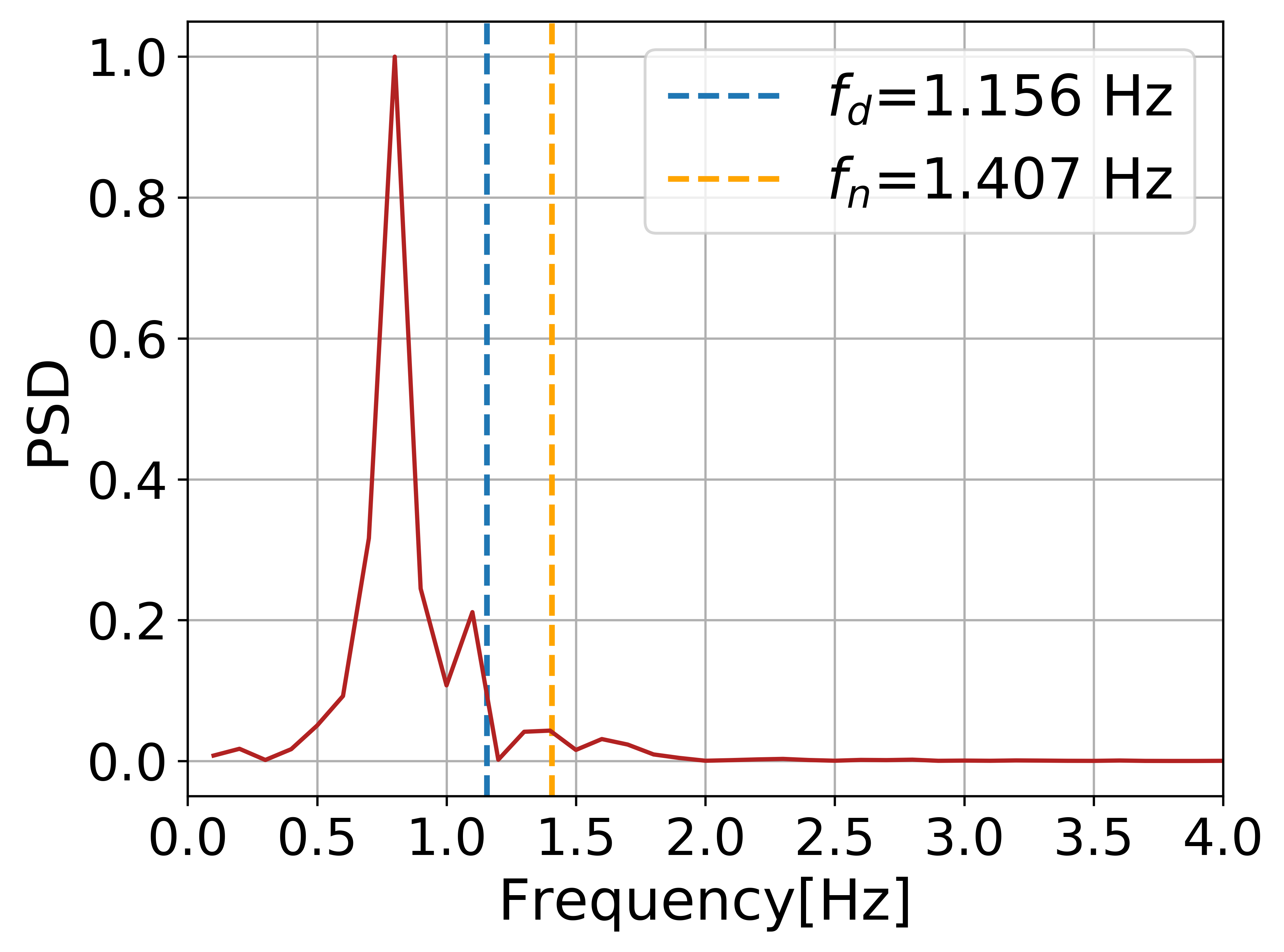}
			\caption{Free surface $K=1.00$}
		\end{subfigure}
	
		\caption{Power Spectral Density (PSD) plots for heave velocity  for various shear rates in free surface simulations. By $f_n$ the natural frequency of the system is denoted and by $f_d$ the damped natural frequency in the heave movement.}
		\label{fig:fs_spectrum_comparison}
	\end{figure}

 In order to identify the emerging frequencies from the existence of the free surface we perform a Fourier analysis. %compare the spectrum of $\dot{y}$ between one phase and free surface simulations of the same shear rate. 
In \autoref{fig:fs_spectrum_comparison} the spectrum of $\dot{y}$ can be seen for the various shear rates. The reason behind choosing $\dot{y}$ lies in the strong relation between power and efficiency, since those quantities are directly proportional to $\dot{y}$ according to \autoref{eq:efficiency_passive} and \autoref{eq:cp_passive}.

In \autoref{fig:fs_spectrum_comparison} we notice the primary frequency of $\dot{y}$ drifting to lower frequencies than the heave damped natural frequency, $f_d$. However for $K=0.2$ this frequency almost coincides with $f_d$ and energy is concentrated in a narrow band close to this frequency. Low energy harmonics vanish from the PSD of $K=0.2$ while they remain present in other shear rates cases including the uniform flow. This observation could help explain the discrete vortex street formed for $K=0.2$ since a unique frequency dominates and vortices shed at a constant rate. We can also assume that for $K=0.2$ a resonance point exists since the primary frequency converges to $f_d$ for this shear rate while it diverges for the remaining values of $K$. In contrast with the one phase simulation PSD plots, \autoref{fig:one_phase_spectrum}, the shear rate strongly affects the primary oscillating frequency. 

Plotting the efficiency and the Power Coefficient for the shear rates examined we confirm that $K=0.2$ is indeed a resonance point for the above performance indicators. The efficiency and the Power Coefficient increase by 24.78\% and 23.52\% respectively in relation with the uniform flow. We further notice that the overall efficiency of the airfoil significantly deteriorates when the airfoil operates under a free surface. For the uniform case, efficiency and Power Coefficient decrease for as much as 20\% and 19.4\% respectively in comparison with one phase simulations. Despite performance plummeting for shear rates larger than $K=0.2$, a plateau seems to form for $K>0.7$ leveling of in a low value of 0.20 for efficiency and 0.52 for Power Coefficient. 

\begin{figure}[H]
	    \centering
		 \includegraphics[width=0.7\textwidth]{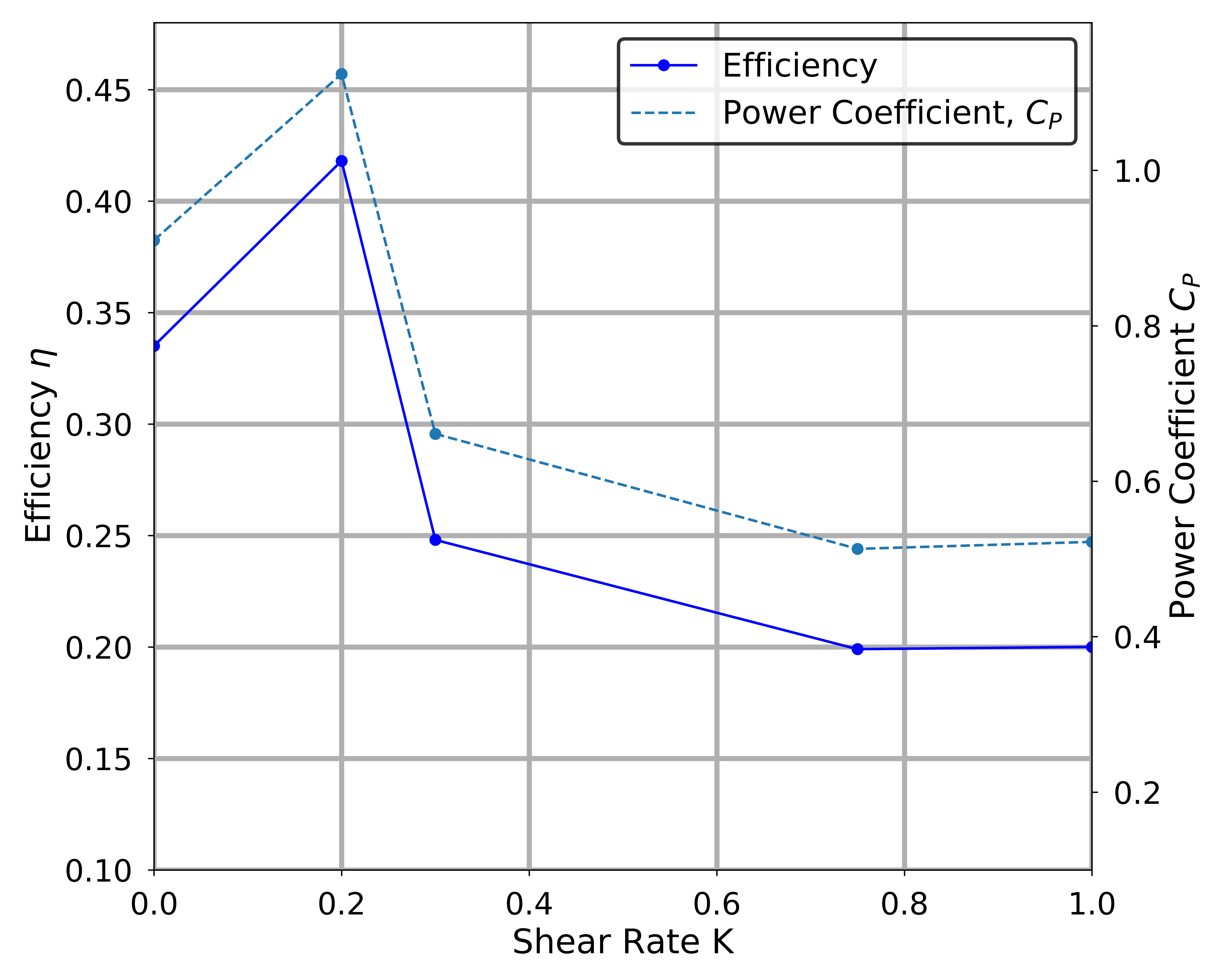}
		\caption{Efficiency and power coefficient for different shear rates in free surface simulations}
		\label{fig:eff_cp_fs}
	\end{figure}

% In the free surface case we notice a high amplitude spike emerging at a frequency slightly higher than the main oscillation frequency of $f=1.2\ Hz$. This amplitude reaches its highest value at $K=0.3$ and then decreases until $K=1.0$. It is also evident that the energy spreads in a wide range of frequencies. This observation indicates that energy is not contained in a few harmonics but spreads over a broad range. This phenomenon is present at every shear rate except $K=0.2$ and $K=0.75$ indicating that these shear rate are favorable for a resonance-like behavior of the vortex street. Although this is obvious in the vorticity contour of $K=0.20$ \nolinebreak \autoref{fig:passive_free_surface_shear_vorticity} this is not immediately visible in the case of $K=0.75$. Identifying the optimum shear rate at which the energy is concentrated in a narrow band of frequencies can lead to more efficient energy extraction.

We proceed to investigate the trends of kinematic quantities and hydrodynamic coefficients with regard to the shear rate.

\begin{figure}[H]
		\begin{subfigure}{0.38\textwidth}
		\includegraphics[width=\textwidth]{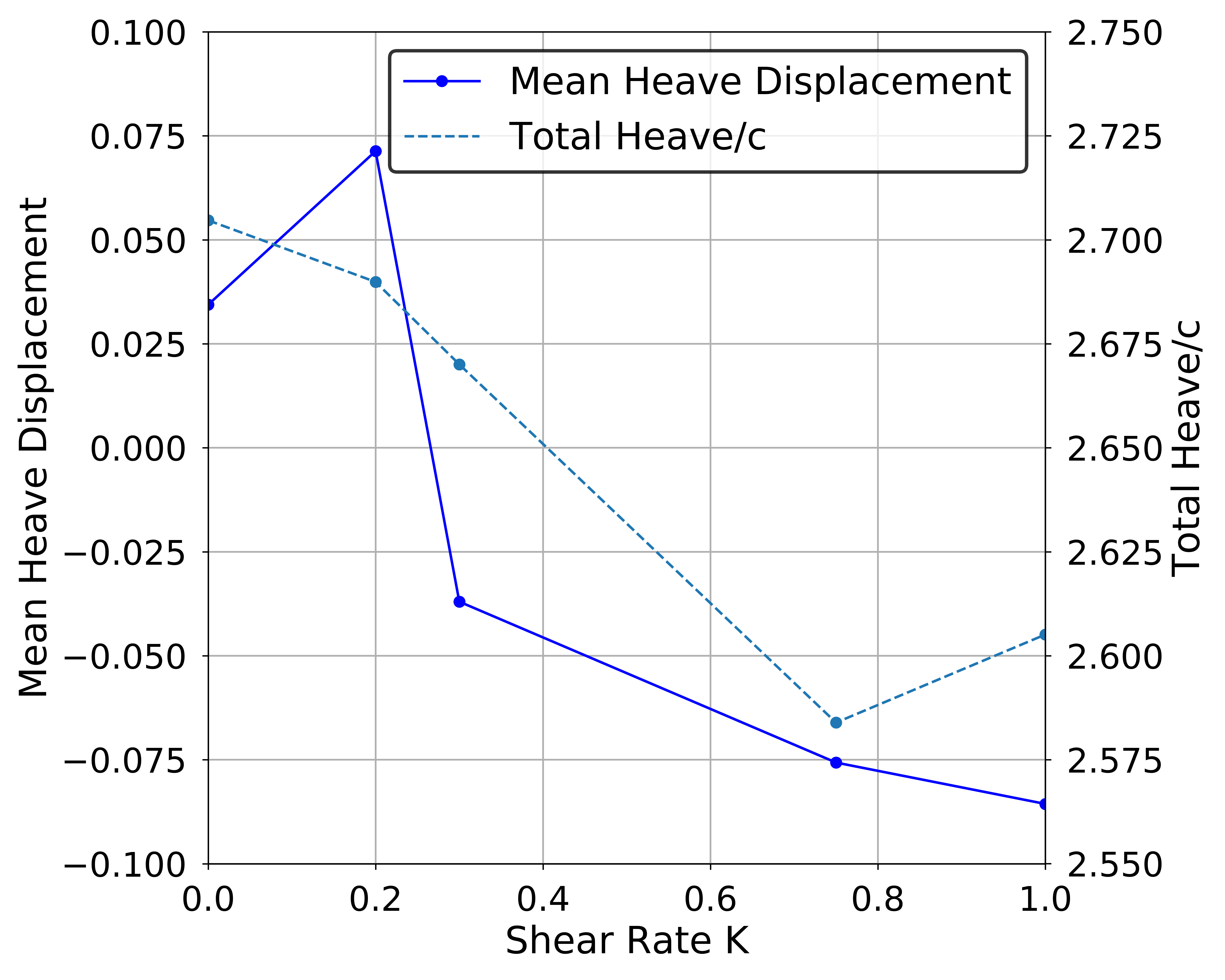}
		\caption{Shift of mean heave position and total heave distance scanned for different shear rates}
		\end{subfigure}\hfill
        \begin{subfigure}{0.30\textwidth}
		\includegraphics[width=\textwidth]{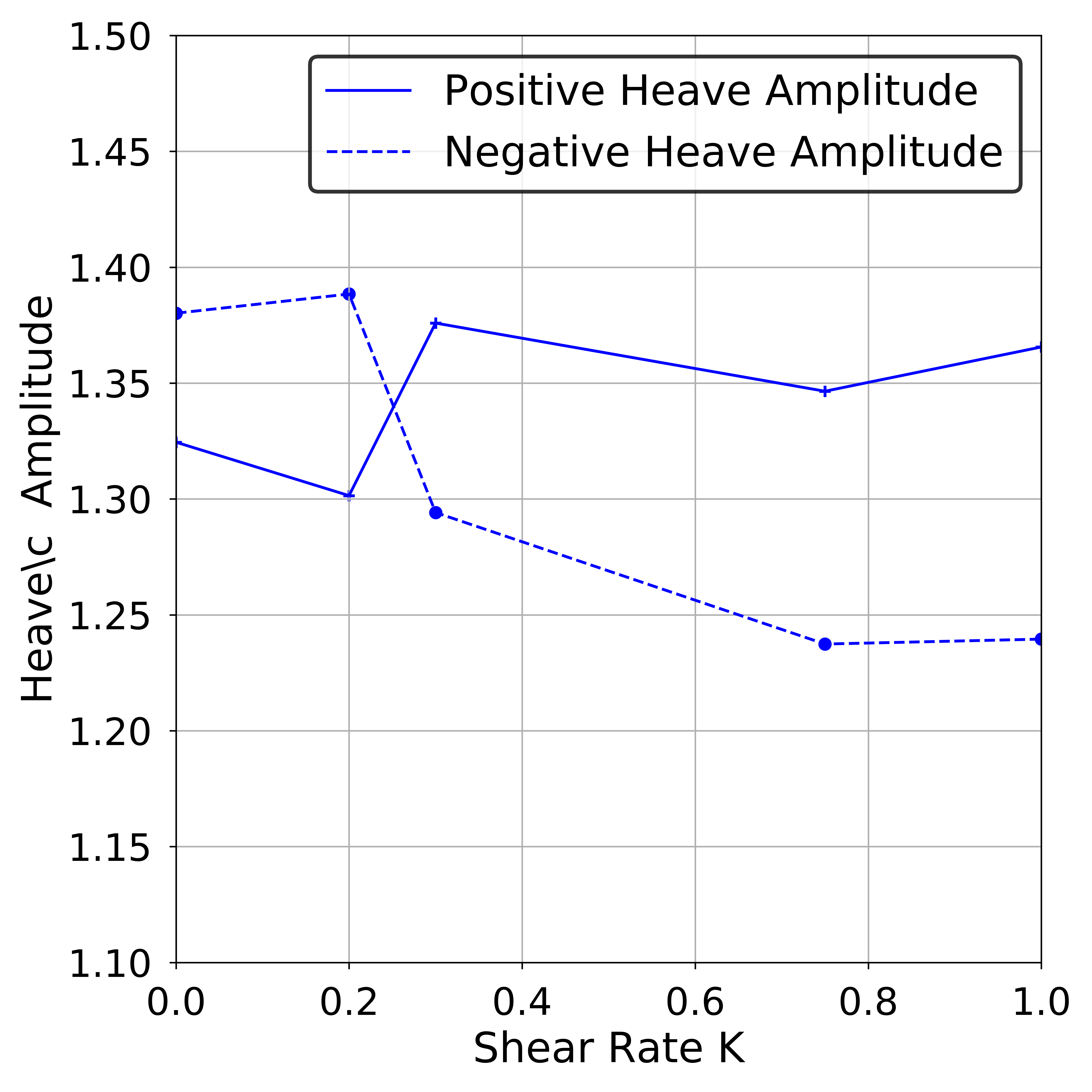}
		\caption{Positive and negative heave amplitudes for different shear rates}
		\end{subfigure}\hfill
		\begin{subfigure}{0.30\textwidth}
		\includegraphics[width=\textwidth]{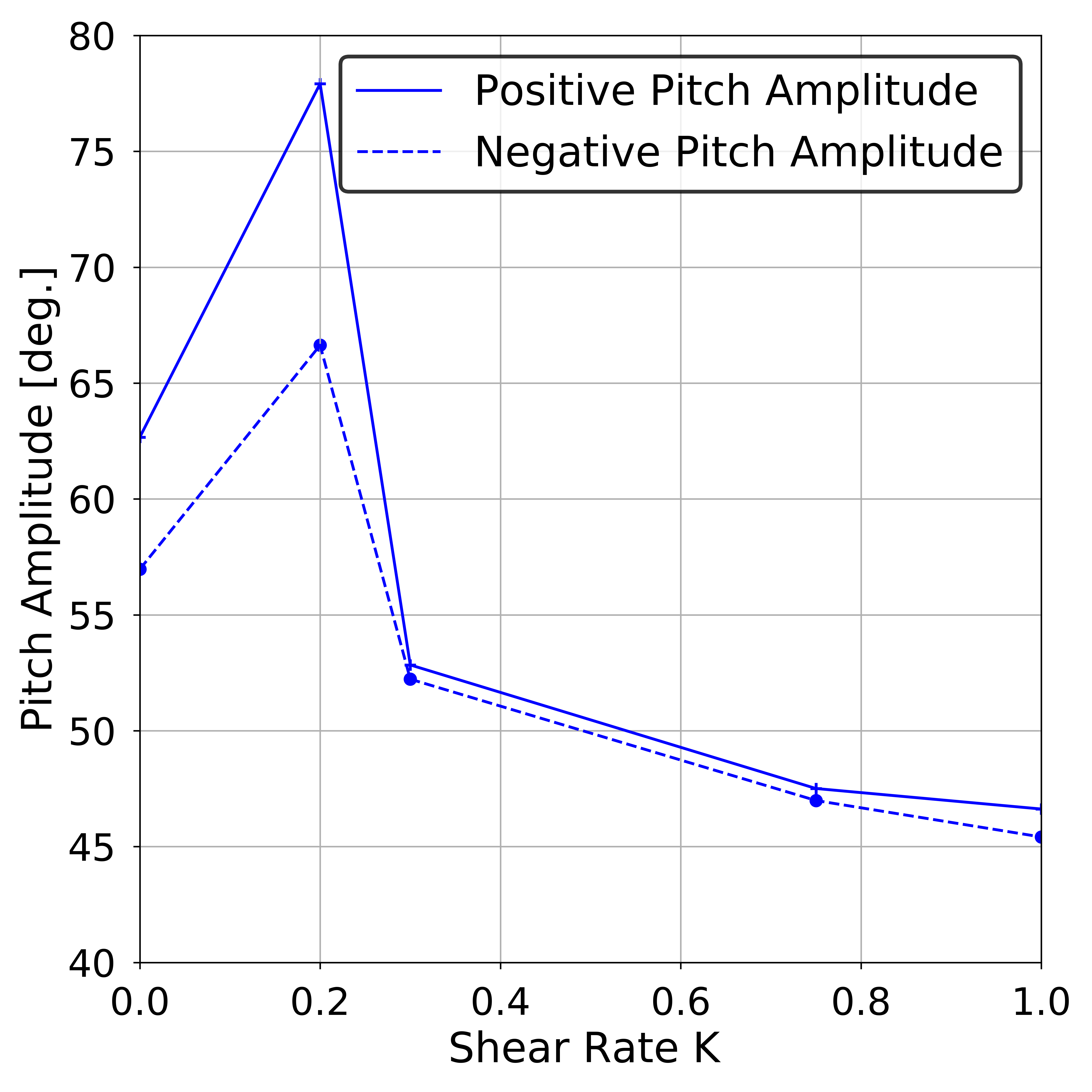}
			\caption{Positive and negative pitch amplitudes for different shear rates}
		\end{subfigure}
			\caption{Comparison of the kinematic quantities of the passive foil for various shear rates in free surface simulations}
		\label{fig:minmax_heave__fs}
\end{figure}

In contrast with \autoref{fig:heave_displacement} we notice an increase of the mean heave position for uniform flow and shear rate of $K=0.2$ possibly due to the free surface creating a suction effect pulling the foil upwards. However, at larger shear rates, greater than $K=0.2$, the mechanism described in \autoref{sec:shear} dominates and drives the foil in negative mean heave values. Concerning the heave amplitudes, the positive amplitude slightly decreases and the negative amplitude slightly increases for $K=0.2$ while for larger shear rates this trend is reversed.  However the total heave distance scanned by the foil decreases steadily until $K=0.75$ and does not exhibit any particular trend change near $K=0.2$.  

Pitch amplitudes in their turn highlight the importance of $K=0.2$ since they reach a peak at that exact shear rate. Larger pitch amplitudes contribute to the initiation of the deep dynamic stall responsible for the creation of the LEV. This LEV possibly triggers a highly unstable wake similar to the one described by \cite{cho2014performance},\cite{schouveiler2005performance},\cite{triantafyllou1991wake},\cite{zhu2011optimal} thus maximizing performance. \linebreak 
It is apparent that the shear rate has a direct impact on the oscillating frequency of the foil and is able to shift this frequency away from the damped natural frequency if the structural parameters of the foil are not re-tuned correctly. 
\section{Conclusions \label{sec:conclusions}}

A fully passive energy harvesting ﬂapping foil has been investigated using the in-house CFD code MaPFlow in uniform and sheared inflow conditions with and without the free surface. Initially MaPFlow predictions for  uniform ﬂow were compared with  experimental and numerical results from \cite{Duarte2021} and good agreement was found.

Afterwards, the study was extended  for sheared  inflow for the same foil configuration. Various sheared rates ($K$) were considered and an optimal shear rate of $K=0.75$ was identiﬁed for the speciﬁc mass-spring-damper set up. Even though the increase in performance was relatively small it is important to stress that optimal behavior does not necessarily occur at uniform ﬂow.

In the final part of this work, two-phase simulations are considered (air-water) and free surface is taken into account. Uniform and sheared inflow is considered for the same mass-spring-damper properties. It was found that when free surface is taken into account a significant deterioration of performance is evident. All the simulations, apart from $K=0.2$, suggested an out of the "lock-in" region operation which had a direct impact on performance. It is noted here, however, that the mass-spring-damper system  {\color{black}was} calibrated in \cite{Duarte2019},\cite{Duarte2021} without taking into account the free surface. A redesign of the structural components of the system might be required when considering such energy harvesting devices operating near the water-air interface.

%Lastly, to investigate the performance of such systems in more realistic conditions,two-phase simulations taking into account both the free surface as well as the seabed must be considered. 

\section*{Acknowledgments}
This work was supported by computational time granted from the Greek Research \& Technology Network (GRNET) in the National HPC facility - ARIS - under project  "SHIPFLOW" with ID  pr010039. The computational grids were generated using the ANSA  pre-processor of BETA-CAE Systems.

\appendix

\section{Mesh Deformation using Radial Basis Functions}

In this work, an RBF  mesh deformation technique\cite{RendallAllen2009} is employed for the mesh to follow the airfoil motion.
Radial Basis Functions (RBF) are used to interpolate the displacements of the inner mesh nodes based on the displacements of the wall nodes of the rigid body. The wall nodes displacement is known from a Rigid Body Dynamics solver and the goal is to deform the surrounding internal nodes in a smooth way.

The interpolated displacement is given in an analytical form :
\begin{align}
s (\mathbf{x}) =\sum_{i=1}^{N} \alpha_i \phi(|| \mathbf{x}-\mathbf{x_{s_i}}||) +p(\mathbf{x})  
\end{align}

In the above formula, $s(x) $ is the the displacement function for the internal node $\mathbf{x}$, $\phi$ is the RBF function of choice and $\mathbf{x}_{s_i}$ are the wall nodes. In this general formula a polynomial term, $p(\mathbf{x})$ is added resulting in full recovery of not only translation but also rotation. As it can be observed the displacement of one internal node depends on the displacement of every wall node, $\mathbf{x}_{s_i}$. In the above summation a coefficient, $\mathbf{a}_i$ serves as a weight for the extend in which each wall node affects the same internal node. Since the RBF in this sum acts upon a norm it is obvious that wall nodes closer to an internal node will affect its motion in a more significant way.

Crucial to the implementation of the RBF method is the calculation of the $\mathbf{a}_i$ coefficients. A linear system is formed after requiring exact recovery of the wall nodes positions when implementing the interpolation scheme to the wall nodes :
\begin{align}
s(\boldsymbol{x}_{s_i}) &=\boldsymbol{X}_{s_i}
\end{align}
An extra requirement is needed when using a polynomial terms and that is :\\

\begin{align}
	\sum_{i=1}^N \alpha_i  q(\mathbf{x}) &=0
	\label{polynomial_condition}
\end{align}

Equation \eqref{polynomial_condition} applies for polynomials with degree less than or equal to the degree of the {RBF} used.

The linear system derived from the above requirements can be written as :
\begin{align}
\boldsymbol{X}_{s_i} &=\boldsymbol{C}_{ss} \boldsymbol{a}_x
\end{align}
, where
\def\arraystretch{1.5}
$$ \mathbf{X_{s_i}}= \left[ \begin{array}{c} 0\\0\\0\\ \mathbf{x_{s_i}} \end{array} \right] ,\,\,\,
\mathbf{x_{s_i}}= \left[ \begin{array}{ccccc} x_{s_1}\\.\\.\\.\\x_{s_N}  \end{array} \right] ,\,\,\,
\mathbf{a_x}= \left[ \begin{array}{ccccccc} \gamma_{0}^x\\ \gamma_{x}^x\\ \gamma_{y}^x\\ \gamma_{z}^x \\ \alpha_{s_1}^x \\.\\.\\.\\ \alpha_{s_N}^x \end{array} \right] $$

and 

$$ \mathbf{C_{ss}} =\left[   \begin{array}{ccccccccc} 0&0&0&1&1&.&.&.&1 \\ 0&0&0&x_{s_1}&x_{s_2}&.&.&.&x_{s_N} \\ 0&0&0&y_{s_1}&y_{s_2}&.&.&.&y_{s_N}  \\ 1& x_{s_1}&y_{s_1}& \phi{_{s{_1}s{_1} }}&\phi{_{s{_1}s{_2} }}&.&.&.&\phi{_{s{_1}s{_N} }} \\ .&.&.&.&.&.&.&.&.\\ .&.&.&.&.&.&.&.&.\\ .&.&.&.&.&.&.&.&. \\ 1& x_{s_N}&y_{s_N}& \phi{_{s{_N}s{_1} }}&\phi{_{s{_N}s{_2} }}&.&.&.&\phi{_{s{_N}s{_N} }}\end{array}  \right]  $$\\

where $\phi{_{s{_i}s{_j} }}=\phi{(||\mathbf{x_{s_i}-x_{s_j} ||)}   }$ and $||.||$ is an appropriate norm of choice.

The matrix $\mathbf{C_{ss}}$ is the coupling matrix between the wall nodes, $x_s{_i}$ and them self, meaning that the position of every wall node is derived from RBF interpolation from the position of every other wall node. The first three rows of $C_{ss}$ constitute the polynomial part of the interpolation and the rest make up for the RBF part.

Having calculated the coefficients $\mathbf{a}_x$ and $\mathbf{a}_y$ we proceed to calculate the interpolation of the position of the internal mesh points, $\mathbf{x}_{a_i}$. A coupling matrix is formed between each internal point and every wall node, $\mathbf{x}_i$ :

$$ \boldsymbol{A}_{as} = \left[ \begin{array}{ccccccccc} 1&x_{\alpha_1}&y_{\alpha_1}&\phi_{a_1s_1}&\phi_{a_1s_2}&.&.&.&\phi_{a_1s_N} \\ .&.&.&.&.&.&.&.&.\\.&.&.&.&.&.&.&.&.\\.&.&.&.&.&.&.&.&.\\1&x_{\alpha_N}&y_{\alpha_N}&\phi_{a_Ns_1}&\phi_{a_Ns_2}&.&.&.&\phi_{a_Ns_N} \end{array} \right]$$
The first three columns constitute the polynomial term and the rest refer to the RBF interpolation.

The new internal points position is then calculated with the RBF interpolation according to the following matrix multiplication :

\begin{align}
	\boldsymbol{x}_a &= \boldsymbol{A}_{as} \boldsymbol{a}_x  \\
	\boldsymbol{y}_a &= \boldsymbol{A}_{as} \boldsymbol{a}_y  
\end{align}
or expressed in terms of $ \mathbf{C}_{ss} $ : 

\begin{align}
	\boldsymbol{x}_a &= \boldsymbol{A}_{as} \boldsymbol{C}_{ss}^{-1} \boldsymbol{X}_s = \boldsymbol{H}\boldsymbol{X}_s  \\
	\boldsymbol{y}_a &= \boldsymbol{A}_{as} \boldsymbol{C}_{ss}^{-1} \boldsymbol{Y}_s =\boldsymbol{H}\boldsymbol{Y}_s 
\end{align}\\
, where $\mathbf{X}_s$ is defined above and  $\mathbf{H} = \mathbf{A}_{as} \mathbf{C}_{ss}^{-1}$ \\
Using the total coupling matrix , $\mathbf{H}$, the above expression can be written as :

\begin{align}
	\left[ \begin{array}{cc} \boldsymbol{x}_a \\ \boldsymbol{y}_a \end{array} \right] = \left[ \begin{array}{cc} \boldsymbol{H}&0 \\ 0&\boldsymbol{H} \end{array} \right]  \left[ \begin{array}{c} \boldsymbol{X}_s \\ \boldsymbol{Y}_s \end{array} \right]
\end{align}
 
In the general case where a polynomial term is used the matrix $\mathbf{C_{ss}}$ can be seen as a block matrix:

\begin{align}
\boldsymbol{C}_{ss}=	\left[ \begin{array}{cc} \boldsymbol{0} &\boldsymbol{P} \\ \boldsymbol{P^T} & \boldsymbol{M} \end{array}   \right]
\end{align}
, where $\mathbf{P}$ is the sub-array containing the polynomial part of the interpolation, $\mathbf{P^T}$ is the transpose of $\mathbf{P}$ and $\mathbf{M}$ is the RBF sub-array containing values of the RBF between different wall nodes. 

Exploiting the zero sub-array that $ \mathbf{C}_{ss} $ exhibits makes its inversion more straightforward. For reasons of further simplicity the coefficients  $\mathbf{a}_x$ and $\mathbf{a}_y$ are decomposed to $\mathbf{a}_x^{polynomial}$ and $\mathbf{a}_x^{RBF}$ so that :

\begin{align}
	\boldsymbol{a}_x=	\left[ \begin{array}{l} \boldsymbol{a}_x^{polynomial} \\ \boldsymbol{a}_x^{RBF}  \end{array}   \right]
\end{align}

The $\mathbf{a}_x^{polynomial}$ and $\mathbf{a}_x^{RBF}$ are calculated as follows, where the inverse of $ \mathbf{C}_{ss} $ is found through its sub-arrays :

\begin{align}
\boldsymbol{a}_x^{Polynomial} &= \boldsymbol{M}_p \boldsymbol{P} \boldsymbol{M}^{-1} \boldsymbol{x}_s\\
\boldsymbol{a}_x^{RBF} &= ( \boldsymbol{M}^{-1} - \boldsymbol{M}^{-1} \boldsymbol{P}^T \boldsymbol{M}_p \boldsymbol{P}\boldsymbol{M}^{-1} ) \boldsymbol{x}_s
\end{align}
, where $\mathbf{M}_p$ is the Schur complement of $\mathbf{M}$ :
\begin{align}
\boldsymbol{M}_p = (\boldsymbol{P} \boldsymbol{M}^{-1} \boldsymbol{P}_T    )^{-1} 
\end{align}                                                       

Using the above expressions we construct the total coupling matrix, $\mathbf{H}$ between the wall nodes and the internal nodes : 
\begin{align}
	\boldsymbol{M}_p = (\boldsymbol{P} \boldsymbol{M}^{-1} \boldsymbol{P}_T    )^{-1} 
\end{align}  

 \begin{align}
\boldsymbol{H} = \boldsymbol{A}_{as} \left[\begin{array}{c}  \boldsymbol{M}_p \boldsymbol{P} \boldsymbol{M}^{-1} \\  \boldsymbol{M}^{-1} - \boldsymbol{M}^{-1} \boldsymbol{P}^T \boldsymbol{M}_p \boldsymbol{P}\boldsymbol{M}^{-1}  \end{array} \right] 
\end{align}   

As mentioned earlier, one of the advantages of RBF interpolation is that it can be easily implemented in every time step by simple matrix multiplications involving the matrix $\mathbf{H}$. $\mathbf{H}$ in his turn is calculated and stored before the simulation starts and thus is always ready for use and no complicated calculations need to take place after that. 
There exist several Radial Basis Functions to be used for the multi-parameter interpolation described above. A common categorization of RBF divides them into functions with compact support and global support. In this work Wedland's $C^2$ function is employed as seen in \autoref{eq:wedc2}

\begin{equation}
    \phi(\xi) = (1-\xi)^4(4\xi+1) 
    \label{eq:wedc2}
\end{equation}
, where $\xi=\frac{x}{r}$ where $r$ is the support radius of our choice.

\bibliographystyle{elsarticle-num} 
\bibliography{main.bib}

\begin{thebibliography}{10}
\expandafter\ifx\csname url\endcsname\relax
  \def\url#1{\texttt{#1}}\fi
\expandafter\ifx\csname urlprefix\endcsname\relax\def\urlprefix{URL }\fi
\expandafter\ifx\csname href\endcsname\relax
  \def\href#1#2{#2} \def\path#1{#1}\fi

\bibitem{Anderson1998}
J.~Anderson, K.~Streilen, D.~Barrett, M.~Triantafyllou, Oscillating foils of
  high propulsive efficiency, Journal of Fluid Mechanics 360 (1998) 41--72.

\bibitem{Wang2000}
Z.~Wang, Vortex shedding and frequency selection in flapping flight, Journal of
  Fluid Mechanics 410 (2000) 323--341.

\bibitem{Triantafyllou2005}
L.~Schouveiler, F.~Hover, M.~Triantafyllou, Performance of flapping foil
  propulsion, Journal of Fluids and Structures 20 (2005) 949--959.

\bibitem{Bose1989}
N.~Bose, Lien.J., Propulsion of a fin whale (balaenoptera physalus): why the
  fin whale is a fast swimmer, Proceedings of the Royal Society of London B 237
  (1989) 175--200.

\bibitem{McKinney1981}
W.~McKinney, J.~DeLaurier, Wingmill: An oscillating-wing windmill, Journal
  Energy 5 (2) (1981) 109--115.

\bibitem{KinseyDumas2008}
T.~Kinsey, G.~Dumas, Parametric study of an oscillating airfoil in power
  extraction regime, AIAA Journal 46 (2008) 1318--1330.

\bibitem{Kinsey2011}
T.~Kinsey, G.~Dumas, G.~Lalande, J.~Ruel, A.~Méhut, P.~Viarouge, J.~Lemay,
  Y.~Jean, Prototype testing of a hydrokinetic turbine based on oscillating
  hydrofoils, Renewable Energy 36 (6) (2011) 1710--1718.

\bibitem{Shimizu2008}
E.~Shimizu, K.~Isogai, S.~Obayashi, Multiobjective design study of a flapping
  wing power generator, Journal of Fluids Engineering 130 (2008).

\bibitem{Peng_and_Zhu2009}
Z.~Peng, Q.~Zhu, Mode coupling and flow energy harvesting by a flapping foil,
  Physics of Fluids 21 (2009).

\bibitem{Stingray2002}
E.~B. Ltd., Research and development of a 150kw tidal stream generator,
  Technical Report (2002).

\bibitem{Zhu_flow_induced2009}
Z.~Peng, Q.~Zhu, Energy harvesting through flow-induced oscillations of a foil,
  Physics of Fluids 21 (12) (2009).

\bibitem{Dumas2017}
J.~Veilleux, G.~Dumas, Numerical optimization of a fully-passive
  flapping-airfoil turbine, Journal of Fluids and Structures 70 (2017)
  102--130.

\bibitem{Duarte2019}
L.~Duarte, N.~Dellinger, G.~Dellinger, A.~Ghenaim, A.~Terfous, Experimental
  investigation of the dynamic behaviour of a fully passive flapping foil
  hydrokinetic turbine, Journal of Fluids and Structures 88 (2019) 1--12.

\bibitem{Simpson2008}
B.~Simpson, F.~Hover, M.~Triantafyllou, Energy exctraction through flapping
  foils, 27th International Conference on Offshore Mechanics and Arctic
  Engineering OMAE-58043 (2008).

\bibitem{Deng2022}
J.~Deng, S.~Wang, P.~Kandel, L.~Teng, {Effects of free surface on a
  flapping-foil based ocean current energy extractor}, Renewable Energy 181
  (2022) 933--944.
\newblock \href {https://doi.org/10.1016/j.renene.2021.09.098}
  {\path{doi:10.1016/j.renene.2021.09.098}}.

\bibitem{Zhu2022}
B.~Zhu, W.~Cheng, J.~Geng, J.~Zhang, {Energy-harvesting characteristics of
  flapping wings with the free-surface effect}, Journal of Renewable and
  Sustainable Energy 14~(1) (2022).

\bibitem{Liu2013}
W.~Liu, Q.~Xiao, F.~Cheng, A bio-inspired study on tidal energy extraction with
  lexible flapping wings, Bioinspiration and Biomimetics 8 (2013) 1--16.

\bibitem{Shoele2013}
K.~Shoele, Q.~Zhu, Performance of a wing with nonuniform flexibility in
  hovering flight, Physics of Fluids 25 (2013).

\bibitem{Cho2014}
H.~Cho, Q.~Zhu, Performance of a flapping foil flow energy harvester in shear
  flows, Journal of Fluids and Structures 51 (2014) 199--20.

\bibitem{Liu2019}
Z.~Liu, H.~Qu, H.~Shi, Performance evaluation and enhancement of a
  semi-activated flapping hydrofoil in shear flows, Energy 189 (2019).

\bibitem{Zhu2012}
Q.Zhu, Energy harvesting by a purely passive flapping foil from shear flows,
  Journal of Fluids and Structures 34 (2012) 157--169.

\bibitem{Duarte2021}
L.~Duarte, G.~Dellinger, N.~Dellinger, A.~Ghenaim, A.~Terfous, Implementation
  and validation of a strongly coupled numerical model of a fully passive
  flapping foil turbine, Journal of Fluids and Structures 102 (2021).

\bibitem{Papadakis2014}
G.~Papadakis, {Development of a hybrid compressible vortex particle method and
  application to external problems including helicopter flows}, Ph.D. thesis,
  National Technical University of Athens (2014).

\bibitem{Diakakis2019}
K.~Diakakis, {Computational analysis of transitional and massively separated
  flows with application to Wind Turbines}, Ph.D. thesis, National Technical
  University of Athens (2019).

\bibitem{Karypis2013}
G.~Karypis, {METIS A Software Package for Partitioning Unstructured Graphs,
  Partitioning Meshes, and Computing Fill-Reducing Orderings of Sparse Matrices
  Version}, Tech. rep., Department of Computer Science \& Engineering
  University of Minnesota Minneapolis (2013).

\bibitem{Chorin1967}
A.~Chorin, A numerical method for solving incompressible visous flow problems,
  Journal of Computational Physics 135 (1967) 118--125.

\bibitem{Hirt1981}
C.~Hirt, B.~Nichols, {Volume of Fluid (VOF) Method for the Dynamics of Free
  Boundaries}, Journal of Computational Physics 39 (1981) 201--225.

\bibitem{JAMESON1991}
A.~Jameson, Time dependent calculations using multigrid, with applications to
  unsteady flows past airfoils and wings, in: 10th Computational Fluid Dynamics
  Conference, 1991.
\newblock \href {https://doi.org/10.13140/2.1.2459.3608}
  {\path{doi:10.13140/2.1.2459.3608}}.

\bibitem{Ntouras2020}
D.~Ntouras, G.~Papadakis, {A Coupled Artificial Compressibility Method for Free
  Surface Flows}, Journal of Marine Science and Engineering 8~(8) (2020) 590.

\bibitem{Dudley2002}
D.~S. Nichols, {Development of a free surface method utilizing an
  incompressible multi-phase algorithm to study the flow about surface ships
  and underwater vehicles}, Ph.D. thesis, Faculty of Mississippi State
  University (2002).

\bibitem{Venkateswaran2001}
S.~Venkateswaran, J.~Lindau, R.~Kunz, C.~Merkle, {Preconditioning algorithms
  for the computation of multi-phase mixture flows}, in: 39th Aerospace
  Sciences Meeting and Exhibit, 2001, pp. 1--14.

\bibitem{Kunz2000}
R.~Kunz, D.~Boger, D.~Stinebring, T.~Chyczewski, J.~Lindau, H.~Gibeling,
  S.~Venkateswaran, T.~Govindan, A preconditioned navier-stokes method for
  two-phase flows with application to cavitation prediction, Computers and
  Fluids 29 (2000) 849--875.

\bibitem{Menter1994}
F.~Menter, Two-equation eddy-viscosity turbulence models for engineering
  applications, AIAA 32 (1994) 1598--1605.

\bibitem{Larsen2018}
B.~E. Larsen, D.~R. Fuhrman, {On the over-production of turbulence beneath
  surface waves in Reynolds-averaged Navier-Stokes models}, Journal of Fluid
  Mechanics 853 (2018) 419--460.

\bibitem{Kamath2015}
A.~Kamath, H.~Bihs, M.~{Alagan Chella}, {\O}.~A. Arntsen, {CFD Simulations of
  Wave Propagation and Shoaling over a Submerged Bar}, Aquatic Procedia 4
  (2015) 308--316.
\newblock \href {https://doi.org/10.1016/j.aqpro.2015.02.042}
  {\path{doi:10.1016/j.aqpro.2015.02.042}}.

\bibitem{Devolder2017}
B.~Devolder, P.~Rauwoens, P.~Troch, Application of a buoyancy-modified k-omega
  sst turbulence model to simulate wave run-up around a monopile subjected to
  regular waves using openfoam, Coastal Engineering 125 (2017) 81--94.

\bibitem{Peric2015}
R.~Peric, M.~Abdel-Maksoud, Reliable damping of free surface waves in numerical
  simulations, Ship Technology Research 63 (05 2015).
\newblock \href {https://doi.org/10.1080/09377255.2015.1119921}
  {\path{doi:10.1080/09377255.2015.1119921}}.

\bibitem{Roe1981}
P.~Roe, Approximate riemann solvers, parameter vectors, and difference schemes,
  Journal of Computational Physics 43 (1981) 357--372.

\bibitem{Queutey2007}
P.~Queutey, M.~Visonneau, An interface capturing method for free-surface
  hydrodynamic flows, Computers and Fluids 36 (2007) 1481--1510.

\bibitem{Leonard1988}
B.~Leonard, Simple high-accuracy resolution program for convective modelling of
  discontinuities, International Journal for Numerical Methods in Fluids 8
  (1988) 1291--1318.

\bibitem{Wackers2011}
J.~Wackers, B.~Koren, H.~Raven, A.~van~der Ploeg, A.~Starke, G.~Deng,
  P.~Queutey, M.~Visonneau, T.~Hino, K.~Ohashi, Free-surface viscous flow
  solution methods for ship hydrodynamics, Archives of Computational Methods in
  Engineering 18 (2011) 1--41.

\bibitem{Biedron2005}
R.~Biedron, V.~Vatsa, H.~Atkins, {Simulation of Unsteady Flows Using an
  Unstructured Navier-Stokes Solver on Moving and Stationary Grids}, 23rd AIAA
  Applied Aerodynamics Conference (2005) 1--17.

\bibitem{de2019conception}
L.~de~Carvalho~Duarte, Conception et optimisation d'un syst{\`e}me hydrolien
  {\`a} aile oscillante passive, Ph.D. thesis, Strasbourg (2019).

\bibitem{cho2014performance}
H.~Cho, Q.~Zhu, Performance of a flapping foil flow energy harvester in shear
  flows, Journal of Fluids and Structures 51 (2014) 199--210.

\bibitem{schouveiler2005performance}
L.~Schouveiler, F.~Hover, M.~Triantafyllou, Performance of flapping foil
  propulsion, Journal of fluids and structures 20~(7) (2005) 949--959.

\bibitem{triantafyllou1991wake}
M.~Triantafyllou, G.~Triantafyllou, R.~Gopalkrishnan, Wake mechanics for thrust
  generation in oscillating foils, Physics of Fluids A: Fluid Dynamics 3~(12)
  (1991) 2835--2837.

\bibitem{zhu2011optimal}
Q.~Zhu, Optimal frequency for flow energy harvesting of a flapping foil,
  Journal of fluid mechanics 675 (2011) 495--517.

\bibitem{RendallAllen2009}
T.~Rendal, C.~Allen, Efficient mesh motion using radial basis functons with
  data reduction algorithms, Journal of Computational Physics 228 (2009)
  6231--6249.

\end{thebibliography}

\end{document}